\documentclass[a4paper,11pt]{article}
\usepackage{jheppub}
\usepackage[T1]{fontenc} 

\usepackage{subfigure}
\usepackage{float}
\def\beq{\begin{equation}}
\def\eeq{\end{equation}}

\usepackage{graphicx}
\usepackage{amsmath}
\usepackage{amssymb}
\usepackage{amsthm}
\usepackage{tikz}
\usepackage{mathrsfs}
\usepackage{multirow}
\usepackage{scalerel}
\usepackage{mathtools}
\usepackage{textcomp}
\usepackage{dsfont}
\usepackage{blkarray}
\usepackage{bm}
\usepackage{todonotes}
\usepackage{hyperref}
\usepackage{pgf-pie}
\usepackage{slashed}
\usepackage[compat=1.1.0]{tikz-feynman}
\allowdisplaybreaks[4]
\graphicspath{{Figures}}
\usepackage{caption}
\usepackage{longtable}

\def\cL{\mathcal{L}}
\def\cM{\mathcal{M}}
\def\cN{\mathcal{N}}
\def\cO{\mathcal{O}}
\def\cS{\mathcal{S}}

\def\cW{\mathcal{W}}

\def\re{\text{Re}}
\def\im{\text{Im}}

\newcommand{\nn}{\nonumber}
\newcommand{\df}{\mathrm{d}}

\definecolor{darkyellow}{rgb}{0.5, 0.5, 0.0}
\definecolor{darkpurple}{rgb}{0.5, 0.2, 0.8}
\definecolor{darkblue}{rgb}{0.0, 0.0, 0.8}
\definecolor{darkgreen}{rgb}{0.0, 0.4, 0.0}
\definecolor{darkred}{rgb}{0.5, 0.0, 0.0}
\definecolor{or}{rgb}{0.88,0.43,0.02}
\definecolor{darkgreen}{rgb}{0.13,0.55,0.13}

\hypersetup{
	linktocpage,
	colorlinks,
	citecolor=darkred,
	linkcolor= darkred,
	urlcolor=darkred
}

\usepackage[normalem]{ulem}
\usepackage{marginnote}

\title{\boldmath Projected Energy Correlators: \\
Two-Loop Jet Functions and NNLL Resummation}

\author[a]{Kyle Lee,}
\author[b]{Yibei Li,}
\author[c]{Zhen Xu,}
\author[d,e]{and Xiaoyuan Zhang}

\affiliation[a]{High Energy Physics Division, Argonne National Laboratory, Lemont, IL 60439, USA}

\affiliation[b]{PRISMA$^+$ Cluster of Excellence Mainz Institute for Theoretical Physics, Johannes Gutenberg University, Staudingerweg 9, 55128 Mainz, Germany}

\affiliation[c]{Deutsches Elektronen-Synchrotron DESY, Notkestr. 85, 22607 Hamburg, Germany}

\affiliation[d]{Department of Physics, Harvard University, Cambridge, MA 02138, USA}

\affiliation[e]{Center for Theoretical Physics - a Leinweber Institute, Massachusetts Institute of Technology, Cambridge, MA 02139, USA}

\emailAdd{kyle@anl.gov}
\emailAdd{yibei.li@uni-mainz.de}
\emailAdd{zhen.xu@desy.de}
\emailAdd{xyz2@mit.edu}

\preprint{\vbox{%
		\hbox{MITP-26-021}
		\hbox{DESY-26-069}
		\hbox{MIT-CTP 6028}
}}

\abstract{
	We present the next-to-next-to-leading logarithmic (NNLL) collinear resummation of projected $N$-point energy correlators (ENCs) up to $N=6$, matched to fixed-order predictions at NLO, in both electron-positron annihilation and Higgs decay to gluons. The key new ingredient is the two-loop jet function for $N=4,5,6$, which we compute semi-analytically using Integration-by-Parts and differential equations.
    We further include the leading non-perturbative corrections for ENCs, described by two universal soft matrix elements ${\overline{\Omega}_{1q},\overline{\Omega}_{1g}}$ of order $\Lambda_{\rm QCD}$, 
    whose evolution is governed by anomalous dimensions for  $(N-1)$-point correlators.
    The matched distributions are compared with parton-shower simulations from \textsc{Pythia8} and \textsc{Herwig7}, and we study the sensitivity of both the absolute spectra and their ratios to the two-point energy correlator under variations of $\alpha_s$ and $\overline{\Omega}_{1q,1g}$. Our results show that higher-point projected energy correlators are now under quantitative control at NNLL accuracy, opening the door to future $\alpha_s$ extractions with complementary systematics.
}

\begin{document} 
\maketitle
\pagebreak

\newpage

\section{Introduction}\label{sec:intro}

An important goal of quantum chromodynamics (QCD) is to identify observables that are simultaneously sensitive to the underlying quark--gluon dynamics and under sufficient theoretical control to enable precision studies. Energy correlators are particularly attractive in this respect. As correlation functions of energy flow operators $\mathcal{E}(\vec{n})$~\cite{Sterman:1975xv,Sveshnikov:1995vi,Tkachov:1995kk,Korchemsky:1999kt}, they probe the flow of energy produced in a collision, are directly measureable in experiment, and admit a clean operator definition in the underlying field theory. This gives them a unique status as observables that connect precision collider phenomenology with fundamental structures in quantum field theory~\cite{Hofman:2008ar,Kologlu:2019mfz}; see Ref.~\cite{Moult:2025nhu} for a recent review.
The two-point energy correlator, or energy-energy correlation (EEC), proposed in the late 1970s~\cite{Basham:1978bw,Basham:1978zq,Basham:1979gh,Basham:1977iq}, measures the correlation of the energy deposited in two detectors as a function of their angular separation, and has played a central role in the development of perturbative QCD calculations. Compared with many classic event-shape observables, the EEC enjoys remarkable analytic structure: it has been computed analytically through next-to-next-to-leading order (NNLO) in $\cN=4$ super Yang-Mills (SYM) theory~\cite{Belitsky:2013ofa,Henn:2019gkr} and through NLO in QCD for both $e^+e^-$ annihilation~\cite{Dixon:2018qgp} and hadronic Higgs decays~\cite{Luo:2019nig,Gao:2020vyx}. This calculability, combined with its operator interpretation, makes the EEC a uniquely transparent probe of QCD dynamics across different scales.

Energy correlators also have a long experimental history. During the LEP era, the EEC was measured by experiments such as ALEPH and DELPHI~\cite{ALEPH:1990vew,DELPHI:1990sof}, establishing it as a classic observable for precision QCD studies. More recently, the subject has entered a qualitatively new stage through the re-analysis of archival LEP data by the Electron-Positron Alliance. In particular, track-based measurements using the ALEPH and DELPHI detectors have achieved unprecedented angular resolution~\cite{Bossi:2024qeu,Bossi:2025xsi,Electron-PositronAlliance:2025fhk,Zhang:2025nlf}, greatly extending the kinematic reach over which energy correlators can be studied. These re-analyses provide direct motivation for sharpening the theoretical description of energy correlators.

A particularly interesting regime of energy correlators is their collinear limit, in which all the energy flow operators have small relative angles. This is the natural domain of the light-ray operator product expansion (OPE)~\cite{Hofman:2008ar,Kologlu:2019mfz}, which dictates the characteristic angular scaling of the correlators in this region. For the two-point energy correlator, the scaling with the angular separation is controlled by the anomalous dimension of twist-two light-ray operators of spin three (up to corrections from the running of the QCD coupling)~\cite{Hofman:2008ar,Dixon:2019uzg,Korchemsky:2019nzm}. This structure generalizes naturally to higher points. For $N$-point correlator, there are $2N-3$ independent variables, which under the collinear limit, can be split into an overall angular scale and $2N-4$ dimensionless cross ratios specifying the shape of the $N$-point configuration. Holding the shape fixed and dialing the overall size produces an angular scaling controlled by the anomalous dimensions of twist-two operators of spin $N+1$. The fully differential three-point correlator has been computed at leading order in $\cN=4$ SYM~\cite{Yan:2022cye}, QCD~\cite{Chen:2019bpb,Yang:2022tgm}, and hadronic Higgs decay~\cite{Yang:2024gcn}. The four-point correlator has been computed to leading order only in $\cN=4$ SYM~\cite{Chicherin:2024ifn}. No higher-point fully differential calculation currently exists though the necessary integrand has been computed to 11-point in $\cN=4$ SYM~\cite{He:2024hbb}. Explicitly deriving the twist-two spin-$(N+1)$ anomalous scaling itself with fixed shape, however, requires NLO computations of these fully differential observables, which have not yet been carried out.

The projected $N$-point energy correlator (ENC), introduced in Ref.~\cite{Chen:2020vvp}, realizes the spin-$(N+1)$ scaling in the simplest way by only measuring the largest pairwise angle $x_L$ of the $N$-point energy correlation and integrating over all other shape variables, which dramatically simplifies higher-order calculations compared to the fully differential observable. The resulting one-dimensional distribution retains the full collinear scaling structure and is convenient for both precision calculations and experimental measurements~\cite{Komiske:2022enw}. In the small-angle limit $x_L \to 0$, incomplete cancellation of infrared singularities generates logarithmically enhanced contributions of the form $\alpha_s^n \ln^n x_L$. Resumming these logarithms to all orders is essential, and up to corrections from the running of the QCD coupling it exponentiates into the spin-$(N+1)$ scaling dictated by the light-ray OPE.

The theoretical framework for this collinear resummation based on QCD factorization is by now well developed. For the EEC, the collinear limit factorization into hard and jet functions was established in Ref.~\cite{Dixon:2019uzg}, and this structure was generalized to projected $N$-point correlators in Ref.~\cite{Chen:2020vvp}.\footnote{The factorization was also extended to the so-called confinement region where $x_L \ll \Lambda_{\rm QCD}/Q$~\cite{Lee:2025okn,Chang:2025kgq,Kang:2025zto}.} In this framework, the hard functions are governed by DGLAP evolution and are universal across the family of projected correlators, whereas the jet functions encode the different $N$-point dependent collinear dynamics. Collinear resummation for the $N=2$ has reached NNLL accuracy in QCD and $\cN=4$ SYM~\cite{Dixon:2019uzg,Kologlu:2019mfz,Korchemsky:2019nzm}. For $N>3$, LL and NLL predictions were developed in Refs.~\cite{Chen:2020vvp,Lee:2022uwt}. The first NNLL result beyond $N=2$ was obtained for the $N=3$ case, where the two-loop jet function was computed in Ref.~\cite{Chen:2023zlx}. 

There are several reasons why extending this program to even more $N$-point projected correlators is important. First, a family of observables with different $N$ provides multiple, correlated probes of the same underlying perturbative dynamics. This makes projected energy correlators a particularly promising framework for precision studies, since one can test the stability of the extracted physics across observables with related but non-identical sensitivity. Second, these observables offer systematics complementary to traditional event shapes in $e^+e^-$ collisions and to standard jet-substructure measurements. This is especially timely in light of the longstanding tension in $\alpha_s$ determinations, where lattice QCD yields $\alpha_s(m_Z)=0.1183(7)$~\cite{FlavourLatticeAveragingGroupFLAG:2024oxs}, while field-theoretic analyses of precision event shapes such as thrust, $C$-parameter, and heavy jet mass prefer smaller values~\cite{Abbate:2010xh,Becher:2008cf,Hoang:2015hka,Hoang:2014wka,Benitez:2024nav,Benitez:2025vsp}; see Refs.~\cite{Huston:2023ofk,dEnterria:2022hzv} for reviews. The CMS collaboration has already extracted $\alpha_s$ from the projected three-point correlator measured inside hadron-collider jets~\cite{CMS:2024mlf}, demonstrating the experimental viability of this program. We view higher-point projected correlators in $e^+e^-$ annihilation and in $H\to gg$ decay, the latter being directly relevant for future Higgs factories such as FCC-ee and CEPC~\cite{FCC:2018evy,CEPCStudyGroup:2018ghi}, as a realistic route toward competitive $\alpha_s$ extractions with genuinely different theoretical and experimental systematics. Third, higher-point projected correlators provide a new handle on non-perturbative physics. Understanding hadronization corrections is essential for turning precision perturbative calculations into precision phenomenology. The leading non-perturbative corrections~\cite{HaoTalkSCET,Lee:2024esz,Chen:2024nyc} for projected energy correlators are governed by the same soft matrix elements $\overline{\Omega}_{1\kappa}$ across different $N$~\cite{Korchemsky:1999kt,Lee:2006fn}, related by a calculable $N$-dependent prefactor. This makes the family of projected correlators a natural setting to extract $\overline{\Omega}_{1\kappa}$ and to help lift their degeneracy with $\alpha_s$.

In this paper, we report the NNLL collinear resummation for projected $ N=2-6$-point correlators in both $e^+e^-$ annihilation and in the hadronic Higgs decay $H\to gg$.
To achieve so, we compute the two-loop jet function for the projected $N$-point energy correlator with $N=4,5,6$. On top of the perturbative resummation, we also include the leading non-perturbative corrections in both the collinear limit and the bulk region, and compute their ratios to EEC. With these results at hand, we also demonstrate the potential applications of projected energy correlators to precision phenomenology.

The outline of this paper is as follows. In Sec.~\ref{sec:pENC}, we review the definition of projected $N$-point energy correlators and present their fixed-order calculations. In Sec.~\ref{sec:factorization}, we review the collinear factorization theorem and summarize the ingredients required for NNLL resummation. In Sec.~\ref{sec:jet_func}, we compute the new two-loop jet functions for $N=4,5,6$ projected energy correlators. In Sec.~\ref{sec:nnll_resum}, we present matched NNLL predictions for $e^+e^-$ annihilation and Higgs decay to gluons, including non-perturbative corrections and parton-shower comparisons. In Sec.~\ref{sec:pheno}, we analyze the sensitivity of these observables to $\alpha_s$ and non-perturbative parameters. We conclude in Sec.~\ref{sec:conclusion}.

\section{Projected $N$-point energy correlators}
\label{sec:pENC}

In this section, we give an explicit definition of the projected $N$-point energy correlator (Sec.~\ref{sec:pENCdef}) and present both the LO (analytic) and NLO (numerical) fixed-order predictions (Sec.~\ref{sec:LO}).

%%%%%%%%%%%%%%%%%%%%%%%%%%%%%%%%%%%%%%%%%%%%%%%%%%%%%%%%%%%%%%%%%%%%%%%%%%
\subsection{Definitions}
\label{sec:pENCdef}
The projected $N$-point energy correlator is the correlation function of $N$ energy flow operators, weighted by the largest pairwise angular separation between them. In the language of detectors placed on the celestial sphere at directions $\vec n_i$, with $x_{ij}\equiv (1-\vec n_i\cdot\vec n_j)/2$, this reads
\begin{align}
\label{eq:enc_operator_def}
\frac{d\sigma^{[N]}}{dx_L}
\;=\;
\int \prod_{i=1}^N d\Omega_{\vec n_i}\,
\delta\!\left(x_L - \max_{1\le i<j\le N} x_{ij}\right)
\frac{1}{Q^N}
\langle 0 | \mathcal{O}^\dagger\, \mathcal{E}(\vec n_1)\cdots \mathcal{E}(\vec n_N)\, \mathcal{O} | 0\rangle\,,
\end{align}
where $\mathcal{E}(\vec n) = \int_0^\infty dt\, \lim_{r\to\infty} r^2 n^i\, T_{0i}(t, r\vec n)$ is the energy flow operator~\cite{Sveshnikov:1995vi,Tkachov:1995kk,Hofman:2008ar}, $\mathcal{O}$ is the source operator for the process under consideration (e.g.\ the electromagnetic current for $e^+e^-\to q\bar q$ or the effective $H gg$ vertex for $H\to gg$), and $Q$ is the total hard scale. For $e^+e^-$ this would be the center-of-mass energy scale, whereas for Higgs decay this would be the mass of the decaying Higgs $m_H$. The $1/Q^N$ normalization makes the distribution dimensionless and ensures the sum rule $\int_0^1 dx_L\, d\sigma^{[N]}/dx_L = \sigma_{\rm tot}$, the total cross-section (or the Higgs to gluon pair decay width).

For an $m$-particle final state $|X_m\rangle$, the energy flow operator acts as $\mathcal{E}(\vec n)|X_m\rangle = \sum_{i\in X_m} E_i\, \delta^{(2)}(\vec n - \vec n_i)|X_m\rangle$. Inserting this into Eq.~\eqref{eq:enc_operator_def} gives the equivalent particle-level definition in perturbation theory,
\begin{align}
\label{eq:enc_particle_def}
\frac{d\sigma^{[N]}}{dx_L}
\;=\;
\sum_m \int d\sigma_{X_m}
\sum_{1\le i_1,\dots,i_N\le m}
\frac{\prod_{a=1}^N E_{i_a}}{Q^N}\,
\delta\!\left(x_L - \max\{x_{i_1 i_2},\, x_{i_1 i_3},\, \dots,\, x_{i_{N-1}i_N}\}\right)\,,
\end{align}
where $x_{ij} = (1-\vec{n}_i\cdot \vec{n}_j)/2$ is the angle between two particles $i$ and $j$. Here, $d\sigma_{X_m}$ is the differential cross-section of producing $m$ final-state particles, in both $e^+e^-$ collision and $H$ decays. The indices $i_a$ run independently over all final-state particles and need not be distinct. For convenience in computation, we can decompose different terms in $\sum_{1\le i_1,\dots,i_N\le m}\prod_{a=1}^N E_{i_a}$ by different configurations. 

The configuration in which all $N$ indices land on the same particle ($i_1 = \cdots = i_N$) has every pairwise angle equal to zero and contributes to a contact term at $x_L = 0$. In general, we write the weight associated with the $N$ indices landing on an $r$-particles as $\mathcal{W}^{[N]}_r$. Such term will be accompanied by the delta function that measures the largest angle amongst $\binom{r}{2}$ pairwise angles. This allows us to rewrite the projected $N$-point correlator as
\begin{align}
\label{eq:projected_exact_decomp}
\frac{d\sigma^{[N]}}{dx_L}
&=
\sum_m\int 
d\sigma_{X_m}\,
\Bigg[
\sum_{1\le i\le m}\cW^{[N]}_1(i)\,\delta(x_L)+\sum_{1\le i<j\le m}\cW^{[N]}_2(i,j)\,\delta(x_L-x_{ij})
\notag\\
&\hspace{1.5cm}
+\sum_{1\le i<j<k\le m}\cW^{[N]}_3(i,j,k)\,
\delta\!\left(x_L-\max\{x_{ij},x_{ik},x_{jk}\}\right)
+\,\cdots
\Bigg]\,,
\end{align}
where the ellipsis denotes terms with four or more distinct measured particles. The weights $\cW_r^{[N]}$ are precisely the integer-$N$ specialization of the recursive weights introduced in Ref.~\cite{Chen:2020vvp}:
\begin{align}
\label{eq:W_definition}
\cW^{[N]}_1(i)
=&
\frac{E_i^N}{Q^N}\,,
\notag\\
\cW^{[N]}_2(i,j)
=&
\frac{(E_i+E_j)^N-E_i^N-E_j^N}{Q^N}\,,
\notag\\
\cW^{[N]}_3(i,j,k)
=&
\frac{(E_i+E_j+E_k)^N-(E_i+E_j)^N-(E_i+E_k)^N-(E_j+E_k)^N+E_i^N+E_j^N+E_k^N}{Q^N}\,,\nn\\
\cdots\cdots\nn\\
&\hspace{-2cm}\cW^{[N]}_M(i_1,i_2,\cdots, i_M)=\frac{\left(\sum_{a=1}^M E_{i_a}\right)^N}{Q^N}-\sum_{1 \leq a_1<a_2<\ldots<a_{M-1} \leq M} \mathcal{W}_{M-1}^{[N]}\left(i_{a_1}, i_{a_2}, \ldots, i_{a_{M-1}}\right)\nn\\
&-\ldots-\sum_{1 \leq a \leq M}^M \mathcal{W}_1^{[N]}\left(i_a\right)\,.
\end{align}
Note that $\cW_r^{[N]} = 0$ when $r>N$. As pointed out in Refs.~\cite{Chen:2020vvp,Budhraja:2024tev}, these weights are valid even for non-integer $N\to \nu$ with $\text{Re}(\nu)>0$ as well. 
For a final state with $m$ partons, summing over all sectors $r=1,2,\dots$ at fixed $m$ reconstructs the binomial expansion $(\sum_i E_i)^N/Q^N = 1$, which is the origin of the sum rule mentioned above.

%%%%%%%%%%%%%%%%%%%%%%%%%%%%%%%%%%%%%%%%%%%%%%%%%%%%%%%%%%%%%
\subsection{Fixed-order results}
\label{sec:LO}
Having defined the projected ENC, we now compute it at the lowest non-trivial order in $\alpha_s$. This begins at the tree-level three-parton final states since the two-parton states $e^+e^-\to q\bar q$ or $H\to gg$ contribute only to contact terms proportional to delta functions. For the $e^+e^-$ process, we consider
\begin{equation}
e^+e^-\to q(p_1)\bar q(p_2)g(p_3)\,,
\end{equation}
and for the Higgs decay to gluons, the resolved final states are
\begin{align}
H\to g(p_1)\, g(p_2)\, g(p_3)\,,\notag\\
H\to q(p_1)\, \bar q(p_2)\, g(p_3)\,.
\end{align}
As anticipated by the decomposition of Eq.~\eqref{eq:projected_exact_decomp}, at LO only the $r=1,2,3$ sectors contribute.
 
To proceed, we parametrize the three-parton final state by the energy fractions
\begin{equation}
y_i\equiv \frac{2E_i}{Q}\,,\qquad 0<y_i<1\,,\qquad y_1+y_2+y_3=2\,,
\end{equation}
where $E_i$ is the energy of the $i$-th parton and $Q$ is the total energy. For massless kinematics, the pairwise angle is
\begin{equation}
x_{12}=\frac{1-\cos\theta_{12}}{2}
=\frac{1-y_3}{y_1y_2}
=\frac{y_1+y_2-1}{y_1y_2}\,,
\end{equation}
and similarly for $x_{13}$ and $x_{23}$ by cyclic permutation. For the $e^+e^-$ process, the normalized tree-level squared amplitude is given as
\begin{equation}
\label{eq:app_mq}
\mathcal{M}^{\rm LO}_{e^+e^-}(y_1,y_2,y_3)
\equiv
\frac{\alpha_sC_F}{2\pi}
\frac{y_1^2+y_2^2}{(1-y_1)(1-y_2)}\,,
\qquad
\frac{1}{\sigma_0}\,d\sigma^{\rm LO}_{e^+e^-}
=
\mathcal{M}^{\rm LO}_{e^+e^-}(y_1,y_2,y_3)\,dy_1\,dy_2\,,
\end{equation}
with $y_3=2-y_1-y_2$. Similar definition is given for the Higgs decay to gluons as well. Here,
$\sigma_0$ is Born cross-section of the given process, $\sigma_0=\sigma_0^{e^{+}e^{-}}$ in $e^+e^-\to q\bar{q}$ or $\sigma_0=\Gamma_0^{H\to gg}$ in  $H\to gg$.
At LO, the bulk of projected $N$-point correlator ($x_L>0$) receives finite contributions only from the $r=2$ and $r=3$ sectors, i.e.
\begin{equation}
\label{eq:enc_LO_decomp}
\frac{1}{\sigma_0}\frac{d\sigma^{[N],\rm LO}}{dx_L}
=
\frac{1}{\sigma_0}\frac{d\sigma^{[N],\rm LO}_{r=2}}{dx_L}
{}+{}
\frac{1}{\sigma_0}\frac{d\sigma^{[N],\rm LO}_{r=3}}{dx_L}\,,
\end{equation}
with weights $\cW_2^{[N]}$ and $\cW_3^{[N]}$ defined in Eq.~\eqref{eq:W_definition}. Using $E_i=Qy_i/2$, these become simple homogeneous polynomials in $y_i$.

To evaluate the pairwise contribution ($r=2$), we solve the $\delta$-function constraint in the measurement function for each particle pair. For instance, we label the contribution from particles with momenta $p_1$ and $p_2$ as $(12)$, for which the measurement delta function reads 
\begin{equation}
\delta\left(\Delta_{12}\right)\equiv\delta\left( x_L-\frac{y_1+y_2-1}{y_1y_2}\right)
\qquad\Longrightarrow\qquad
y_1=\frac{1-y_2}{1-x_Ly_2}\,,
\end{equation}
with the associated Jacobian
\begin{equation}
J_{12}\equiv \left|\frac{\partial\Delta_{12}}{\partial y_1}\right|
=\frac{(1-x_Ly_2)^2}{y_2(1-y_2)}\,,
\qquad
\frac{1}{J_{12}}=\frac{y_2(1-y_2)}{(1-x_Ly_2)^2}\,.
\end{equation}
The Jacobians $J_{13}$ and $J_{23}$ follow from the same formula after relabeling the indices. The two-particle contribution is therefore
\begin{align}
\label{eq:lo_pair_master}
\frac{1}{\sigma_0}\frac{d\sigma^{[N],\rm LO}_{r=2}}{dx_L}
&=
\int_0^1dy_2\,
\left[
\mathcal{M}^{\rm LO}(y_1,y_2,y_3)
\cW_2^{[N]}(1,2)\,
\frac{1}{J_{12}}
\right]_{y_1=(1-y_2)/(1-x_Ly_2)}
\notag\\
&\quad+
\int_0^1dy_3\,
\left[
\mathcal{M}^{\rm LO}(y_1,y_2,y_3)
\cW_2^{[N]}(1,3)\,
\frac{1}{J_{13}}
\right]_{y_1=(1-y_3)/(1-x_Ly_3)}
\notag\\
&\quad+
\int_0^1dy_3\,
\left[
\mathcal{M}^{\rm LO}(y_1,y_2,y_3)
\cW_2^{[N]}(2,3)\,
\frac{1}{J_{23}}
\right]_{y_2=(1-y_3)/(1-x_Ly_3)} \,.
\end{align}
In the first, second, and third terms the remaining energy fraction is fixed by $y_3=2-y_1-y_2$, $y_2=2-y_1-y_3$, and $y_1=2-y_2-y_3$, respectively.
 
For the genuinely resolved three-particle term ($r=3$), we now need to compare the three angles $x_{12}, x_{13},x_{23}$ and measure the largest angle. For the contribution in which the largest angle is given by $x_L=x_{12}$, solving the inequalities
\begin{equation}
x_{12}>x_{13}\,,\qquad x_{12}>x_{23}\,.
\end{equation}
together with the on-shell constraints give
\begin{equation}
\label{eq:lo_3p_region}
\frac{3}{4}<x_L<1\,,
\qquad
\frac{1}{2x_L}<y_2<\frac{2x_L-1}{x_L}\,.
\end{equation}
This is the origin of the step function that multiplies every resolved three-particle contribution. Writing
\begin{equation}
y_1=\frac{1-y_2}{1-x_Ly_2}\,,
\qquad
y_3=2-y_1-y_2=\frac{y_2(1-x_L)}{1-x_Ly_2}\,,
\end{equation}
the full $r=3$ contribution is obtained by summing the three contributions where different pairs are the largest angle,
\begin{align}
\label{eq:lo_3p_master}
\frac{1}{\sigma_0}\frac{d\sigma^{[N],\rm LO}_{r=3}}{dx_L}
&=
\theta\!\left(x_L-\frac{3}{4}\right)
\int_{1/(2x_L)}^{(2x_L-1)/x_L}\!dy_2\,
\left[
\mathcal{M}^{\rm LO}\,
\cW_3^{[N]}(1,2,3)\,
\frac{1}{J_{12}}
\right]_{y_1=(1-y_2)/(1-x_Ly_2)}
\notag\\
&\hspace{-1cm}+
\theta\!\left(x_L-\frac{3}{4}\right)
\int_{1/(2x_L)}^{(2x_L-1)/x_L}\!dy_3\,
\left[
\mathcal{M}^{\rm LO}\,
\cW_3^{[N]}(1,3,2)\,
\frac{1}{J_{13}}
\right]_{y_1=(1-y_3)/(1-x_Ly_3)}
\notag\\
&\hspace{-1cm}+
\theta\!\left(x_L-\frac{3}{4}\right)
\int_{1/(2x_L)}^{(2x_L-1)/x_L}\!dy_3\,
\left[
\mathcal{M}^{\rm LO}\,
\cW_3^{[N]}(2,3,1)\,
\frac{1}{J_{23}}
\right]_{y_2=(1-y_3)/(1-x_Ly_3)}\,.
\end{align}
In both Eqs.~\eqref{eq:lo_pair_master} and~\eqref{eq:lo_3p_master}, the remaining integrand is a rational function of the corresponding integration variable and $x_L$, so the final integrations only contain rational terms together with $\ln(1-x_L)$ and, in the $r=3$ sector, $\ln 2$ from the upper endpoint.

Putting together, we obtain the analytical expressions of the LO ENC for both processes. 
We can write the result as
\begin{align}
\label{eq:enc_LO_form}
\frac{1}{\sigma_0}\frac{d\sigma_{e^+e^-}^{[2],\rm LO}}{dx_L}&= \frac{\alpha_s C_F}{4\pi x_L^N (1-x_L) }\times \left[q_1^{[N]}(x_L)+\theta\left(x_L-\frac{3}{4}\right)q_2^{[N]}(x_L) \right]\,,\nn\\
\frac{1}{\sigma_0}\frac{d\sigma_H^{[2],\rm LO}}{dx_L}&= \frac{\alpha_s}{4\pi x_L^N (1-x_L) }\times \left[g_1^{[N]}(x_L)+\theta\left(x_L-\frac{3}{4}\right)g_2^{[N]}(x_L) \right] \,,
\end{align}
with $q_1^{[N]}$, $q_2^{[N]}$, $g_1^{[N]}$ and $g_2^{[N]}$ functions of $x_L$. Note that $q_2^{[2]}=g_2^{[2]}=0$. Taking $N=4$ $e^+e^-$ collision as an illustration example, we find
\begin{align}
    q_1^{[4]}&=\left(6 x_L+\frac{555}{2 x_L}-\frac{282}{x_L^2}+\frac{90}{x_L^3}-95\right) \log
   \left(1-x_L\right)-\frac{3 x_L^2}{2}-\frac{111 x_L}{4}-\frac{237}{x_L}+\frac{90}{x_L^2}+\frac{333}{2}\,,\notag\\
q_2^{[4]}&=\left(-48 x_L^2+336 x_L+\frac{630}{x_L}-\frac{180}{x_L^2}-738\right) \log \left(4-4 x_L\right)\,.
\end{align}
The full analytic expressions for both channels and all $N=2,\dots,6$ are collected in App.~\ref{app:lo_enc}.
 
Expanding Eq.~\eqref{eq:enc_LO_form} around $x_L = 0$ exposes the singular collinear structure. For all integer $N\ge 2$, one finds
 
\begin{equation}
\label{eq:LO_small_xL}
\frac{1}{\sigma_0}\frac{d\sigma^{[N],\rm LO}}{dx_L}
\;\xrightarrow{x_L\to 0}\;
\frac{\alpha_s}{4\pi}\,\frac{c^{[N]}_{1}}{x_L}\,
+\,(\text{less singular})\,,
\end{equation}
i.e.\ a single $1/x_L$ enhancement times a constant coefficient $c^{[N]}_{1}$. Careful regulation would also promote this into a plus distribution $[1/x_L]_+$. Once integrated up to a cumulant, this gives a single logarithm of $x_L$, consistent with the $\cL^{j}(x_L) = [\ln^j(x_L)/x_L]_+$ plus-distribution structure of the singular expansion discussed in Sec.~\ref{sec:factorization} that requires resummation. 

For the EEC ($N=2$), the NLO QCD calculation of $d\sigma/dx_L$ has been carried out analytically in both $e^+e^-$ annihilation~\cite{Dixon:2018qgp} and hadronic Higgs decay~\cite{Luo:2019nig,Gao:2020vyx}, and the NNLO numerical prediction for $e^+e^-$ annihilation is computed in \textsc{Colorfulnnlo}~\cite{Somogyi:2006da,Somogyi:2006db,Aglietti:2008fe}. For the higher-point projected correlators $N\ge 3$, no analytic NLO calculation is currently available. Instead, we obtain numerical NLO results for $e^+e^-\to q\bar q$ from the dipole-subtraction program \textsc{Event2}~\cite{Catani:1996jh,Catani:1996vz}, which provides the highest fixed order to which our resummed predictions can be matched in this channel. For $H\to gg$, we use \textsc{Eerad3}~\cite{Gehrmann-DeRidder:2014hxk,Aveleira:2025svg} to calculate the NLO distributions numerically.

%%%%%%%%%%%%%%%%%%%%%%%%%%%%%%%%%%%%%%%%%%%%%%%%%%%%%%%%%%%%%%%%%%%%%%%%
\section{Collinear factorization of projected energy correlators}
\label{sec:factorization}
The fixed-order results, both analytic (LO) and numerical (NLO from \textsc{Event2} and \textsc{Eerad3}), are singular in the collinear limit as already pointed out in Eq.~\eqref{eq:LO_small_xL}.
At $\cO(\alpha_s^L)$, this singularity in the projected ENC is captured by a tower of plus distributions $[\ln^j(x_L)/x_L]_+$ with $0\le j\le L-1$, corresponding to a single-logarithmic series in the differential distribution. Explicitly, we observe 
\begin{align}
	\frac{d\sigma^{[N]}}{dx_L}&= \sum_{L=1}^\infty \sum_{j=-1}^{L-1} \left(\frac{\alpha_s(\mu)}{4 \pi} \right)^L e_{L,j} \cL^j (x_L) +\ldots \,,\, \nn\\
    &\text{where} \qquad \cL^{-1}(x_L)=\delta(x_L),\, \qquad \cL^{j}(x_L)= \left[\ln^j(x_L)/x_L \right]_+\,,
\end{align}
where the ellipsis denotes terms that are less singular as $x_L\to 0$. 
These collinear logarithms become large when $\alpha_s \ln(1/x_L) \sim 1$, 
and must be resummed to all orders in $\alpha_s$. 
The natural framework for this resummation is provided by collinear factorization, to which we now turn.

It is convenient to formulate the resummation at the level of the cumulant,
\begin{align}
	\label{eq:cumulant}
	\Sigma^{[N]}\left(x_L, \ln\frac{Q^2}{\mu^2}\right) =  \frac{1}{\sigma_0} \int_{0}^{x_L}	dx_L^{\prime} \, \frac{ d\sigma^{[N]}}{d x_L^{\prime}}\left(x_L^{\prime}, \ln\frac{Q^2}{\mu^2}\right) \, , 
\end{align}
since the cumulant maps the plus distributions into ordinary logarithms and makes the logarithmic counting transparent. In particular, $[\ln^j(x_L)/x_L]_+$ integrates to $\ln^{j+1}x_L$ and $\mathrm{N}^k\mathrm{LL}$ resummation accuracy corresponds to resumming the logarithmic tower $\alpha_s^n \ln^{m} x_L$ up to $n \geq m \geq n-k$.

At leading power, the cumulant factorizes into a convolution of a process-dependent hard function and a universal jet function,
\begin{align}
	\label{eq:fac_nu}
	\Sigma_{ee,H}^{[N]}\left(x_L, \ln\frac{Q^2}{\mu^2}\right) = \int_0^1  d x\,
	x^N \vec{J}^{[N]}\left(\ln \frac{x_L x^2 Q^2}{\mu^2}\right) \cdot \vec{H}_{ee,H}\left(x, \ln \frac{Q^2}{\mu^2}\right)\, ,
\end{align}
where the subscript $ee$ refers to $e^+e^-$ annihilation and $H$ refers to the Higgs decay process $H\to gg$. The factor $x^N$ reflects the energy weighting carried by the projected $N$-point correlator. 
The hard function $\vec{H}_{ee,H}$ describes the production of a parent parton with energy fraction $x$ and can be identified with that of a 
semi-inclusive fragmentation function~\cite{Rijken:1996vr,Rijken:1996ns,Rijken:1996npa,Mitov:2006wy,Mitov:2006ic,Moch:2007tx,Almasy:2011eq}, while the jet function $\vec{J}^{[N]}$ captures the subsequent collinear dynamics probed by the measurement. Both quantities are two-component vectors in quark--gluon flavor space. The important point for the present work is that the process dependence is entirely encoded in the hard function, whereas the dependence on the projected $N$-point measurement resides in the jet function.

The renormalization group (RG) equation of the hard function is timelike DGLAP evolution,
\begin{equation}
	\label{eq:hard_evo}
	\frac{d \vec{H}(x, \ln\frac{Q^2}{\mu^2})}{d \ln \mu^2}
	= 
	- \int_x^1\! \frac{dy}{y} \widehat{P}(y) \cdot \vec{H} \left( \frac{x}{y},
	\ln\frac{Q^2}{\mu^2}\right) \,,
\end{equation}
where $\widehat{P}(y)$ is the singlet timelike splitting matrix, 
now known in full through three loops~\cite{Chen:2020uvt,Almasy:2011eq} 
and partially at four-loop~\cite{Gehrmann:2023cqm,Falcioni:2023luc,Falcioni:2023vqq,Falcioni:2024xyt,Falcioni:2024qpd,Falcioni:2025hfz}.
Requiring the factorization formula in Eq.~\eqref{eq:fac_nu} to be RG invariant then fixes the evolution of the jet function. One finds a modified timelike DGLAP equation,
\begin{align}
	\label{eq:jet_evo}
	\frac{d \vec{J}^{[N]}( \ln \frac{x_L Q^2}{\mu^2})}{d \ln\mu^2} = \int_0^1  dy\, y^N \vec{J}^{[N]}\left( \ln \frac{x_L y^2 Q^2}{\mu^2}\right)
	\cdot 
	\widehat{P}(y) \,. 
\end{align}
Compared with the evolution of the hard function, the jet equation carries an additional factor of $y^N$ and a shifted logarithmic argument. These terms encode how the projected observable transforms under a collinear splitting. The factorization theorem above holds for arbitrary $N$ at leading power, and thus we only need to compute the jet boundary constants for $N$-point correlators.

The solutions to these RGEs will evolve the hard (jet) function from its typical scale $\mu_h$ ($\mu_j$) to the common scale $\mu$. We can formally write the solutions as
\begin{align}
    \vec{H}\left(x,\mu,\ln\frac{Q^2}{\mu^2}\right)&=U^H(\mu,\mu_h,x)\otimes\vec{H}\left(x,\mu_h,\ln\frac{Q^2}{\mu_h^2}\right)\,,\\
    \vec{J}^{[N]}\left(\mu,\ln\frac{x_L x^2 Q^2}{\mu^2}\right)&=U^J(\mu,\mu_j,x)\otimes\vec{J}^{[N]}\left(\mu_j,\ln\frac{x_L x^2 Q^2}{\mu_j^2}\right)\,,
\end{align}
with $U^H(\mu,\mu_h,x)$ and $U^J(\mu,\mu_j,x)$ the RG evolution kernels.
Since the physical observable, i.e. the LHS of Eq.~\eqref{eq:fac_nu}, should not depend on the factorization scale $\mu$, we can choose its value arbitrarily. We set $\mu=\mu_h$ for convenience, and the resummed cumulant becomes
\begin{equation}
    \Sigma^{[N]}(x_L)=\int_0^1 dx x^N \left[U^J(\mu_h,\mu_j,x)\otimes \vec{J}^{[N]}\left(\mu_j,\ln\frac{x^2 x_LQ^2}{\mu_j^2}\right)\right]\vec{H}\left(x,\mu_h,\ln\frac{Q^2}{\mu_h^2}\right) \,.
\end{equation}
Note that when convolving $\vec{H}\left(x,\mu_h,\ln\frac{Q^2}{\mu_h^2}\right)$ with $\vec{J}^{[N]}\left(\mu_j,\ln\frac{x_L x^2 Q^2}{\mu_j^2}\right)$, we should truncate the product to some order in $\alpha_s(\mu_h)\sim\alpha_s(\mu_j)\sim \lambda $, such that equal scale $\mu_h=\mu_j$ will give precisely the singular expansion in the collinear limit of the fixed-order calculations we match to.

We use the expanded solution to the jet function RGE in this work. Similarly to Refs.~\cite{Dixon:2019uzg,Chen:2023zlx}, we write an ansatz
\begin{equation}
\label{eq:jetansatz}
\vec{J}^{[N]}=\underbrace{\sum_{i=0}^{\infty} \alpha_s^i(\mu) L^i \vec{c}^{\,[N]}_{i,i}}_{\text{LL}}+\underbrace{\sum_{i=1}^{\infty} \alpha_s^i(\mu) L^{i-1} \vec{c}^{\,[N]}_{i,i-1}}_{\text{NLL}}+\underbrace{\sum_{i=2}^{\infty} \alpha_s^i(\mu) L^{i-2} \vec{c}^{\,[N]}_{i,i-2}}_{\text{NNLL}}+\cdots \, ,
\end{equation}
with $L\equiv\ln(x_L Q^2/\mu^2)$ and $c^{[N]}_{i,j}$ unknown constants. Note that we always normalize the jet function to have $\vec{c}^{[N]}_{0,0}=\{2^{-N},2^{-N}\}$, which is natural due to $1/Q^N$ normalization in Eq.~\eqref{eq:enc_particle_def}. Combining with the $\beta$-RGE,
\begin{equation}
	\label{eq:beta}
	\frac{\mathrm{d}\alpha_s(\mu)}{\mathrm{d}\ln \mu}= \beta (\alpha_s(\mu)),\quad 
	\beta (\alpha)=- 2 \alpha\, \left[ \left( \frac{\alpha}{4 \pi} \right) \beta_0 + \left(
	\frac{\alpha}{4 \pi} \right)^2 \beta_1 + \left( \frac{\alpha}{4 \pi}
	\right)^3 \beta_2 + \cdots \right]\,,
\end{equation}
we can form linear equations at each order in $\alpha_s$ and $L$, and determine the $c^{[N]}_{i,j}$ coefficients. In practice, we need to truncate the sum to some order $\mathcal{O}(\alpha_s^k)$ in Eq.~\eqref{eq:jetansatz}.
We find that $k=10\sim 15$ gives more than sufficient convergence in the perturbative regime at $Q \sim 100\,\mathrm{GeV}$, which is the scale we consider.
The expressions for $c_{i,j}$ depend on the jet constants and anomalous dimensions, defined as
\begin{equation}
\label{eq:jet_ana_dim}
\gamma_T(\nu)\equiv -\int_0^1 dy\, y^\nu \hat P(y)=\left(\frac{\alpha_s}{4\pi}\right)\gamma_T^{(0)}+\left(\frac{\alpha_s}{4\pi}\right)^2 \gamma_T^{(1)}+\cdots \,,
\end{equation}
using the timelike splitting kernel $\hat P(y)$. We have summarized the values of these anomalous dimensions in App.~\ref{app:anomalous_dim}.
\begin{table}[!ht]
\begin{center}
\begingroup
\renewcommand{\arraystretch}{1.2}
  \begin{tabular}{|c|c|c|c|c|c|}  \hline
   resummation order &     $\hat{P}(y)$ &    $\vec{H}$, $\vec{J}\, \text{ constants}$ &   $\beta[\alpha_s]$ & fixed-order matching\\  
    \hline
    LL  & tree & tree & 1-loop & -- \\
    \hline
    NLL  & 1-loop & 1-loop & 2-loop & LO \\
    \hline
    NNLL & 2-loop & 2-loop & 3-loop & NLO \\
    \hline
    N$^k$LL & $k$-loop & $k$-loop & $(k+1)$-loop & N$^{k-1}$LO \\
    \hline
\end{tabular}
\endgroup
\end{center}
\vspace{-0.2cm}
\caption{Definition of the resummation order and their corresponding fixed-order matching.}
\label{tab:ords}
\end{table}

With the expanded solutions, the convolution with hard functions will only give rise to the following integrals
\begin{equation}
    \int_0^1 dx \,x^N\, \ln^M(x^2) \,\vec{H}\left(x,\mu_h,\ln\frac{Q^2}{\mu_h^2}\right) = \sum_{L}\left(\frac{\alpha_s(\mu_h)}{4\pi}\right)^L \vec{h}_{L} (N,M)\,,
\end{equation}
with $\vec{h}_{L} (N,M)$ the Mellin moments of hard functions. These integrals can be evaluated analytically or numerically. Putting everything together, we obtain the resummed distributions for projected $N$-point correlators. In Tab.~\ref{tab:ords}, we summarize the convention for N${}^k$LL resummation. In Sec .~\ref {sec:nnll_resum}, we will describe in detail how we carry out the fixed-order matching.

%%%%%%%%%%%%%%%%%%%%%%%%%%%%%%%%%%%%%%%%%%%%%%%%%%%%%%%%%%%%%%%%%%%%%%%%%%%%%%%%%%%%%%%%%%%%%%%%%%%%%%%%%%%%%%%%%%%%%%%%%%%%%%%%%%%%%%%%%%%%%%%%%%%%%%%%%%%%%%%%%%%%%%%%%%%%%%%%%%%%%%%%%%%%%%%%%%%%%%%%%%%%%%%
\section{Calculation of the two-loop jet functions}\label{sec:jet_func}

Given the factorization in Eq.~\eqref{eq:fac_nu} and the jet ansatz in 
Eq.~\eqref{eq:jetansatz}, the only missing ingredient for NNLL resummation is the 
two-loop boundary constant $c_{2,0}^{[N]}$ of the jet function. The $N=2,3$ cases were 
obtained in Refs.~\cite{Dixon:2019uzg,Chen:2023zlx}; here we extend the calculation 
to $N=4,5,6$. Following Ref.~\cite{Chen:2023zlx}, we extract $c_{2,0}^{[N]}$ by matching 
the singular $x_L\to 0$ limit of the fixed-order projected correlator at 
$\mathcal{O}(\alpha_s^2)$ onto Eq.~\eqref{eq:fac_nu}, using $e^+e^-$ annihilation 
for quark jets and the corresponding $H\to gg$ channels in the Higgs effective field theory for 
gluon jets. We decompose the calculation into different measurement sectors 
(Sec.~\ref{sec:meas_decomp}), and then organize the calculations into contact terms (Sec.~\ref{sec:contact}), and 
three-particle terms (Sec.~\ref{sec:3p}). Finally we collect the results in 
Sec.~\ref{sec:jet_results}.

\subsection{Measurement decomposition}
\label{sec:meas_decomp}

Recall from Eq.~\eqref{eq:projected_exact_decomp} and Eq.~\eqref{eq:W_definition}, we write the definition of projected $N$-point correlator in perturbation theory in terms of different measurement sectors $\cW_r^{[N]}$. To NLO $\mathcal{O}(\alpha_s^2)$, we can have at most four final-state particles. For $N\leq 3$, we will at most measure three particles; starting from $N=4$, we can have detectors on all four particles. However, momentum conservation forbids all final-state particles from sitting inside an arbitrarily small cone, which places a lower bound on $x_L$ whenever an energy flow operator is placed on every produced particle. Such configurations therefore do not contribute in the $x_L\to 0$ limit.
Following this spirit, since NLO has up to four particles in the final state, we only need to consider $\cW_r^{[N]}$ with $r=1,2,3$ at NLO for the collinear limit.

Following the definition, each term in $\cW_r^{[N]}$ takes the form
\begin{align}
\sim \frac{E_{i_1}^{a_1}E_{i_2}^{a_2}\cdots E_{i_r}^{a_r}}{Q^{N}}\,,
\qquad a_1+a_2+\cdots+a_r = N\,,
\end{align}
i.e. a monomial of total degree $N$ in the parton energies. For the calculation it is 
useful to organize $\cW_r^{[N]}$ further by grouping together monomials that share 
the same multiset of exponents. Writing
\begin{align}
\cW_2^{[N]}(i,j)
&=
\sum_{\lambda\in\Lambda_N^{(2)}}\cW_\lambda^{[N]}(i,j)\,,
\notag\\
\cW_3^{[N]}(i,j,k)
&=
\sum_{\lambda\in\Lambda_N^{(3)}}\cW_\lambda^{[N]}(i,j,k)\,,
\end{align}
where $\Lambda_N^{(r)}$ denotes the set of unordered partitions of $N$ into $r$ 
positive parts. For each partition $\lambda=\{a,b\}\in\Lambda_N^{(2)}$ or 
$\lambda=\{a,b,c\}\in\Lambda_N^{(3)}$, the symmetrized monomial is
\begin{align}
\cW^{[N]}_{\{a,b,c\}}(i,j,k)
&\equiv
\frac{N!}{a!\,b!\,c!}
\sum_{(a',b',c')\in \mathrm{Perm}(a,b,c)}
\frac{E_i^{a'}E_j^{b'}E_k^{c'}}{Q^N}\,,
\notag\\
\cW^{[N]}_{\{a,b\}}(i,j)
&\equiv
\frac{N!}{a!\,b!}
\sum_{(a',b')\in \mathrm{Perm}(a,b)}
\frac{E_i^{a'}E_j^{b'}}{Q^N}\,,\notag\\
\cW^{[N]}_{\{N\}}(i)&\equiv \frac{E_i^N}{Q^N}\,,
\end{align}
where $\mathrm{Perm}(\cdots)$ denotes the distinct permutations of the multiset of 
exponents, and the multinomial prefactor counts the number of ways to distribute 
the $N$ detector insertions over the chosen particles. For the observables considered 
here, the relevant partitions with $N\le 6$ are
\begin{align}
\Lambda^{(3)}_3&=\bigl\{\{1,1,1\}\bigr\}\,,&
\Lambda^{(2)}_3&=\bigl\{\{2,1\}\bigr\}\,,\notag\\
\Lambda^{(3)}_4&=\bigl\{\{2,1,1\}\bigr\}\,,&
\Lambda^{(2)}_4&=\bigl\{\{3,1\},\{2,2\}\bigr\}\,,\notag\\
\Lambda^{(3)}_5&=\bigl\{\{3,1,1\},\{2,2,1\}\bigr\}\,,&
\Lambda^{(2)}_5&=\bigl\{\{4,1\},\{3,2\}\bigr\}\,,\notag\\
\Lambda^{(3)}_6&=\bigl\{\{4,1,1\},\{3,2,1\},\{2,2,2\}\bigr\}\,,&
\Lambda^{(2)}_6&=\bigl\{\{5,1\},\{4,2\},\{3,3\}\bigr\}\,.
\end{align}

Specializing Eq.~\eqref{eq:projected_exact_decomp} to the two-loop singular cross section, the decomposition is most transparent in the $N=4$ case:
\begin{align}
\label{eq:E4C_decomp}
\frac{d\sigma^{[4]}}{dx_L}
&=
\sum_{1\leq i<j<k\leq 4}
\int \df\mathrm{LIPS}_4\,|\mathcal{M}_{4}|^2\,
\cW_3^{[4]}(i,j,k)\,
\delta\left(x_L-\max\{x_{ij},x_{ik},x_{jk}\}\right)
\notag\\
&\quad+
\sum_{m\in\{3,4\}}\sum_{1\leq i<j\leq m}
\int \df\mathrm{LIPS}_m\,|\mathcal{M}_{m}|^2\,
\cW_2^{[4]}(i,j)\,
\delta\left(x_L-x_{ij}\right)
\notag\\
&\quad+
\sum_{m\in\{2,3,4\}}\sum_{1\leq i\leq m}
\int \df\mathrm{LIPS}_m\,|\mathcal{M}_{m}|^2\,
\cW_1^{[4]}(i)\,
\delta(x_L)\,,
\end{align}
with $\df\mathrm{LIPS}_m$ the Lorentz-invariant phase space for $m$ final-state particles and $|\mathcal{M}_{m}|^2$ the corresponding squared matrix elements. 
The first line involves only the double-real $m=4$ contribution. Notice that we dropped configurations in which detectors are placed on every final-state particle, e.g.
we do not consider $\cW_3^{[N]}$ on three-particle final states or $\cW_2^{[N]}$ on two-particle final states, as they do not contribute to the $x_L\to 0$ limit.
The measurement weights are explicitly given as
\begin{align}
\cW_3^{[4]}(i,j,k)
&=
\cW^{[4]}_{\{2,1,1\}}(i,j,k)
=
\frac{12}{Q^4}
\left(E_i^2E_jE_k+E_iE_j^2E_k+E_iE_jE_k^2\right)\,,
\notag\\
\cW_2^{[4]}(i,j)
&=
\cW^{[4]}_{\{3,1\}}(i,j)+\cW^{[4]}_{\{2,2\}}(i,j)
=
\frac{4}{Q^4}\left(E_i^3E_j+E_iE_j^3\right)
+\frac{6}{Q^4}E_i^2E_j^2\,,
\notag\\
\cW_1^{[4]}(i)&=\cW_{\{4\}}^{[4]}(i)=\frac{E_i^4}{Q^4}\,.
\end{align}
The same notation keeps the higher-point cases compact. For $N=5$ one has
\begin{align}
\cW_3^{[5]}(i,j,k)
&=
\cW^{[5]}_{\{3,1,1\}}(i,j,k)+\cW^{[5]}_{\{2,2,1\}}(i,j,k)\,,
\notag\\
\cW^{[5]}_{\{3,1,1\}}(i,j,k)
&=
\frac{20}{Q^5}
\left(E_i^3E_jE_k+E_iE_j^3E_k+E_iE_jE_k^3\right)\,,
\notag\\
\cW^{[5]}_{\{2,2,1\}}(i,j,k)
&=
\frac{30}{Q^5}
\left(E_i^2E_j^2E_k+E_i^2E_jE_k^2+E_iE_j^2E_k^2\right)\,,\notag\\
\cW_2^{[5]}(i,j)
&=
\cW^{[5]}_{\{4,1\}}(i,j)+\cW^{[5]}_{\{3,2\}}(i,j)\,,
\notag\\
%\cW^{[5]}_{\{4,1\}}(i,j)
%&=
%\frac{5}{Q^5}\left(E_i^4E_j+E_iE_j^4\right)\,,
%\qquad
%\cW^{[5]}_{\{3,2\}}(i,j)=\frac{10}{Q^5}\left(E_i^3E_j^2+E_i^2E_j^3\right)\,,
%\notag\\
&=\frac{5}{Q^5}\left(E_i^4E_j+E_iE_j^4\right)+\frac{10}{Q^5}\left(E_i^3E_j^2+E_i^2E_j^3\right)\,,\notag\\
\cW_1^{[5]}(i)
&=
\cW_{\{5\}}^{[5]}(i)=\frac{E_i^5}{Q^5}\,,
\end{align}
and for $N=6$,
\begin{align}
\cW_3^{[6]}(i,j,k)
&=
\cW^{[6]}_{\{4,1,1\}}(i,j,k)+\cW^{[6]}_{\{3,2,1\}}(i,j,k)+\cW^{[6]}_{\{2,2,2\}}(i,j,k)\,,
\notag\\
\cW^{[6]}_{\{4,1,1\}}(i,j,k)
&=
\frac{30}{Q^6}
\left(E_i^4E_jE_k+E_iE_j^4E_k+E_iE_jE_k^4\right)\,,
\notag\\
\cW^{[6]}_{\{3,2,1\}}(i,j,k)
&=
\frac{60}{Q^6}
\sum_{(a',b',c')\in\mathrm{Perm}(3,2,1)}
E_i^{a'}E_j^{b'}E_k^{c'}\,,
\notag\\
\cW^{[6]}_{\{2,2,2\}}(i,j,k)
&=
\frac{90}{Q^6}E_i^2E_j^2E_k^2\,,\notag\\
\cW_2^{[6]}(i,j)
&=
\cW^{[6]}_{\{5,1\}}(i,j)+\cW^{[6]}_{\{4,2\}}(i,j)+\cW^{[6]}_{\{3,3\}}(i,j)\,,
\notag\\
&=\frac{6}{Q^6}\left(E_i^5E_j+E_iE_j^5\right)+\frac{15}{Q^6}\left(E_i^4E_j^2+E_i^2E_j^4\right)+\frac{20}{Q^6}E_i^3E_j^3\,,\notag\\
\cW_1^{[6]}(i)
&=
\cW_{\{6\}}^{[6]}(i)=\frac{E_i^6}{Q^6}\,.
\end{align}
This decomposition is the direct two-loop truncation of the exact observable definition in Ref.~\cite{Chen:2020vvp}. The three-particle terms are the only pieces with support on the projected three-angle measurement involving multiple nontrivial angles, $\delta[x_L-\max\{x_{ij},x_{ik},x_{jk}\}]$. All other remaining pieces collapse either to a single pairwise angle $\delta(x_L-x_{ij})$ or to $\delta(x_L)$ itself. Below we will discuss how to compute these two parts separately.

\subsection{Contact terms}
\label{sec:contact}

In this subsection, we describe the analytic calculations of the contact terms. As described above, the contact terms contain both one-particle partition and two-particle partition $\Lambda_N^{(2)}$. The one-particle partition gives the self-correlation contribution, which has support only at $x_L=0$. In the two-particle partition, all $N$ detector insertions are carried by only two resolved final-state particles, so only depends on a single pairwise angle. Therefore, the contact term share the same type of integrals as ordinary EEC in Refs.~\cite{Dixon:2018qgp,Luo:2019nig}, with the EEC weight $2E_iE_j/Q^2$ replaced by the polynomial weight $\cW^{[N]}_{\lambda}(i,j)\propto E_i^aE_j^b+E_i^bE_j^a$. Explicitly, we will consider the following integrals
\begin{equation}
\label{eq:contact_pair_def}
\frac{d\sigma^{[N]}_{2}}{dx_L}
=
\sum_{m\in\{3,4\}}
\sum_{1\le i<j\le m}
\int \df\mathrm{LIPS}_m\,|\mathcal M_m|^2\,
\cW^{[N]}_{2}(i,j)\,
\delta(x_L-x_{ij})\,,
\end{equation}
and
\begin{equation}
\label{eq:contact_self_def}
\frac{d\sigma^{[N]}_{1}}{dx_L}
=
\sum_{m\in\{2,3,4\}}
\sum_{1\le i\le m}
\int \df\mathrm{LIPS}_m\,|\mathcal M_m|^2\,
\cW^{[N]}_{\{N\}}(i)\,
\delta(x_L)\,.
\end{equation}
The pairwise terms in Eq.~\eqref{eq:contact_pair_def} receive contributions from $m=3$ and $m=4$ final states, corresponding to real-virtual and double-real corrections. The self-correlation term in Eq.~\eqref{eq:contact_self_def} also contains the $m=2$ double-virtual contribution, but is otherwise simpler because it never involves a nontrivial angular measurement. 
We will denote the sum of these two integral as
\begin{equation}
    \frac{d\sigma^{[N],\mathrm{2\mbox{-}loop}}_{C}}{dx_L}=\frac{d\sigma^{[N]}_{2}}{dx_L}+\frac{d\sigma^{[N]}_{1}}{dx_L}\,.
\end{equation}

In order to extract both quark and gluon jet functions, we consider the $e^+e^-$ annihilation and Higgs decay into gluons up to NLO, which involves the following subprocesses:
\begin{align}
	{\bf e^+e^- \textbf{annihilation }}\qquad\qquad& \textbf{Higgs decays}\nonumber\\
	\gamma^*\rightarrow q\bar{q}+VV \qquad\qquad& H\rightarrow gg+VV\nonumber\\
	\gamma^*\rightarrow q\bar{q}g+V \qquad\qquad& H\rightarrow ggg+V\nonumber\\
	& H\rightarrow q\bar{q}g+V\nonumber\\
	\gamma^*\rightarrow q\bar{q}gg \qquad\qquad& H\rightarrow gggg\nonumber\\
	\gamma^*\rightarrow q\bar{q}q\bar{q} \qquad\qquad& H\rightarrow q\bar{q}gg\nonumber\\
	\gamma^*\rightarrow q\bar{q}q'\bar{q}' \qquad\qquad& H\rightarrow q\bar{q}q\bar{q}\nonumber\\
	& H\rightarrow q\bar{q}q'\bar{q}'
\end{align}
where $V$ and $VV$ denote one-loop and two-loop virtual corrections. We generate all amplitudes with \textsc{Qgraf}~\cite{NOGUEIRA1993279} and perform the Dirac and color algebra in \textsc{Form}~\cite{Vermaseren:2000nd,Kuipers:2012rf,Ruijl:2017dtg}, together with the \textsc{Color} package~\cite{vanRitbergen:1998pn}. We also repeat the calculations in \textsc{FeynCalc}~\cite{Mertig:1990an,Shtabovenko:2016sxi,Shtabovenko:2020gxv,Shtabovenko:2023idz} as a crosscheck. After inserting the energy weights $\cW_{\lambda}^{[N]}$ and the angle measurements, we manage to write all ingredients in terms of Lorentz scalars of loop and external momenta.

The phase-space integrals in Eq.~\eqref{eq:contact_pair_def} and Eq.~\eqref{eq:contact_self_def} are then converted into cut loop integrals by reverse unitarity, followed by the relation
\begin{equation}
\delta_+(p^2)\longrightarrow \frac{1}{2\pi i}
\left(
\frac{1}{p^2-i0}-\frac{1}{p^2+i0}
\right)\,,
\end{equation}
while for the two-particle partition part, the measurement for the chosen particles $(i,j)$ becomes
\begin{align}
&\delta\left(x_L-\frac{1-\cos\theta_{ij}}{2}\right)
=
\frac{p_i\!\cdot p_j}{x_L}\,
\delta\!\left[2x_L(p_i\!\cdot Q)(p_j\!\cdot Q)-p_i\!\cdot p_j\right]
\notag\\
&=
\frac{1}{2\pi i}\frac{p_i\!\cdot p_j}{x_L}
\left\{
\frac{1}{\left[2x_L(p_i\!\cdot Q)(p_j\!\cdot Q)-p_i\!\cdot p_j\right]-i0}
\,-\,
\frac{1}{\left[2x_L(p_i\!\cdot Q)(p_j\!\cdot Q)-p_i\!\cdot p_j\right]+i0}
\right\}\,.
\end{align}
The second line introduces a nonlinear propagator that carries the $x_L$ dependence of the projected measurement.
We then perform the so-called topology classification procedure: classify these integrals into families by their denominator structure after loop-momentum relabeling and shifts, and external momenta exchanging. In this representation the dependence on the observable is isolated in the numerator weight $\cW_{\lambda}^{[N]}(i,j)$ and in the single nonlinear propagator above.

We use \textsc{LiteRed}~\cite{Lee:2012cn,Lee:2013mka} to generate the standard Integration-by-Part (IBP) equations. Since the additional measurement propagator is nonlinear in the loop momenta, the family is first initialized without it; the nonlinear denominator is then inserted directly into the propagator list and scalar-product replacement rules before generating the IBP system. In addition to the standard IBP relations, one needs the identity
\begin{equation}
\label{eq:extra_IBP_contact}
\left(2x_L\,p_i\!\cdot Q\,p_j\!\cdot Q-p_i\!\cdot p_j\right)
\left[\delta\!\left(\mathcal K_{ij}(x_L)\right)\right]^k
=
\left[\delta\!\left(\mathcal K_{ij}(x_L)\right)\right]^{k-1}\,,
\end{equation}
\begin{equation}
\delta\!\left(\mathcal K_{ij}(x_L)\right)
\equiv
\frac{1}{2\pi i}
\left(
\frac{1}{2x_L\,p_i\!\cdot Q\,p_j\!\cdot Q-p_i\!\cdot p_j-i0}
\,-\,
\frac{1}{2x_L\,p_i\!\cdot Q\,p_j\!\cdot Q-p_i\!\cdot p_j+i0}
\right)\,,
\end{equation}
which lowers the power of the measurement propagator and closes the reduction. The resulting system is exported to \textsc{FIRE6}~\cite{Smirnov:2019qkx}, which performs the Laporta reduction after the cut propagators are specified. The self-correlation sector $\lambda=\{N\}$ is simpler because its measurement is already proportional to $\delta(x_L)$ and therefore involves no nonlinear propagator; it reduces to inclusive cut integrals with polynomial energy numerators.
After reduction, the differential cross-section can be written as 
\begin{equation}
\label{eq:contact_MI_expansion}
\frac{d\sigma^{[N]}_{2}}{dx_L}
=
\sum_k \mathcal C_{k}^{[N]}(x_L,\epsilon)\,
\widetilde{\mathcal I}_k(x_L,\epsilon)\,,
\end{equation}
where the coefficient functions $\mathcal C_{k}^{[N]}$ are rational in $x_L$ and $\epsilon$, and the nontrivial analytic dependence is contained in the master integrals $\widetilde{\mathcal I}_k$. We derive differential equations for the master integrals by differentiating with respect to $x_L$ and reducing the resulting integrals back to the master basis. The differential systems are transformed to canonical form with \textsc{Canonica}~\cite{Meyer:2017joq} or \textsc{Libra}~\cite{Lee:2014ioa,Lee:2020zfb}, and then solved iteratively in $\epsilon$. With natural boundary conditions, we obtain the analytic expressions for $0<x_L<1$ in terms of iterated integrals, which in the present problem reduce to harmonic polylogarithms and classical polylogarithms. 

Lastly, we also need to calculate the $\delta(x_L)$ part for the jet function. In this case, we make use the $x_L\to 0$ singular limit. By performing the asymptotic expansion at the level of the differential equation, we find that the leading power takes the following form 
\begin{equation}\label{eq:resum_E2EC}
x_L^{-1-\epsilon}A_1(\epsilon)+x_L^{-1-2\epsilon}A_2(\epsilon)  \,,
\end{equation}
with unknown $\epsilon$-series $A_1(\epsilon)$ and $A_2(\epsilon)$. Then we can match this formula to the analytic expression in the bulk $0<x_L<1$ region, by expanding the latter in $x_L$. This amounts to giving the $\epsilon$ expansion of $A_{1,2}(\epsilon)$. Then we can use the definition of plus distribution to obtain the $\delta(x_L)$ endpoint contribution at the level of differential cross-section, 
\begin{equation}
\label{eq:xLplus}
x_L^{-1-a\epsilon}
=
-\frac{\delta(x_L)}{a\epsilon}
+\left[\frac{1}{x_L}\right]_+
-a\epsilon\left[\frac{\ln x_L}{x_L}\right]_+
+\cO(\epsilon^2)\,.
\end{equation}
This avoids explicitly calculating the asymptotic expansion of individual master integral. Putting everything together, we obtain the analytic solution for the contact terms.

\subsection{Three-particle terms}
\label{sec:3p}
We now turn to the genuinely new part of the higher-point calculation, namely the three-particle sector. For $N=4,5,6$ this corresponds to the partitions in $\Lambda_N^{(3)}$, and at two loops it is generated entirely by the double-real contribution. 
Since we are only interested in extracting the jet functions, we can take the triple-collinear limit. Under this limit, the tree-level four-parton matrix element factorizes into the Born hard process times the $1\to3$ timelike splitting kernel, so this sector can be computed starting from the fully differential triple-collinear distribution.

For a given partition $\lambda=\{a,b,c\}\in\Lambda_N^{(3)}$ we define the measurement operator
\begin{equation}
\widehat{\cM}^{[N]}_{\lambda}(x_1,x_2,x_3)
\equiv
\sum_{i,j,k}
\cW^{[N]}_{\lambda}(i,j,k)\,
\delta(x_1-x_{jk})\,
\delta(x_2-x_{ik})\,
\delta(x_3-x_{ij})\,,
\end{equation}
where the sum runs over all ordered triples of distinct final-state particles. The corresponding three-particle term that contributes to the jet function is
\begin{equation}
\label{eq:Jnonid_general}
J^{[N]}_{\mathrm{3p},\lambda}
=
\int \df\Phi_c^{(3)}
\left(\frac{\mu^2e^{\gamma_E}}{4\pi}\right)^{2\epsilon}
\frac{4g^4}{s_{123}^2}
\sum_{i,j,k}P_{ijk}\,
\widehat{\cM}^{[N]}_{\lambda}\,,
\end{equation}
with $\df\Phi_c^{(3)}$ the standard triple-collinear phase space and $P_{ijk}$ the $1\to3$ splitting function~\cite{Campbell:1997hg,Catani:1998nv,Ritzmann:2014mka}. The full three-particle term is obtained by summing Eq.~\eqref{eq:Jnonid_general} over $\lambda\in\Lambda_N^{(3)}$: $J^{[N],\mathrm{2\mbox{-}loop}}_{\mathrm{3p}}=\sum_{\lambda\in\Lambda_N^{(3)}} J^{[N],\mathrm{2\mbox{-}loop}}_{\mathrm{3p},\lambda}$.

As in the E3C case, it is convenient to parametrize the three measured angles by
\begin{equation}
x_1=x_L z\bar z\,,\qquad
x_2=x_L(1-z)(1-\bar z)\,,\qquad
x_3=x_L\,,
\end{equation}
so that the projected measurement is implemented by fixing the largest angle to be $x_3=x_L$. For each partition $\lambda$, the differential integrand takes the form of
\begin{multline}
\label{eq:nonid_expansion}
\frac{dJ^{[N],\mathrm{2\mbox{-}loop}}_{\mathrm{3p},\lambda}}
{dx_L\,d\re(z)\,d\im(z)}
=
\left(\frac{\mu^2}{Q^2}\right)^{2\epsilon}
\frac{\alpha_s^2}{\pi^3}
\frac{e^{2\epsilon\gamma_E}}{\Gamma(1-2\epsilon)}
\frac{1}{x_L^{1+2\epsilon}}
\frac{1}{(2\im z)^{2\epsilon}}\\
\times \Bigl[
G_{\lambda}(z,\bar z)
+\epsilon F_{\lambda}(z,\bar z)
+\epsilon^2 H_{\lambda}(z,\bar z)
+\cO(\epsilon^3)
\Bigr]\,.
\end{multline}
The dependence on $x_L$ is universal due to the triple collinear limit, so after integrating over $z$ we only need the coefficient of Eq.~\eqref{eq:nonid_expansion} through $\cO(\epsilon)$ to obtain the finite $\delta(x_L)$ term via Eq.~\eqref{eq:xLplus}.

The integration region is identical to the E3C case. Using the $S_3$ symmetry of the three-parton configuration space, the full $z$ plane can be reduced to the fundamental domain
\begin{equation}
\cS
=
\left\{
z\in\mathbb{C}\,\bigg|\,
0\le \im(z)\le \frac{\sqrt{3}}{2}\,,
\qquad
\frac{1}{2}\le \re(z)\le \sqrt{1-\im^2(z)}
\right\},
\end{equation}
and the total three-particle contribution becomes
\begin{equation}
\frac{dJ^{[N]}_{\mathrm{3p}}}{dx_L}
=
\left(\frac{\mu^2}{Q^2}\right)^{2\epsilon}
\frac{\alpha_s^2}{\pi^3}
\frac{e^{2\epsilon\gamma_E}}{\Gamma(1-2\epsilon)}
\frac{6}{x_L^{1+2\epsilon}}
\sum_{\lambda\in\Lambda_N^{(3)}} A_\lambda(\epsilon)\,,
\end{equation}
with
\begin{equation}\label{eq:Alambda_def}
A_\lambda(\epsilon)
\equiv
\int_{\cS}\frac{d\re(z)\,d\im(z)}{(2\im z)^{2\epsilon}}
\Bigl[
G_{\lambda}(z,\bar z)
+\epsilon F_{\lambda}(z,\bar z)
+\epsilon^2 H_{\lambda}(z,\bar z)
+\cO(\epsilon^3)
\Bigr]\,.
\end{equation}

The difficulty is that the integrand develops squeezed singularities when two partons become much closer to each other than to the third one. In the $z$ parametrization, these singular regions sit at $z\to1$ and at its $S_3$ images. To avoid the divergence, we construct the local subtraction term by expanding the $G_{\lambda}(z,\bar z),F_{\lambda}(z,\bar z),H_{\lambda}(z,\bar z)$ in the squeeze $z\to 1$ limit. Note that in the complex plane, there is path dependence when approaching the squeeze limit. To capture that, we introduce $1-z=r \exp(i\theta), 1-\bar z = r \exp(-i\theta)$, take $r\to 0$ for generic $\theta$, and then convert it back to $z,\bar z$. 
For $G_{\lambda}(z,\bar z)$, the $N=3$ result is computed in Ref.~\cite{Chen:2019bpb} and the higher points are bootstrapped recently in Ref.~\cite{Gong:2025jqi}. We can perform the $r$ expansion directly. For $F_{\lambda}(z,\bar z)$ and $H_{\lambda}(z,\bar z)$, in principle, we can also compute them analytically; however, it is more convenient to perform the expansion in the middle of phase space integrations. Notice that a naive expansion in the splitting function will lead to the wrong result since it does not commute with the integration, so instead, we perform the first two energy integrals in $\df\Phi_c^{(3)}$ in Eq.~\eqref{eq:Jnonid_general}.
Eventually, for each partition $\lambda$ we denote them as 
\begin{equation}
G_{\lambda}^{\mathrm{sq}}(z,\bar z)
+\epsilon F_{\lambda}^{\mathrm{sq}}(z,\bar z)
+\epsilon^2 H_{\lambda}^{\mathrm{sq}}(z,\bar z)\,,
\end{equation}
and rewrite Eq.~\eqref{eq:Alambda_def}
\begin{align}
A_\lambda(\epsilon)
&=
\int_{\cS}\frac{d\re(z)\,d\im(z)}{(2\im z)^{2\epsilon}}
\Bigl[
G_{\lambda}^{\mathrm{sq}}
+\epsilon F_{\lambda}^{\mathrm{sq}}
+\epsilon^2 H_{\lambda}^{\mathrm{sq}}
\Bigr]
\notag\\
&\quad+
\int_{\cS}\frac{d\re(z)\,d\im(z)}{(2\im z)^{2\epsilon}}
\Bigl[
\bigl(G_{\lambda}-G_{\lambda}^{\mathrm{sq}}\bigr)
+\epsilon\bigl(F_{\lambda}-F_{\lambda}^{\mathrm{sq}}\bigr)
\Bigr]
\,+\,\cO(\epsilon^2)\,.
\end{align}
The first integral is evaluated analytically in $d=4-2\epsilon$ dimensions. The second one is finite and can therefore be expanded in $\epsilon$ before integration. As mentioned above, we do not have the analytic form for $G_\lambda$ and $F_\lambda$ for higher points, and thus we also need to evaluate the phase space integrals numerically. In practice, we implement the second term in \textsc{C++} and perform the Monte-Carlo integration with \textsc{Cuba}~\cite{Hahn:2004fe}. We also repeat the E3C calculation as a cross-check.

\begin{figure}[!htbp]
	\centering
	\includegraphics[width=0.43\textwidth]{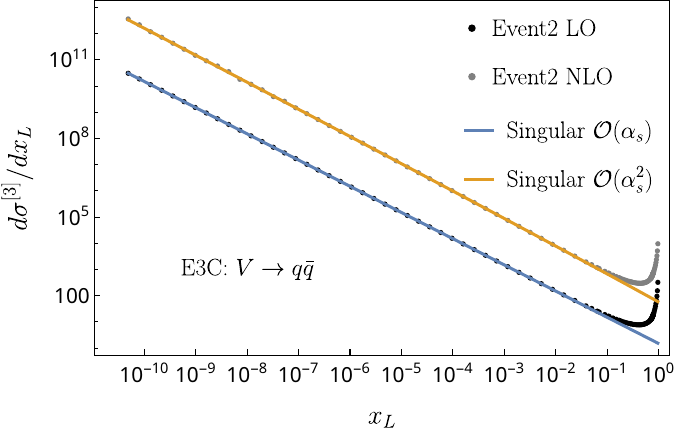}
	\includegraphics[width=0.43\textwidth]{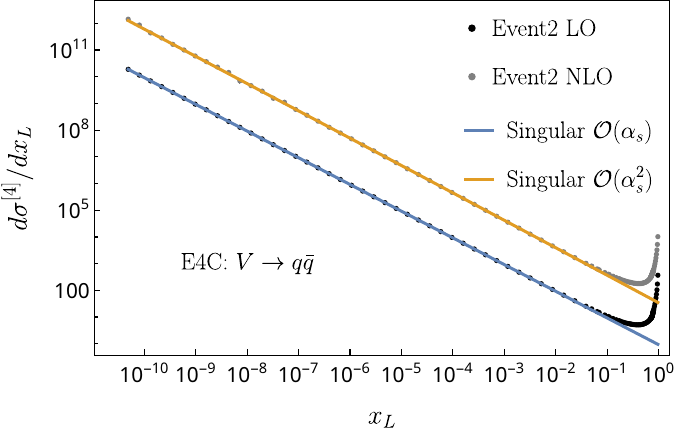}\\
	\vspace{0.2cm}
	\includegraphics[width=0.43\textwidth]{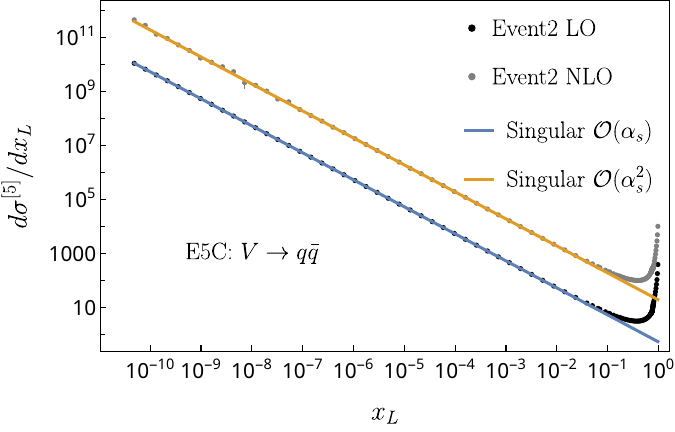}
	\includegraphics[width=0.43\textwidth]{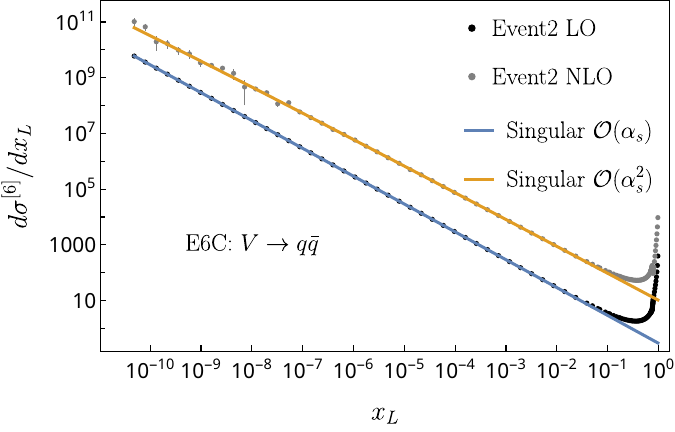}
	\caption{Good agreement between our two-loop calculation for the collinear limit of E3C to E6C and numerical results from \textsc{Event2}. The distribution is normalized to the Born cross-section $\sigma_0^{e^{+}e^{-}}$. Note that here LO and NLO correspond to perturbative coefficients at $\mathcal{O}(\alpha_s / (4\pi))$ and $\mathcal{O}(\alpha_s^2 / (4\pi)^2)$, respectively.
}
	\label{fig:compareEvent2}
\end{figure}

\begin{figure}[!htbp]
	\centering
	\includegraphics[width=0.43\textwidth]{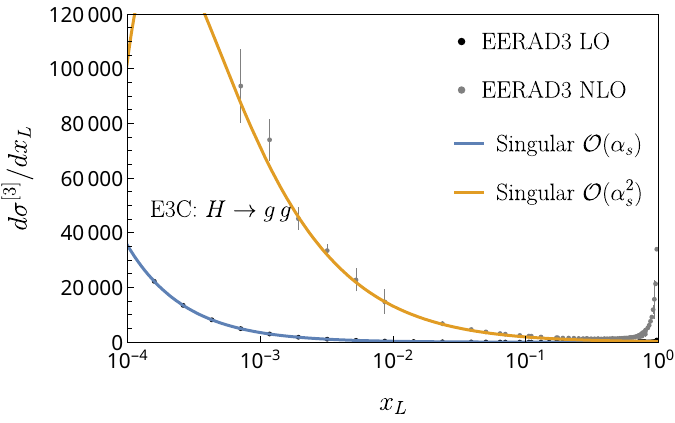}
	\includegraphics[width=0.43\textwidth]{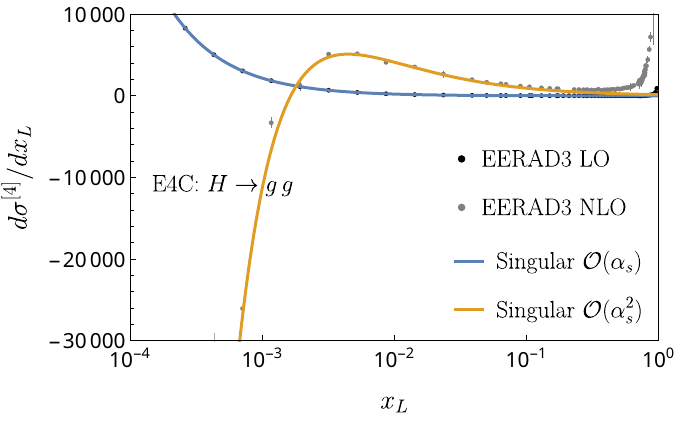}\\
	\vspace{0.2cm}
	\includegraphics[width=0.43\textwidth]{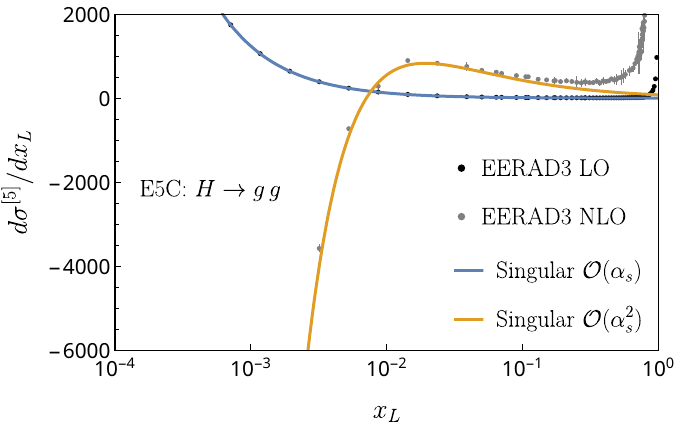}
	\includegraphics[width=0.43\textwidth]{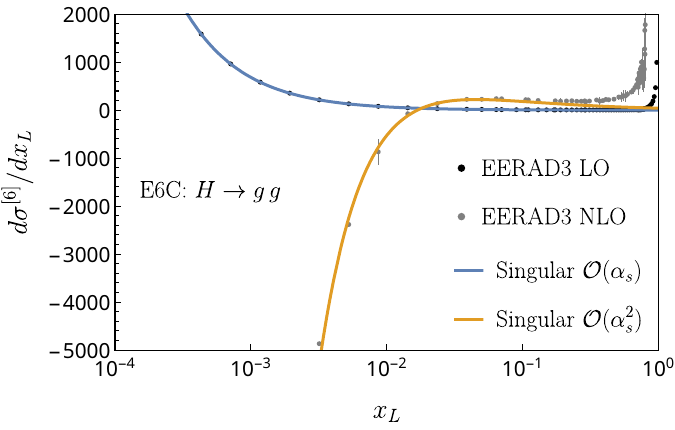}
	\caption{Good agreement between our two-loop calculation for the collinear limit of E3C to E6C and numerical results from \textsc{Eerad3}. The distribution is normalized to the Born cross-section $\Gamma_0^{H\to gg}$.}
	\label{fig:compareEERAD3}
\end{figure}
\subsection{Results}
\label{sec:jet_results}
Combining the contact terms and the three-particle terms, the singular two-loop cross section takes the form
\begin{align}
\label{eq:Nloop_match_q}
\frac{1}{\sigma_0}\frac{d\sigma^{[N],\mathrm{2\mbox{-}loop}}_{e^+e^-,H}}{dx_L}
&=
2\,\frac{dJ^{[N],\mathrm{2\mbox{-}loop}}_{\mathrm{3p}}}{dx_L}
\,+\,
\frac{1}{\sigma_0}\frac{d\sigma^{[N],\mathrm{2\mbox{-}loop}}_{C}}{dx_L}\,.
\end{align}
Again the Born cross-section $\sigma_0=\sigma_0^{e^{+}e^{-}}$ in $e^+e^-\to q\bar{q}$ and $\Gamma_0^{H\to gg}$ in  $H\to gg$.
The factor of $2$ appears because the fixed-order processes contain two primary jets, while the jet function in the factorization theorem describes a single parent parton. The two-loop jet constant $j_2^{i,[N]}$ (which is equal to $\vec{c}^{\,[N]}_{2,0}$ in the jet ansatz in Eq.~\eqref{eq:jetansatz}) is then read off from the coefficient of $\delta(x_L)$ after matching to the hard functions. In every channel the explicit $\mu$ dependence agrees with the prediction of the jet evolution equation in Eq.~\eqref{eq:jet_evo}, and the cancellation of all infrared $1/\epsilon$ poles provides a strong check of the calculation.

Our final results for the new higher-point jet constants are
\begin{align}
	{\color{blue} 2^4 j_2^{q,[4]}}&=C_F T_F n_f (124.967567\pm 0.002166) + C_F^2 (254.438849 \pm 0.018860)\notag\\
	&+ C_F C_A (-269.979451 \pm 0.009042)\,,\notag\\
	{\color{blue} 2^5 j_2^{q,[5]}}&=C_F T_F n_f (147.891331 \pm 0.002602) + C_F^2 (339.713439 \pm 0.021279)\notag\\
	&+ C_F C_A (-323.170146 \pm 0.009702)\,,\notag\\
	{\color{blue} 2^6 j_2^{q,[6]}}&=C_F T_F n_f (168.459145 \pm 0.003037) + C_F^2 (426.020756 \pm 0.024441)\notag\\
	&+ C_F C_A (-371.824995 \pm 0.011163)\,,\\
	{\color{red} 2^4 j_2^{g,[4]}}&=C_F T_F n_f ( -21.592489 \pm 0.001800 )+ C_A T_F n_f ( 182.634493 \pm 0.003985 )\notag\\
	&+C_A^2 ( -27.054523 \pm 0.022288 )+ n_f^2 T_F^2 \left( \frac{52689656}{3472875}-\frac{608 \pi ^2}{945} \right)\,,\notag\\
	{\color{red} 2^5 j_2^{g,[5]}}&=C_F T_F n_f ( -25.583483 \pm 0.001906 )+ C_A T_F n_f ( 219.793744 \pm 0.007128 )\notag\\
	&+C_A^2 ( 2.207608 \pm 0.036991 )+ n_f^2 T_F^2 \left( \frac{11872972}{694575}-\frac{136 \pi ^2}{189} \right)\,,\notag\\
	{\color{red} 2^6 j_2^{g,[6]}}&=C_F T_F n_f ( -28.753959 \pm 0.002008 )+ C_A T_F n_f ( 253.496395 \pm 0.005500 )\notag\\
	&+C_A^2 ( 37.086330 \pm 0.026721 )+ n_f^2 T_F^2 \left( \frac{348270772}{18753525}-\frac{440 \pi ^2}{567} \right)\,.
\end{align}
The quoted uncertainties are purely the Monte-Carlo uncertainty from \textsc{Cuba} in the three-particle calculations, 
except for the $n_f^2 T_F^2$ channel in the gluon jet, where the integration is done analytically. 
As a cross-check, our setup reproduces the known $N=2$ constants of Ref.~\cite{Dixon:2019uzg} 
and $N=3$ constants of Ref.~\cite{Chen:2023zlx}. In addition, for the quark channel, 
we can compare our prediction in Eq.~\eqref{eq:Nloop_match_q} with \textsc{Event2}~\cite{Catani:1996jh,Catani:1996vz} in the small $x_L$ region. 
For the gluon channel, we compare with \textsc{Eerad3}~\cite{Gehrmann-DeRidder:2014hxk,Aveleira:2025svg} as well. 
Note that we only calculate up to $x_L\sim 10^{-4}$ in the latter case because the numerical result becomes unstable with smaller $x_L$.
In Fig.~\ref{fig:compareEvent2} and Fig.~\ref{fig:compareEERAD3}, 
we compare the numerical coefficients from the fixed-order programs with our two-loop calculations and find good agreement.
In App.~\ref{app:singularNNLO}, we provide the analytic expression of singular expansion up to $\mathcal{O}(\alpha_s^3)$.

%%%%%%%%%%%%%%%%%%%%%%%%%%%%%%%%%%%%%%%%%%%%%%%%%%%%%%%%%%%%%%%%%%%%%%%%%%%%%%%%%%%%%%%%%%%%%%%%%%%%%%%%%%%%%%%%%%%%%%%%%%%%%%%%%%%%%%%%%%%%%%%%%%%%%%%%%%%%%%%%%%%%%%%%%%%%%%%%%%%%%%%%%%%%%%%%%%%%%%%%%%%%%%%

\section{NNLL collinear resummation}\label{sec:nnll_resum}

With the two-loop jet constants determined in Sec.~\ref{sec:jet_func}, all ingredients required for NNLL resummation up to six-point projected energy correlators are now available. In this section we present the resummation for both $e^+e^-\to q\bar q$ and $H\to gg$ with matching to available fixed-order results. In addition, we also include the leading non-perturbative corrections proportional to $\overline{\Omega}_{1q,1g}$.

\subsection{Matching procedure and nonperturbative power corrections}

To obtain precise predictions from NNLL resummation, we need to determine the resummation scales and estimate uncertainties from missing higher orders. To eliminate the large logarithms associated with the hard and jet function boundaries, we choose the canonical scales
\begin{equation}
\label{eq:can_scales}
\mu_h=e_h Q\,,
\qquad
\mu_j=e_h e_j Q\sqrt{x_L}\,,
\end{equation}
with $(e_h,e_j)=(1,1)$ for the central values. Here, $e_h\in[1/2,2]$ and $e_j\in[1/2,2]$ control the hard scale and the jet scale variations, respectively. In this work, we choose the envelope of the following seven-point scale variations:
\begin{equation}
\label{eq:can_scale_var}
(e_h,e_j)\in
\left\{
(1,1),\,
(2,1),\,
\left(\frac{1}{2},1\right),\,
(1,2),\,
\left(1,\frac{1}{2}\right),\,
(2,2),\,
\left(\frac{1}{2},\frac{1}{2}\right)
\right\}\,.
\end{equation}
In principle, we will also need to consider a profile function to transition 
from non-perturbative scales $\gtrsim\Lambda_{\text{QCD}}$ in very small $x_L$ 
to canonical scales in Eq.~\eqref{eq:can_scales}. 
This in particular requires a prescription in the canonical scale 
to avoid $\mu_j$ being too small and going out of the perturbative regime. Here, we will not implement such transition and our perturbative curves should not be trusted in the non-perturbative region. The onset of where the transition occurs will be discussed below and will depend actually on the process and choice of $N$-point projected energy correlators.

We also need to include the non-singular contribution from the fixed-order calculation,
\begin{equation}
\label{eq:enc_match}
\frac{1}{\sigma_{\rm tot}}\frac{d\hat{\sigma}^{[N]}_{\rm match}}{dx_L}
=
\frac{1}{\sigma_{\rm tot}}\frac{d\hat{\sigma}^{[N]}_{\rm resum}}{dx_L}
+
\frac{1}{\sigma_{\rm tot}}\frac{d\hat{\sigma}^{[N]}_{\rm ns}}{dx_L}
\,,
\end{equation}
where $\frac{d\hat{\sigma}^{[N]}_{\rm ns}}{dx_L}$ becomes increasingly important as one approaches the bulk angular region of the distribution. It is defined as
\begin{equation}
\label{eq:match}
d\hat{\sigma}^{[N]}_{\rm ns}\equiv d\hat{\sigma}^{[N]}_{\rm FO}-d\hat{\sigma}^{[N]}_{\rm sing}\,,
\end{equation}
with $d\hat{\sigma}^{[N]}_{\rm sing}$ the singular expansion of the resummed distribution through the order in $\alpha_s$ to which we match. Standard logarithmic counting will require matching N${}^k$LL to N${}^{k-1}$LO, 
i.e. $\mathcal{O}(\alpha_s^k)$. 
In principle, the transition from the resummation region to the fixed-order region should also include turning off the resummation via profile functions~\cite{Abbate:2010xh}. In practice, we find the naive matching in Eq.~\eqref{eq:match} sufficient, since the resummation effects already taper off in the region where the non-singular contribution becomes important. Note that we have been using the hat symbol in $\hat{\sigma}$ to denote purely perturbative predictions. The full prediction requires including nonperturbative power corrections:
\begin{align}
\frac{1}{\sigma_{\rm tot}}\frac{d\sigma^{[N]}}{dx_L} = \frac{1}{\sigma_{\rm tot}}\frac{d\hat{\sigma}^{[N]}_{\rm match}}{dx_L} + \frac{1}{\sigma_{\rm tot}}\frac{d\sigma^{[N]}_{\rm match, NP}}{dx_L}\,,
\label{eq:full}
\end{align}
where $d\sigma^{[N]}_{\rm match, NP}$ is the additive leading-power NP corrections which we now discuss. 

The leading non-perturbative power corrections~\cite{Lee:2024esz,Chen:2024nyc} for projected energy correlators are given by the representation-dependent soft matrix element
\begin{equation}
\label{eq:Omega1_def}
\Omega_{1\kappa}\equiv
\frac{1}{N_\kappa}
\langle 0|
\mathrm{tr}\!\left[
\overline{Y}_{\bar n}^{\dagger \kappa}
Y_n^{\dagger \kappa}
\mathcal{E}_T(0)
Y_n^\kappa
\overline{Y}_{\bar n}^\kappa
\right]
|0\rangle\,,
\qquad
\kappa=q,g\,,
\end{equation}
with $N_q=N_c$ and $N_g=N_c^2-1$. 
In the fixed-order region, the distribution with leading non-perturbative power correction 
in the $\overline{\rm MS}$ scheme takes the form
\begin{equation}
\label{eq:np_fixed}
\frac{d\sigma^{[N]}_{\rm FO}}{dx_L}
=
\frac{d\hat{\sigma}^{[N]}_{\rm FO}}{dx_L}
+
\sigma_0\,\frac{N}{2^N}\frac{\overline{\Omega}_{1\kappa}}{Q\,[x_L(1-x_L)]^{3/2}}\,,
\end{equation}
which exhibits enhancement as we approach the kinematic endpoints $x_L \to 0,1$. For $e^+e^-$ and $H\to gg$ processes, the soft matrix elements are in $\kappa = q,g$ representation at LO, respectively. 
For the collinear resummation, the leading non-perturbative power correction is realized 
in the jet boundary 
\begin{equation}
\label{eq:np_jet}
2^N J^{\kappa[N]}\!\left(\ln\frac{x_L x^2 Q^2}{\mu_j^2},\mu_j\right)
=
2^N \hat{J}^{\kappa[N]}\!\left(\ln\frac{x_L x^2 Q^2}{\mu_j^2},\mu_j\right)
-
\frac{N\,\overline{\Omega}_{1\kappa}}{x Q\sqrt{x_L}}\,
2^{N-1}
\hat{\mathcal J}^{\kappa[N-1]}\!\left(\ln\frac{x_L x^2 Q^2}{\mu_j^2},\mu_j\right)\,,
\end{equation}
where the prefactor $2^N$ follows from our jet-function normalization convention. The important point is that the power-correction term is proportional to
the perturbative coefficient $\hat{\mathcal J}^{\kappa[N-1]}$, whose logarithmic evolution
is the same as that of the $(N-1)$-point jet function, up to different boundary constants. Therefore, the RG of the jet function associated with nonperturbative power corrections can also be solved using the ansatz given in Eq.~\eqref{eq:jetansatz}. At LL accuracy, only the leading matching coefficient $\overline{\Omega}_{1\kappa}$ is required. Although Eq.~\eqref{eq:np_jet} appears as a product of $\overline{\Omega}_{1\kappa}$ and
$\hat{\mathcal J}^{\kappa[N-1]}$, $\overline{\Omega}_{1\kappa}$ should be treated as the zeroth-order
jet constant in the RG evolution, i.e. $\vec{c}_{0,0}^{[N-1]}$ in Eq.~\eqref{eq:jetansatz}.
Throughout this paper, we keep the $\overline{\rm MS}$ form in Eq.~\eqref{eq:np_jet}
and do not perform renormalon subtraction~\cite{Schindler:2023cww,Lee:2024esz}. 

The leading singular expansion of the resummed non-perturbative corrections is simply given by 
\begin{align}
\frac{N}{2^N}\frac{\overline{\Omega}_{1\kappa}}{Q\,x_L^{3/2}}\,.
\end{align}
Therefore, the corresponding non-singular contribution of the nonperturbative corrections is given as
\begin{equation}
\frac{d\sigma^{[N]}_{\rm ns,NP}}{dx_L}\equiv \sigma_0\,
\frac{N}{2^N}\frac{\overline{\Omega}_{1\kappa}}{Q}
\left[\frac{1}{[x_L(1-x_L)]^{3/2}}-\frac{1}{x_L^{3/2}}\right]\,.
\end{equation}
Combining with the collinear resummation for the nonperturbative corrections, the matched contribution is given as
\begin{equation}
    \label{eq:enc_match_NP}
\frac{1}{\sigma_{\rm tot}}\frac{d\sigma^{[N]}_{\rm match, NP}}{dx_L}
=
\frac{1}{\sigma_{\rm tot}}\frac{d\sigma^{[N]}_{\rm resum,NP}}{dx_L}+\frac{1}{\sigma_{\rm tot}}\frac{d\sigma^{[N]}_{\rm ns,NP}}{dx_L}\,.
\end{equation}

Finally, as an observable, projected ENC obeys the sum rule
\begin{align}
\int_0^1 dx_L \frac{d\sigma^{[N]}}{dx_L} = \sigma_{\rm tot}\,.
\end{align}
This sum rule makes it natural to normalize projected energy correlators by its total cross-section or total decay width.\footnote{Although we have been using $\sigma_{\rm tot}$ in this section, it should be interpreted as $\Gamma_{\rm tot}^{H\to gg}$ for the Higgs to gluon decay case.} 
This sum rule is obeyed order-by-order in the fixed order calculations, though with resummation and matching it is less obvious to make the full distribution obey the sum rule. We normalize our projected ENC distributions by the total cross section (decay width) computed to the same order in $\alpha_s$. Since the differential ENC starts at $\mathcal{O}(\alpha_s)$ while the total rate starts at $\mathcal{O}(\alpha_s^0)$, an N$^k$LL+N$^{k-1}$LO distribution, whose fixed-order content reaches $\mathcal{O}(\alpha_s^{k})$, is normalized by the N$^{k}$LO total cross section, which is likewise of $\mathcal{O}(\alpha_s^{k})$.

The total cross-section $\sigma_{\rm tot}^{e^+e^-}$ and total decay width $\Gamma^{H\to gg}$ are known to N$^4$LO accuracy and are given in terms of their Born-level results and perturbative correction factor $K = 1 + a_s K^{(1)}+ a_s^2 K^{(2)}+ a_s^3 K^{(3)}+ a_s^4 K^{(4)}+\cdots $ with $a_s=\alpha_s(\mu)/(4\pi)$ as
\begin{align}
\sigma_{\rm tot}^{e^+e^-} &\equiv \sigma_{0}^{e^+e^-} K_{e^+e^-}(\mu)\,,\nn\\
\Gamma^{H\to gg}_{\rm tot} &\equiv  \Gamma^{H\to gg}_0\, K_{H\to gg}(\mu)\,.
\end{align}
For $e^+e^-\to q\bar q$, the correction $K_{e^+e^-}(\mu)$ at scale $\mu=m_Z$ is given as
\begin{align}
    K_{e^+e^-}(m_Z) &= 1 + 4\, a_s + a_s^2 \left( 170 - \frac{368}{3}\zeta_3 \right)+ a_s^3 \left( \frac{745802}{81} - \frac{2116}{27}\pi^2 - \frac{2810096}{297}\zeta_3 + \frac{18400}{9}\zeta_5 \right) \nn \\
  &+ a_s^4 \left( \frac{4616777951}{8019} - \frac{296470}{27}\pi^2 - \frac{593880640}{891}\zeta_3 + \frac{194672}{27}\pi^2\zeta_3 \right. \nn \\
  &\qquad\qquad \left. {} + \frac{10800800}{99}\zeta_3^{\,2} + \frac{27925880}{297}\zeta_5 - \frac{244720}{9}\zeta_7 \right) \,.
\end{align}
For $H\to gg$, if we choose the $\overline{\text{MS}}$ scheme, the ratio $K_{H\to gg}(\mu)$ at $\mu=m_H$ is given as
\begin{align}
    K_{H\to gg}(m_H)&=1+\frac{149}{3}a_s+a_s^2 \left(\frac{148409}{54}-\frac{529 \pi ^2}{9}-890 \zeta_3\right)+a_s^3 \left(\frac{115266007}{729}-\frac{192326 }{27}\pi ^2\right.\nn\\
    &\hspace{-0.5cm}\left.-\frac{2903752 }{27}\zeta_3+\frac{241880}{9} \zeta_5\right)+a_s^4 \left(\frac{84364936609}{8748}-\frac{481238662 }{729}\pi ^2-\frac{2399426306 }{243}\zeta_3\right.\nn\\
    &\hspace{-0.5cm}\left.+\frac{4374047 }{1215}\pi ^4+\frac{11416400 }{3}\zeta_5+\frac{4708100}{27}\pi ^2 \zeta_3+\frac{8713100 }{9}\zeta_3^2-\frac{24107300}{27}\zeta_7\right)\,.
    \label{eq:KHgg}
\end{align}
In both cases, we have plugged in the $\rm SU(3)$ color factors. Note that in the $K_{H\to gg}$ expression, we do not include the matching coefficient $|C(m_t,\mu)|^2$ as in Ref.~\cite{Luo:2019nig}. This coefficient also enters the numerator of the projected energy correlator. Since we normalize our distributions to the decay width, $|C(m_t,\mu)|^2$ cancels between numerator and denominator, and can be dropped from both provided this is done consistently\footnote{Note that in \textsc{Eerad3}, the $|C(m_t,\mu)|^2$ contribution is included in the numerical results, so we remove it manually in this work.}.  Finally, we can combine all these ingredients according to Eq.~\eqref{eq:full}.

\begin{figure}[!hbp]
    \centering
    \includegraphics[width=0.32\linewidth]{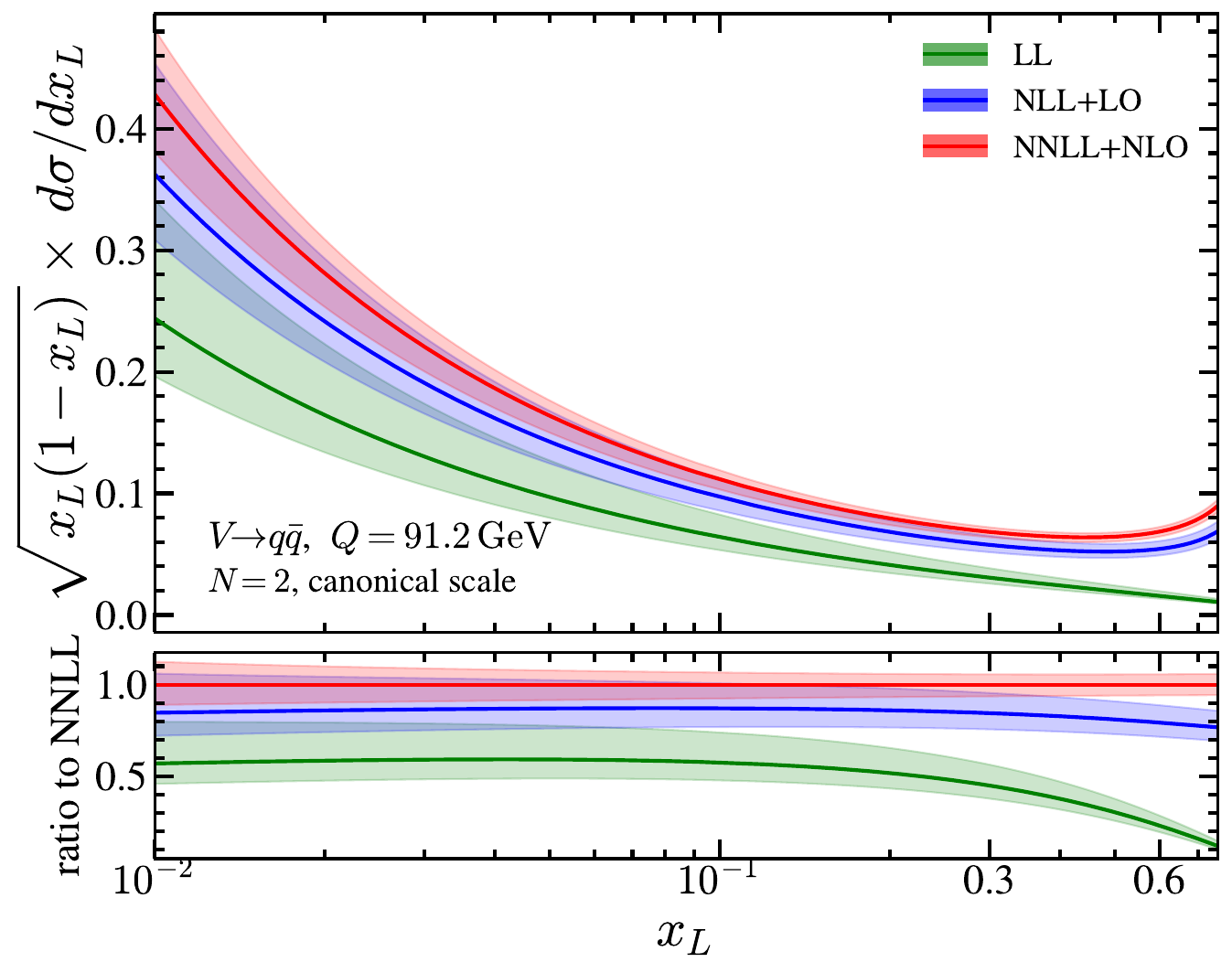}
    \includegraphics[width=0.32\linewidth]{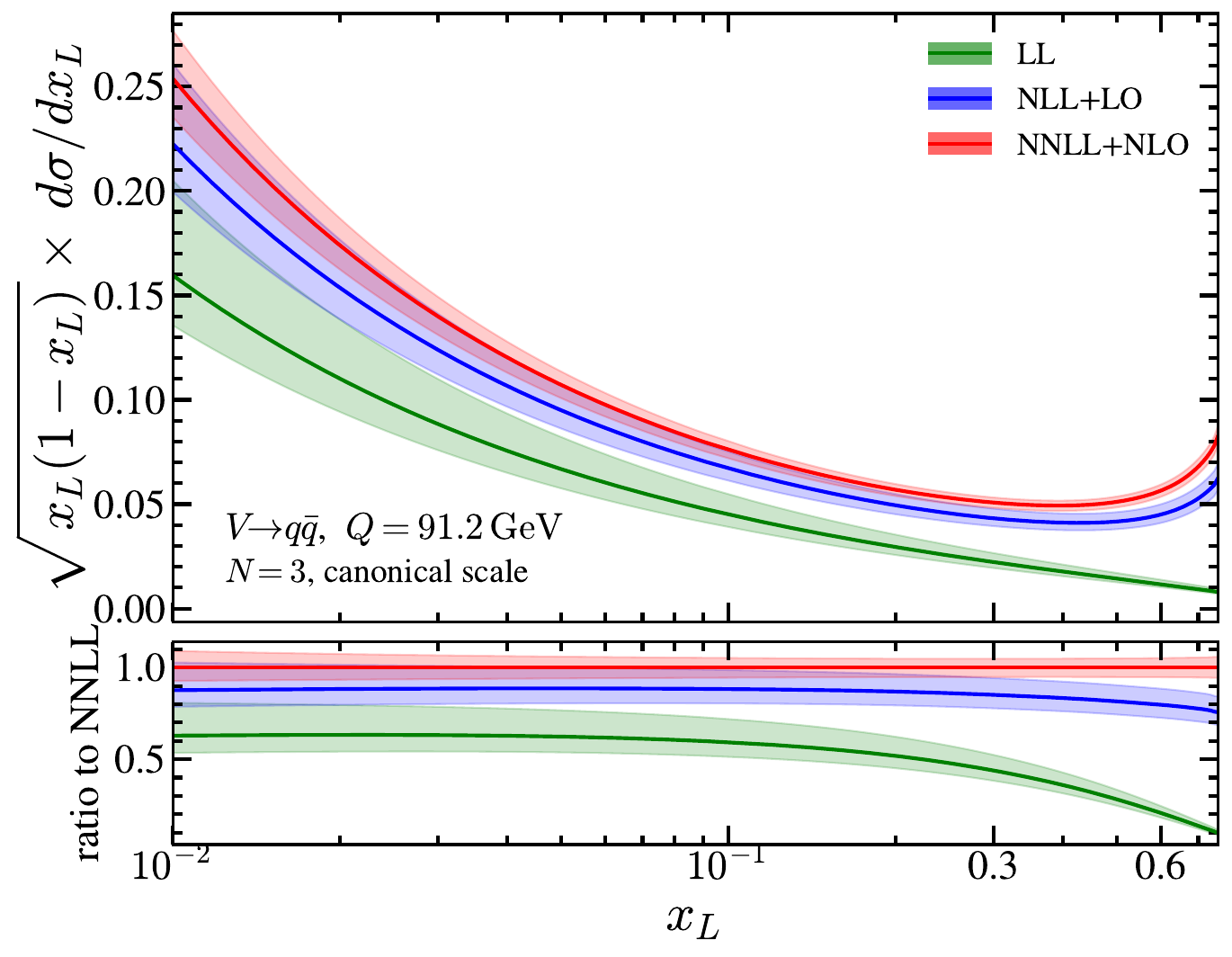}
    \includegraphics[width=0.32\linewidth]{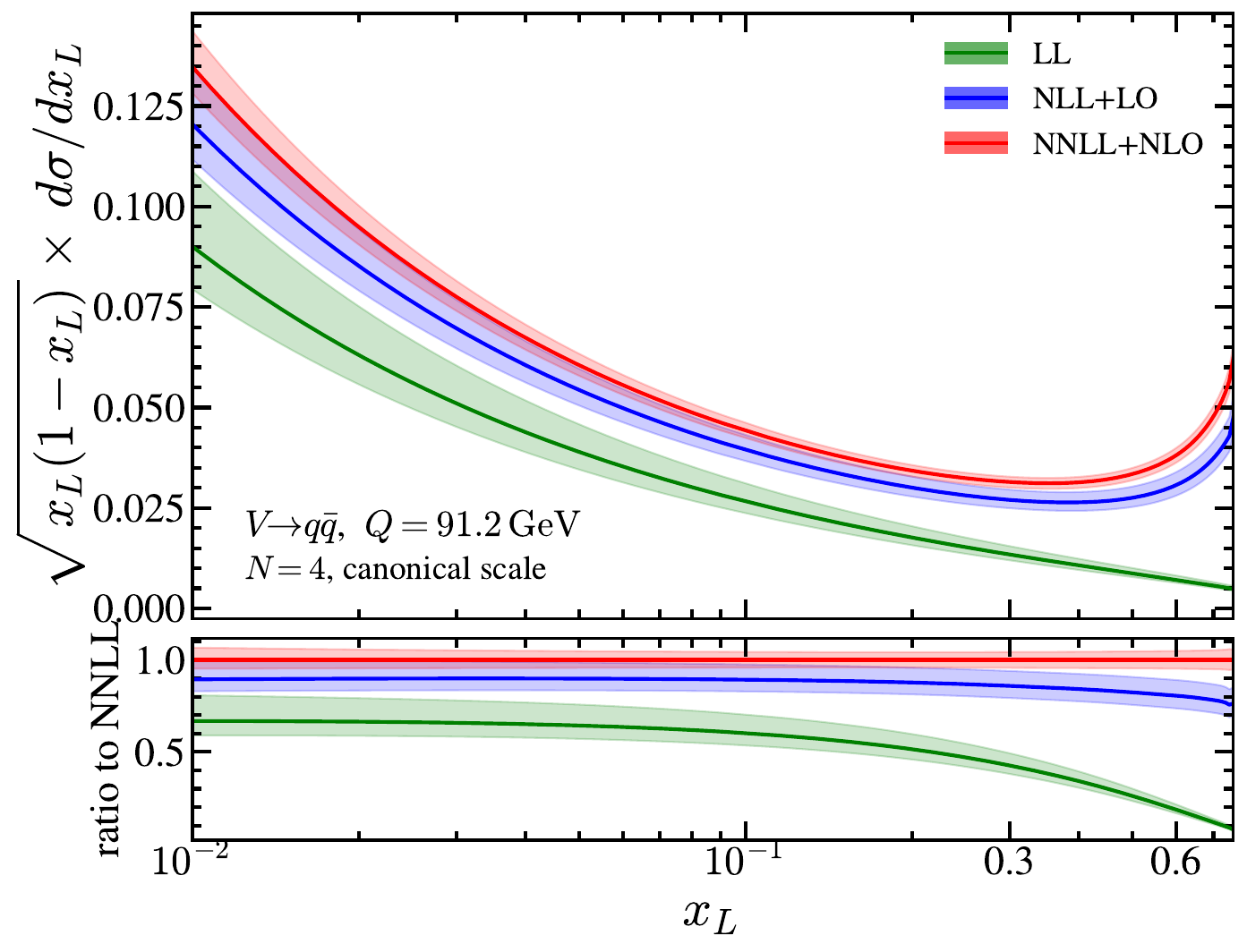}\\
    \includegraphics[width=0.32\linewidth]{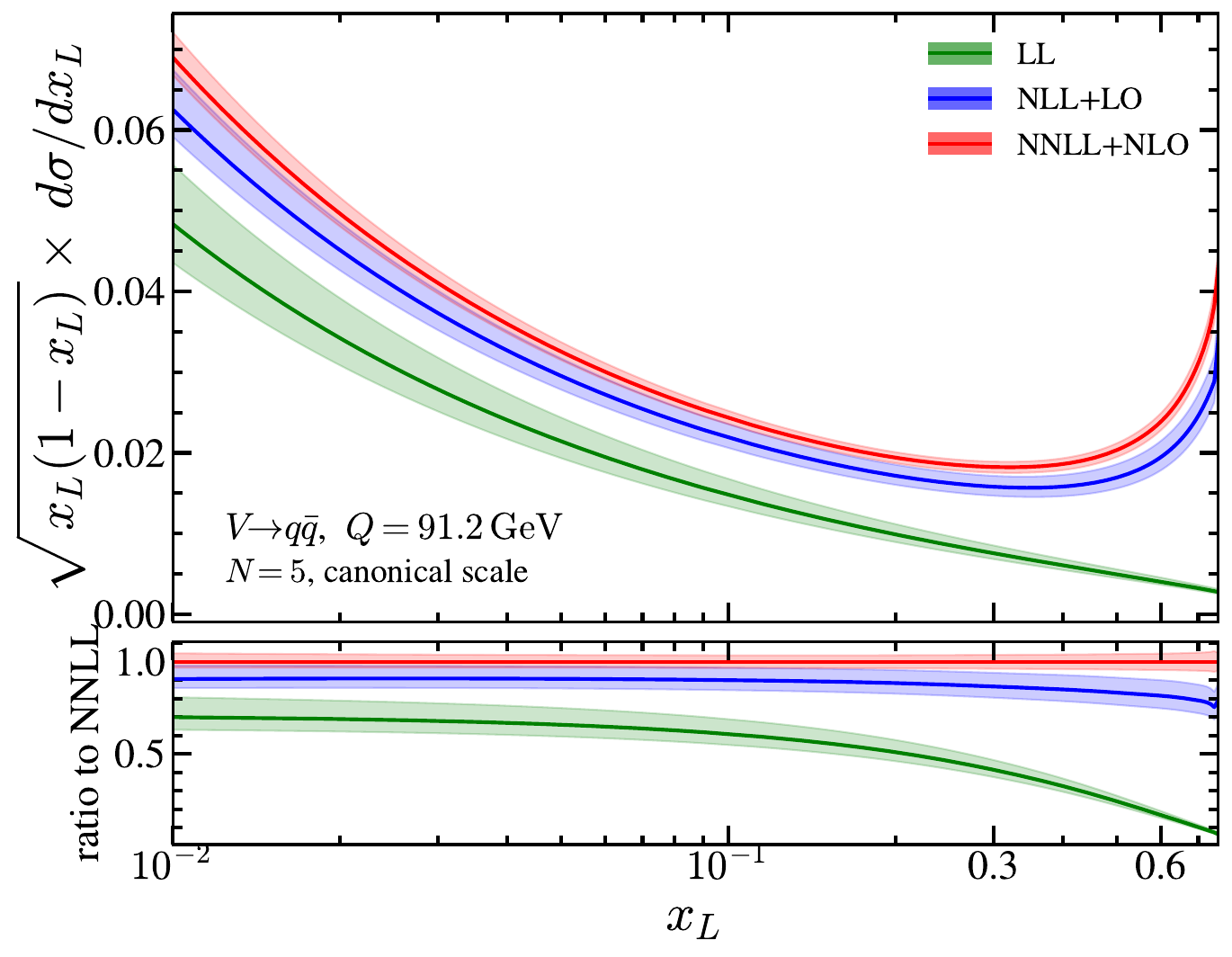}
    \includegraphics[width=0.32\linewidth]{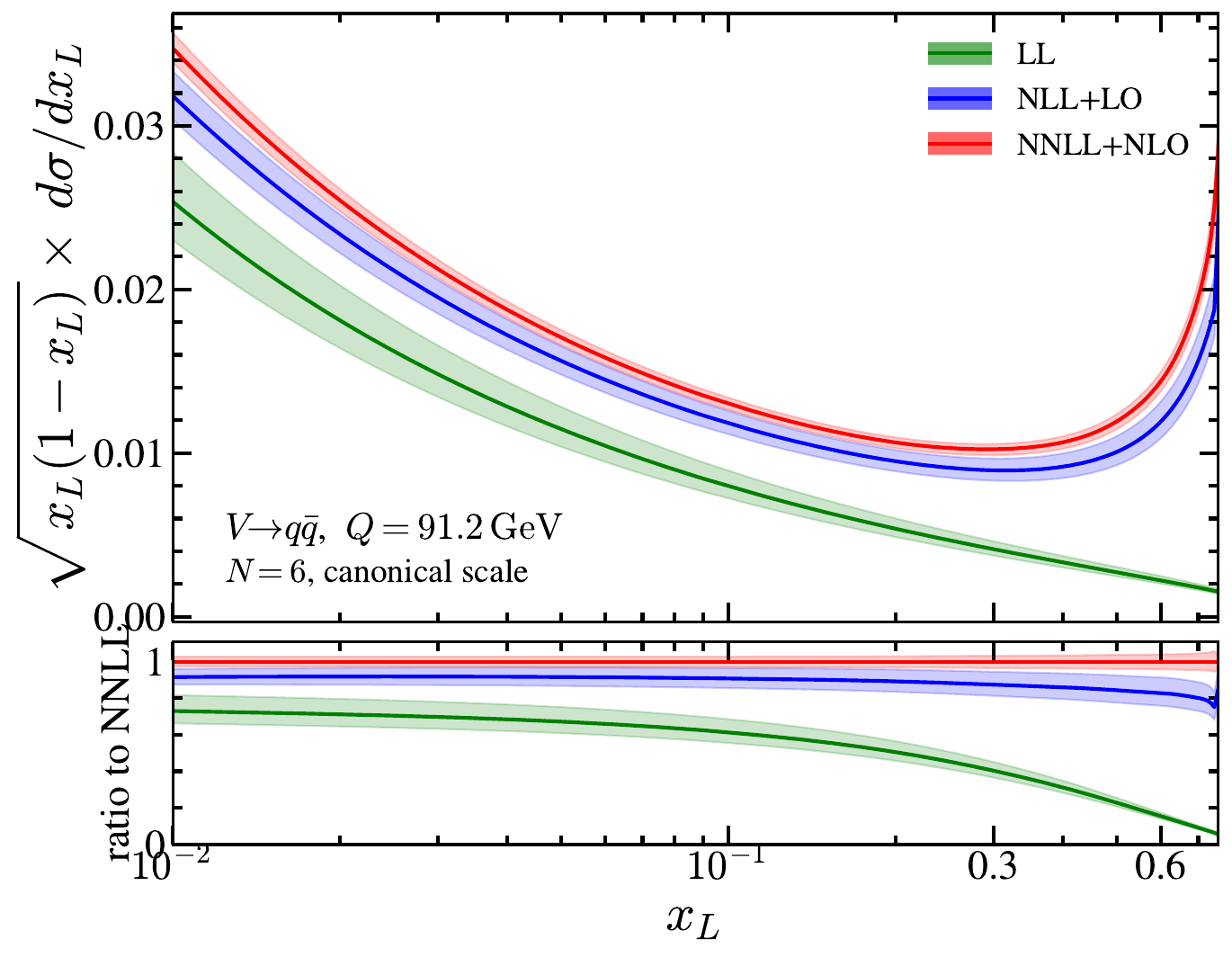}
    \includegraphics[width=0.32\linewidth]{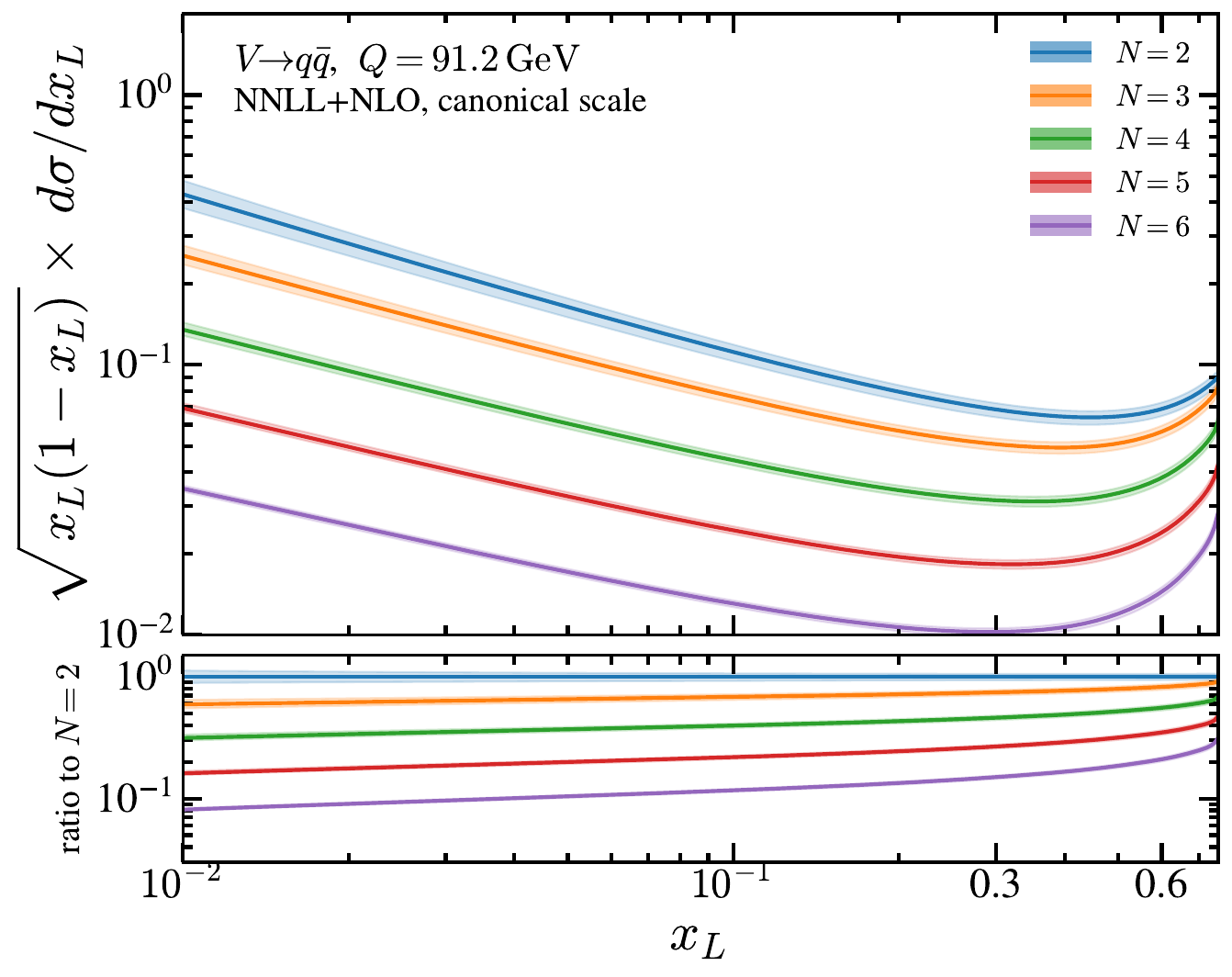}
    \caption{Matched predictions for E2C to E6C in $e^+e^-\to q\bar q$. In the first five figures, the upper panels show the spectra with the sequence LL, NLL+LO, and NNLL+NLO and the lower panels show the ratio to NNLL+NLO, respectively. The uncertainty band is the envelope of the seven-point scale variation in Eq.~\eqref{eq:can_scale_var}. In the last figure, the upper panel shows the highest order for all $N$ together and the lower panel shows their ratios to EEC.}
    \label{fig:match_can_pert_q}
\end{figure}

\begin{figure}[!hbp]
    \centering
    \includegraphics[width=0.32\linewidth]{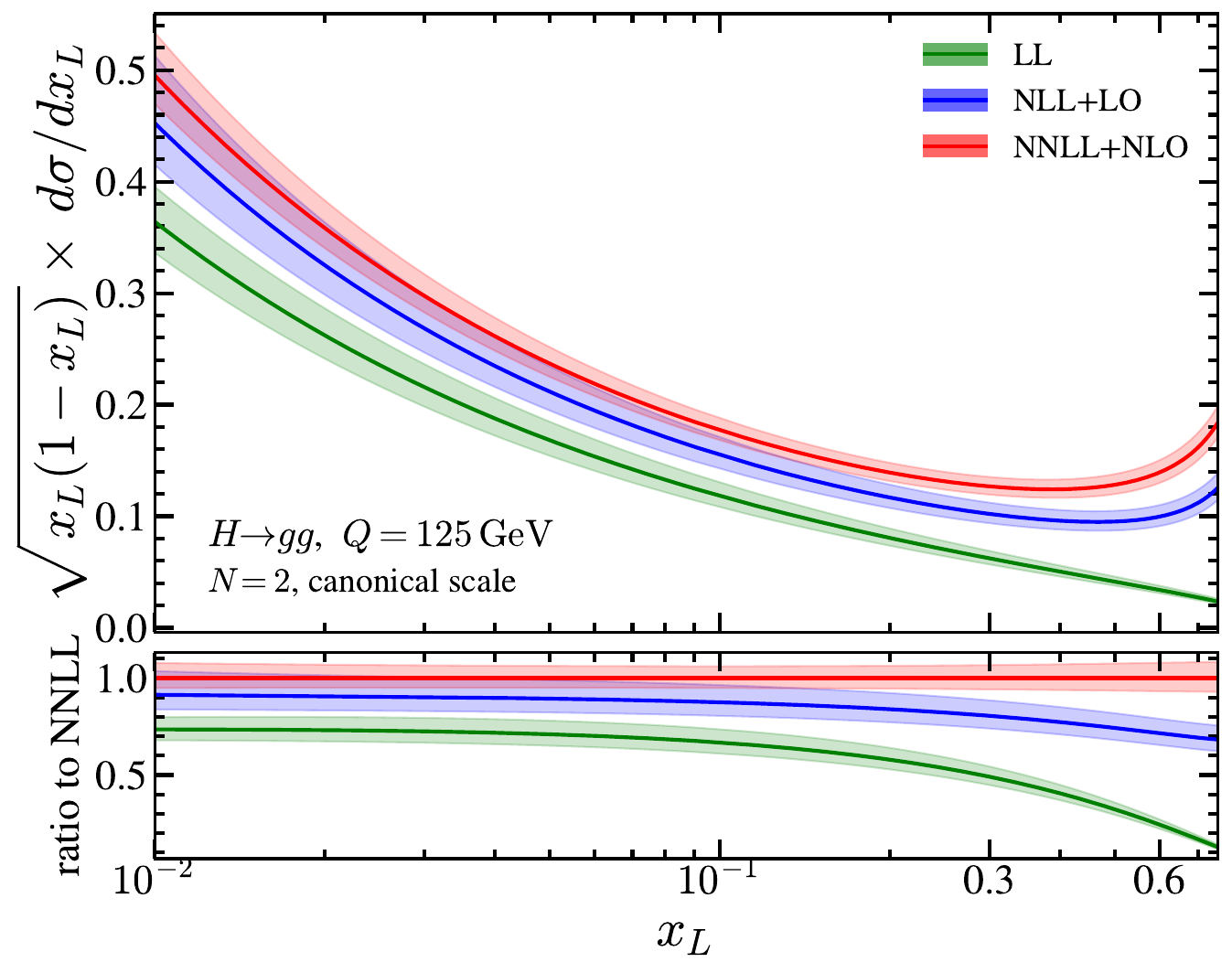}
    \includegraphics[width=0.32\linewidth]{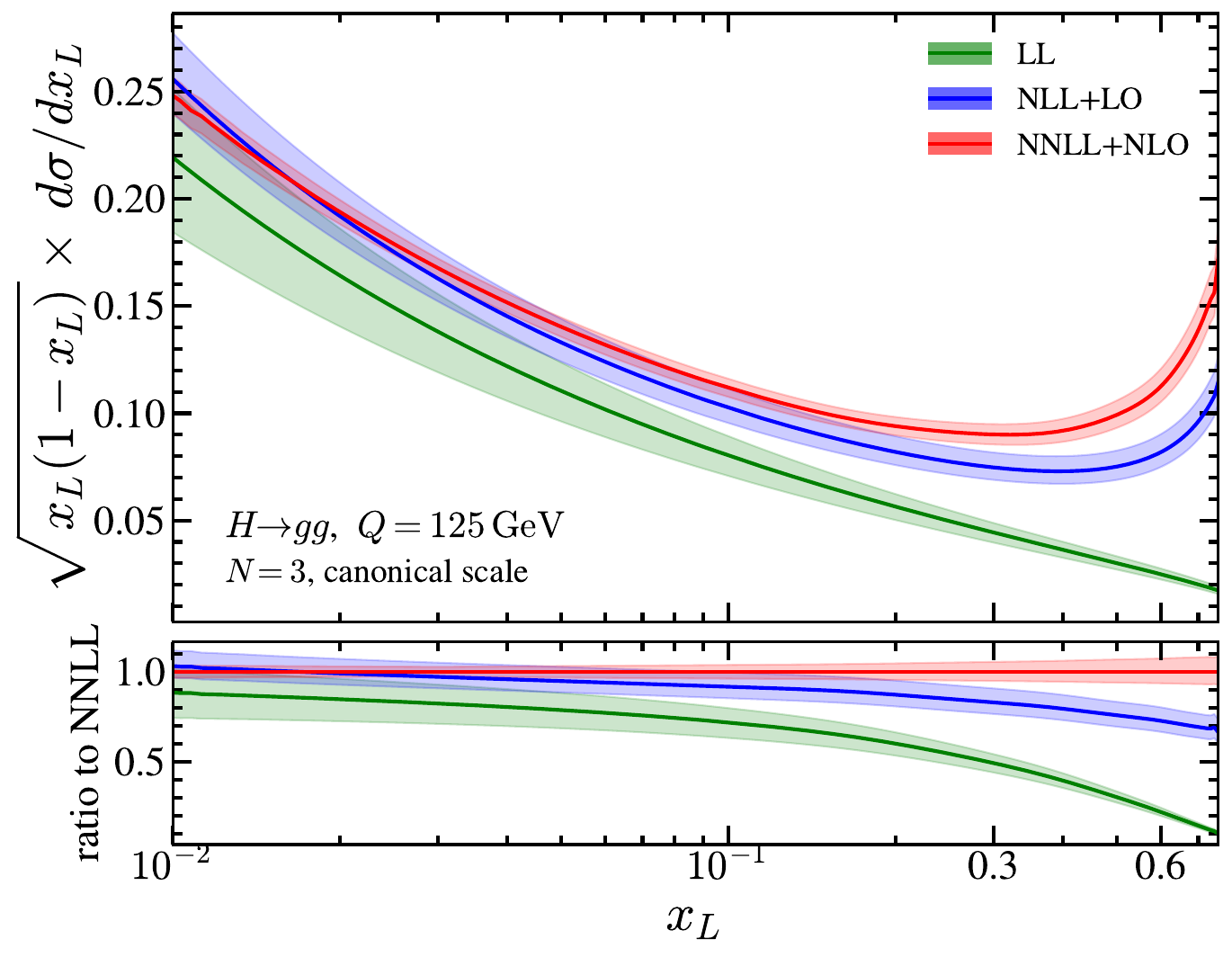}
    \includegraphics[width=0.32\linewidth]{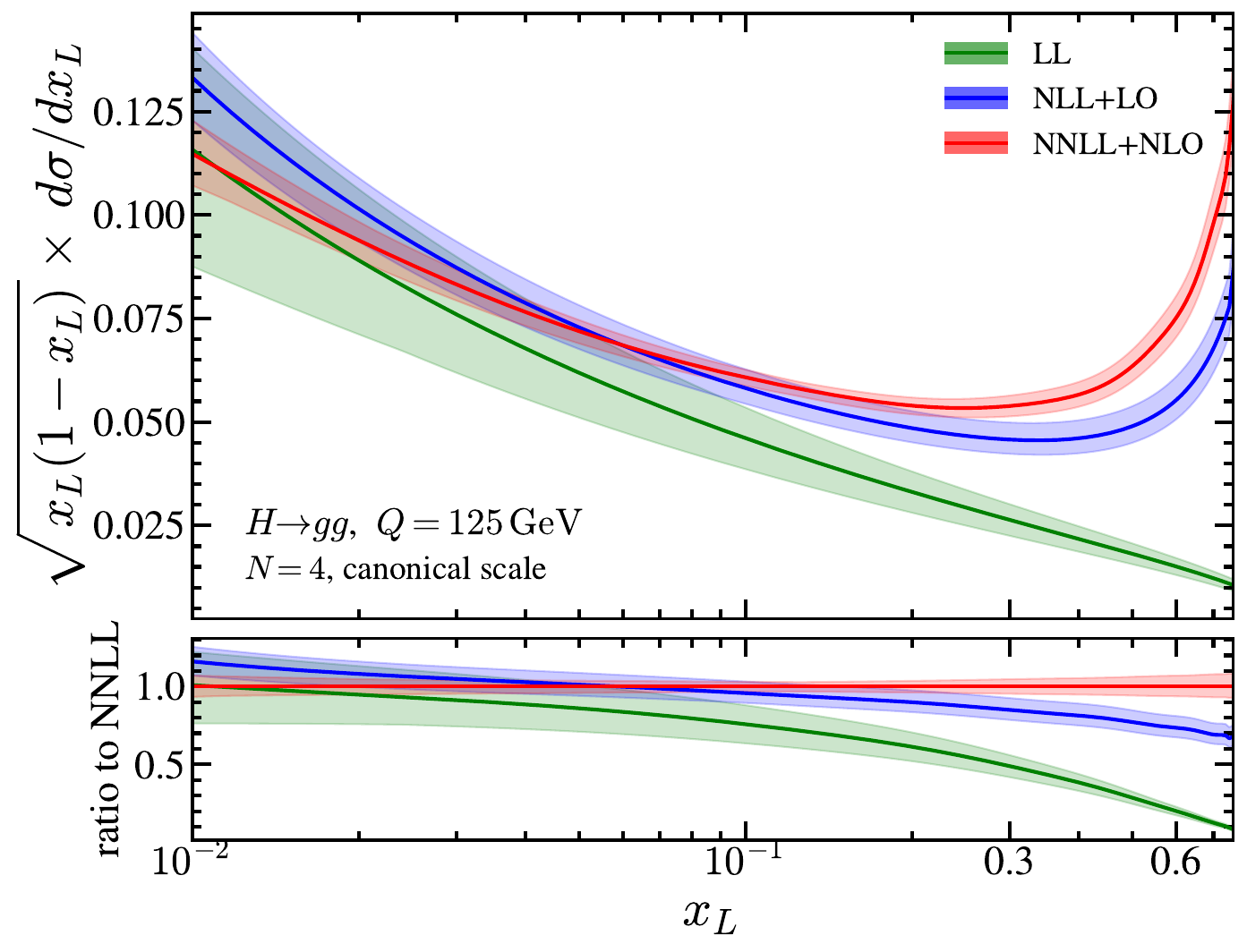}\\
    \includegraphics[width=0.32\linewidth]{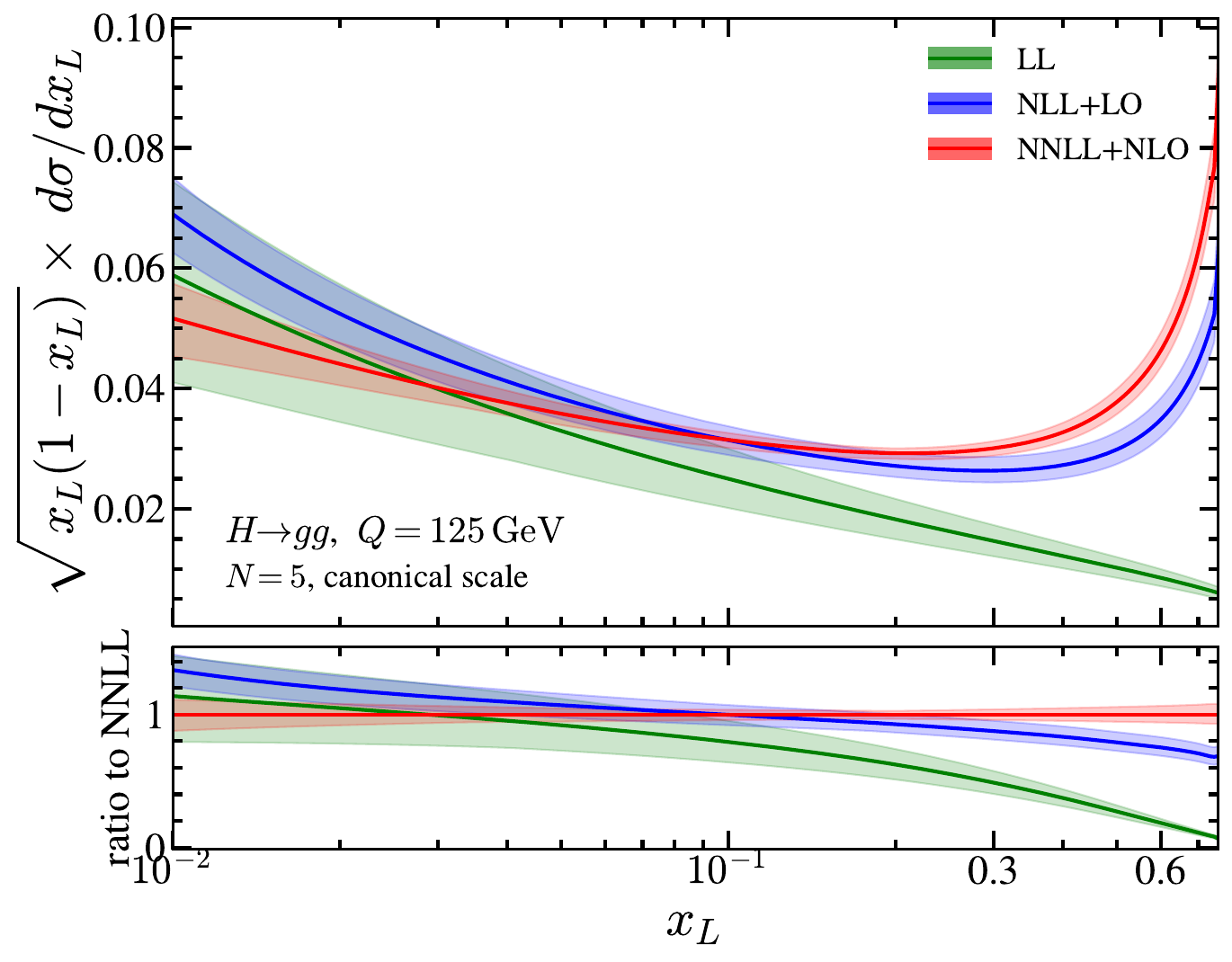}
    \includegraphics[width=0.32\linewidth]{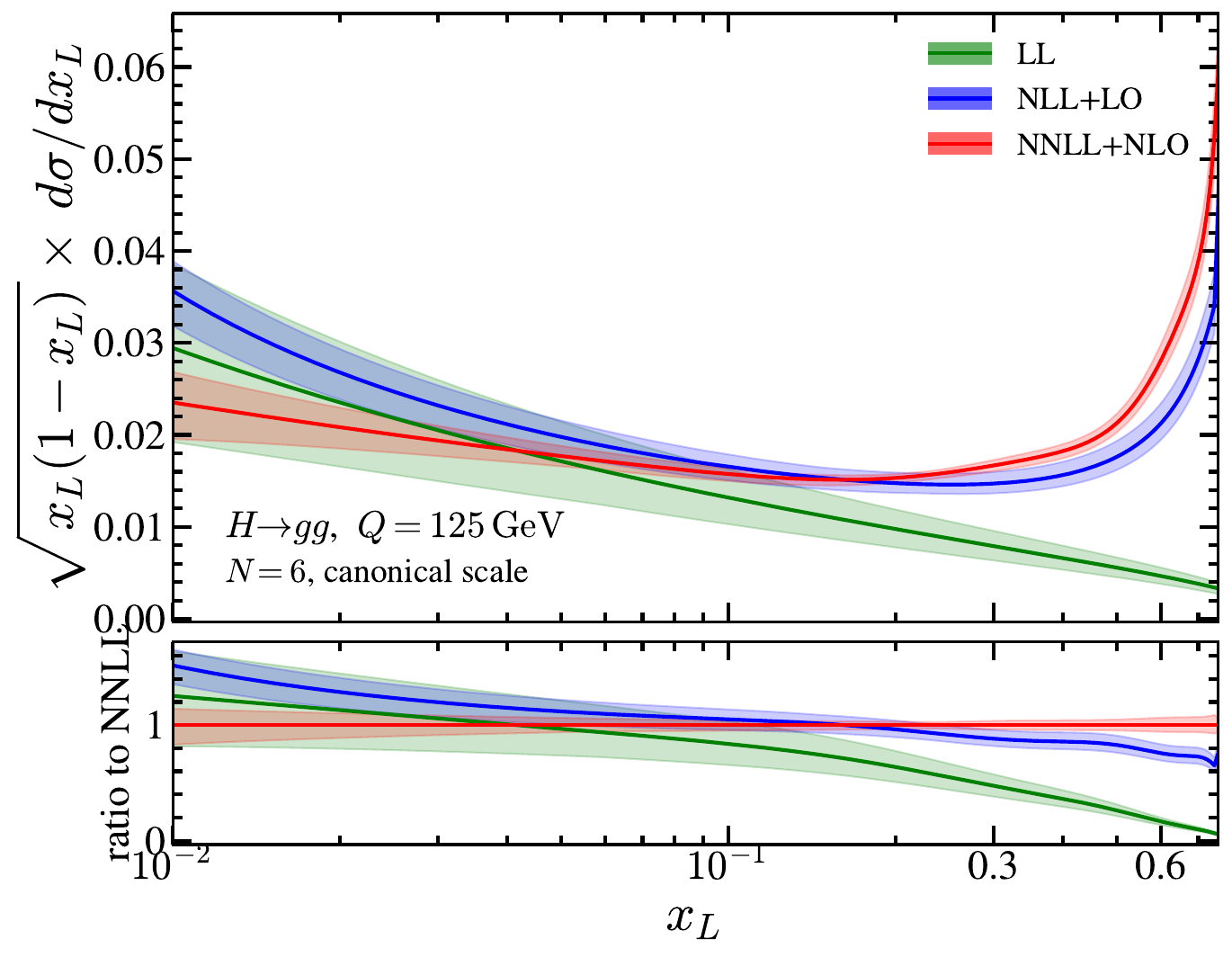}
    \includegraphics[width=0.32\linewidth]{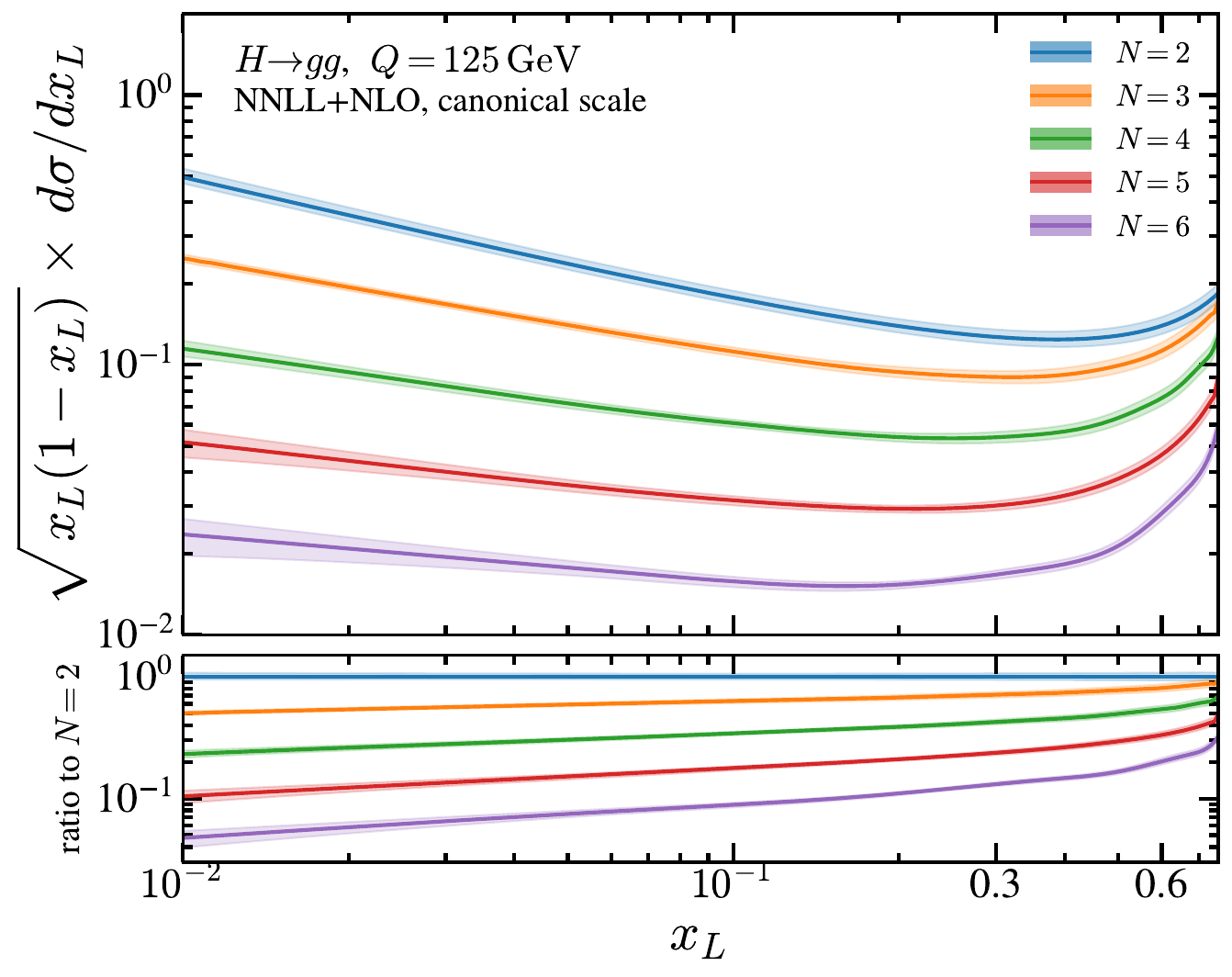}
    \caption{Matched predictions for E2C to E6C in $H\to gg$. In the first five figures, the upper panels show the spectra with the sequence LL, NLL+LO, and NNLL+NLO and the lower panels show the ratio to NNLL+NLO, respectively. The uncertainty band is the envelope of the seven-point scale variation in Eq.~\eqref{eq:can_scale_var}. In the last figure, the upper panel shows the highest order for all $N$ together and the lower panel shows their ratios to EEC.}
    \label{fig:match_can_pert_g}
\end{figure}

\subsection{Matched distributions of projected energy correlators}

We now present the matched distributions of projected energy correlators up to $N=6$ points. Although $N=2$ and $N=3$ cases were already presented in 
the literature~\cite{Dixon:2019uzg,Chen:2023zlx}, we present them here again as well for comparison. The matched perturbative predictions of
\begin{align}
    \sqrt{x_L(1-x_L)}\times\frac{1}{\sigma_{\rm tot}}\frac{ d\hat{\sigma}^{[N]}_{\rm match}}{dx_L} =  \frac{1}{\sigma_{\rm tot}}\frac{ d\hat{\sigma}^{[N]}_{\rm match}}{d\theta_L}\,, 
\end{align}  for $e^+e^-$ and $H\to gg$ are shown, respectively, in Fig.~\ref{fig:match_can_pert_q}
and Fig.~\ref{fig:match_can_pert_g}.  Here, $\theta_L$ is related to $x_L$ as $x_L = (1-\cos\theta_L)/2$.

\begin{figure}[!htbp]
    \centering
    \includegraphics[width=0.48\linewidth]{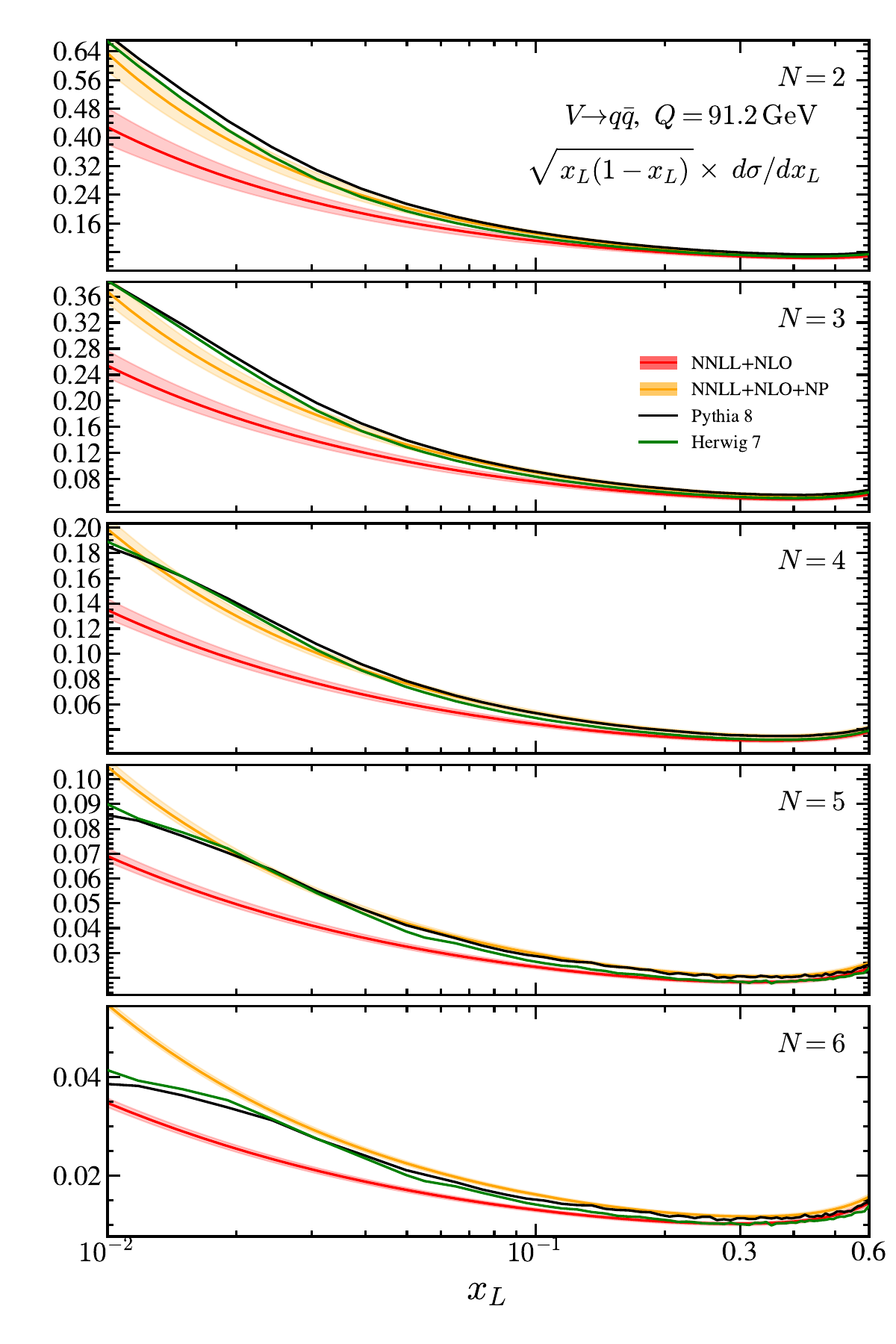}
    \includegraphics[width=0.48\linewidth]{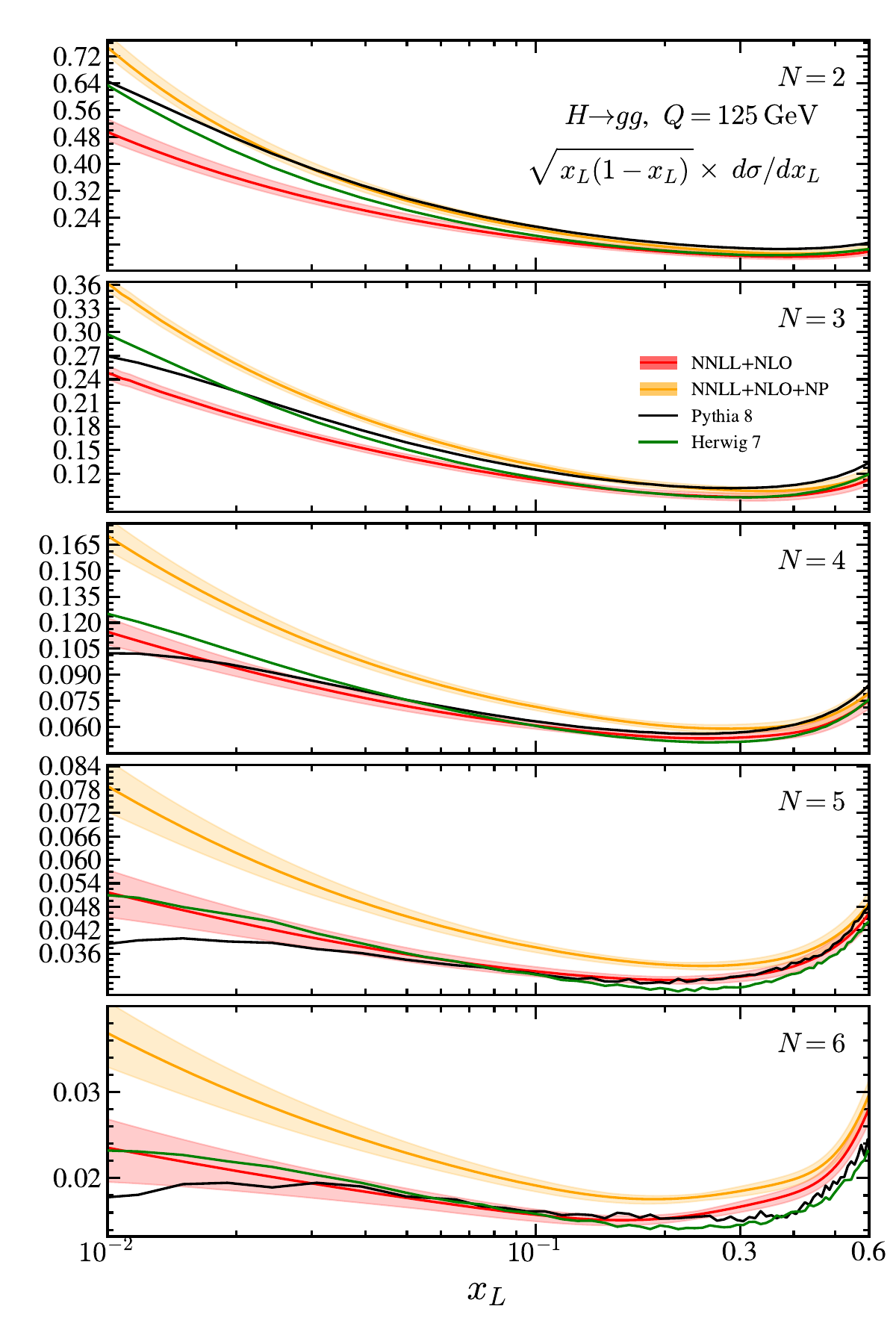}
    \caption{Matched NNLL+NLO spectra for projected $N$-point energy correlators in $e^+e^-\to q\bar q$ at $Q=m_Z$ and $H\to gg$ at $Q=m_H$, with $N=2$ to $6$ from top to the bottom. The red band is the perturbative result and the orange band includes the leading non-perturbative correction. We use $\overline{\Omega}_{1q}=0.305$ GeV and $\overline{\Omega}_{1g}=0.686$ GeV.  }
    \label{fig:matched_NP_spectra}
\end{figure}
For $e^+e^-$ annihilation and Higgs decay we set the hard scale to $Q=m_Z=91.2$ GeV and $Q=m_H=125$ GeV, respectively. For Higgs bosons produced with nonzero boost at hadron or lepton colliders, the boost kernel developed in Refs.~\cite{Gao:2026xuq,Holguin:2026vld} can be used to obtain the corresponding distributions in the boosted frame. While LL captures the qualitative shape of the resummed distribution, its scale variation, formed via the envelope of Eq.~\eqref{eq:can_scale_var}, underestimates the true theoretical uncertainty. From NLL to NNLL we observe good convergence across most of the perturbative resummation window, with deviations setting in only at very small $x_L$. Such deviations 
appear at progressively larger $x_L$ as $N$ grows, and at larger $x_L$ in $H\to gg$ than in $e^+e^-\to q\bar q$.
In the large $x_L$ region, the deviation arises as the fixed-order result dominates and there are other large logarithmic enhancements, which are beyond the scope of this paper. 

This pattern is consistent with the structure of the leading non-perturbative correction. As discussed in Ref.~\cite{HaoTalkSCET,Lee:2024esz,Chen:2024nyc} and visible in Eq.~\eqref{eq:np_fixed}, the perturbative term scales as $1/x_L$ in the collinear region while the leading hadronization correction scales has additional relative scaling $N\overline{\Omega}_{1\kappa}/(Q\sqrt{x_L})$, and thus the non-perturbative expansion is valid only when $N\overline{\Omega}_{1\kappa}\ll Q\sqrt{x_L}$. Near $N\overline{\Omega}_{1\kappa}\sim Q\sqrt{x_L}$ the projected correlators begin transitioning to hadronic degrees of freedom, as confirmed by the Monte Carlo comparisons below; in this region one also expects $\mathcal{O}(\Lambda_{\rm QCD})$ renormalon ambiguities to spoil the perturbative convergence of any prediction without explicit renormalon subtraction~\cite{Schindler:2023cww,Lee:2024esz}, which we do not include in this paper. The observed onset of poor perturbative convergence at larger $x_L$ for larger $N$, and at larger $x_L$ in $H\to gg$ than in $e^+e^-$ due to $\overline{\Omega}_{1g}>\overline{\Omega}_{1q}$, matches this expectation.

Fig.~\ref{fig:matched_NP_spectra} compares the highest perturbative result with the one including leading non-perturbative corrections according to Eq.~\eqref{eq:full}. We recall that ${\overline{\Omega}_{1q},\overline{\Omega}_{1g}}$ are defined in the $\overline{\text{MS}}$ scheme. Throughout this subsection we adopt $\overline{\Omega}_{1q}=0.305\,\text{GeV}$, taken from the global fit of $e^+e^-$ N${}^3$LL${}^\prime+$NNLO thrust~\cite{Abbate:2010xh,Benitez:2024nav}.\footnote{The $\Omega_{1q}$ extracted from the thrust global fit is in the R-gap scheme; its conversion to $\overline{\text{MS}}$ is worked out in Ref.~\cite{Schindler:2023cww}.} The gluon parameter $\overline{\Omega}_{1g}$ is currently unconstrained; for definiteness we adopt the naive Casimir-rescaled value $\overline{\Omega}_{1g}=(C_A/C_F)\overline{\Omega}_{1q}$. In Sec.~\ref{sec:pheno}, we will revisit sensitivity study of different $\overline{\Omega}_{1g}$ choices.

For comparison, we also plot Monte Carlo simulations from \textsc{Pythia8}~\cite{Bierlich:2022pfr} and \textsc{Herwig7}~\cite{Bellm:2015jjp} at the hadron level, with initial-state radiation turned off. We generate 500k events for $N=2,3,4$ in both processes, 10k events for $N=5,6$ in $e^+e^-$ case and $N=5$ in $H\to gg$ case, and 3k events for $N=6$ in $H\to gg$ case, since the observable becomes computationally expensive at higher $N$. As anticipated from the $1/x_L^{3/2}$ enhancement in Eqs.~\eqref{eq:np_fixed} and~\eqref{eq:np_jet}, the $\overline{\Omega}_{1\kappa}$ contribution becomes dominant at small $x_L$, and including it is essential for reproducing the simulation behavior in $e^+e^-$ annihilation. For $H\to gg$ decay, the transition to the non-perturbative regime sets in at considerably larger $x_L$ than in $e^+e^-$, especially at higher $N$, as expected from $\overline{\Omega}_{1g}>\overline{\Omega}_{1q}$. This is also consistent with the onset of poor perturbative convergence. Even with this enhancement, however, we observe significant discrepancies between our predictions and the Monte Carlo results for $N>4$. Sizable deviations also appear between \textsc{Pythia8} and \textsc{Herwig7} themselves, plausibly reflecting the lack of precision $H\to gg$ data to tune against.

\begin{figure}[!hbp]
    \centering
    \includegraphics[width=0.46\linewidth]{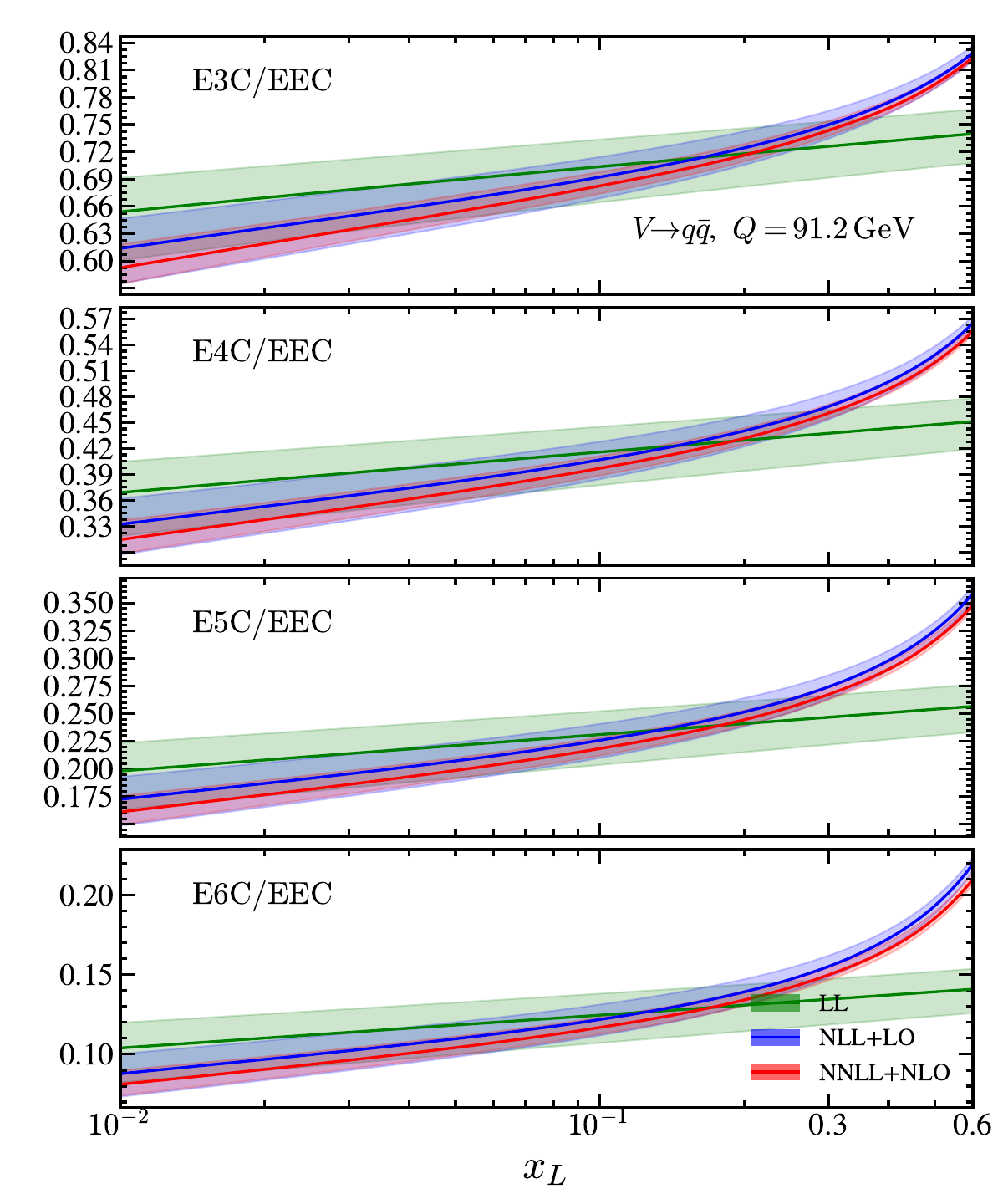}
    \includegraphics[width=0.46\linewidth]{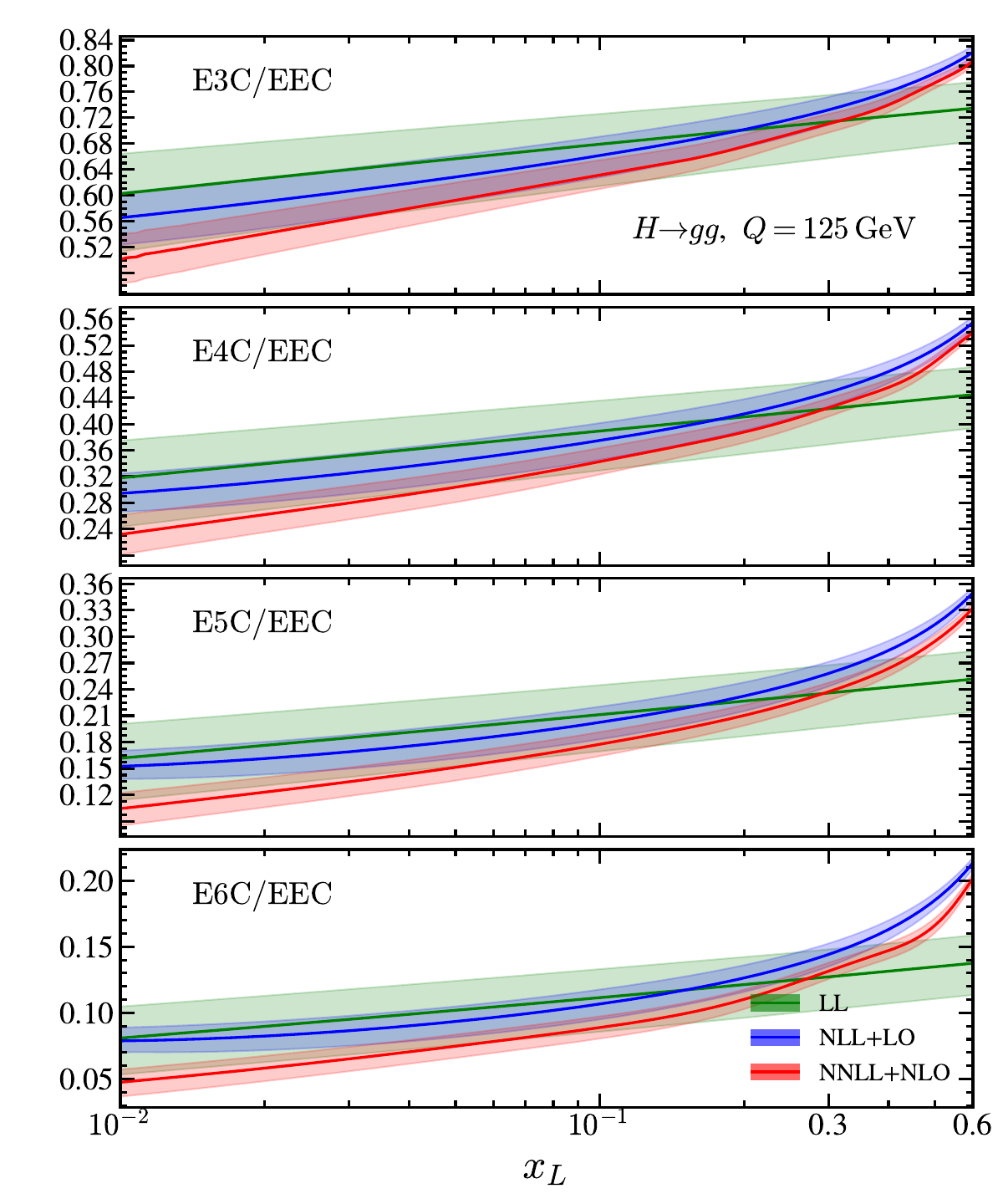}
    \caption{The perturbative resummation of ratio projected energy correlators ENC/EEC for $N=3,4,5,6$ for $e^+e^-\to q\bar q$ and $H\to gg$.}
    \label{fig:ratio_pert_spectra}
\end{figure}

\begin{figure}[!hbp]
    \centering
    \includegraphics[width=0.46\linewidth]{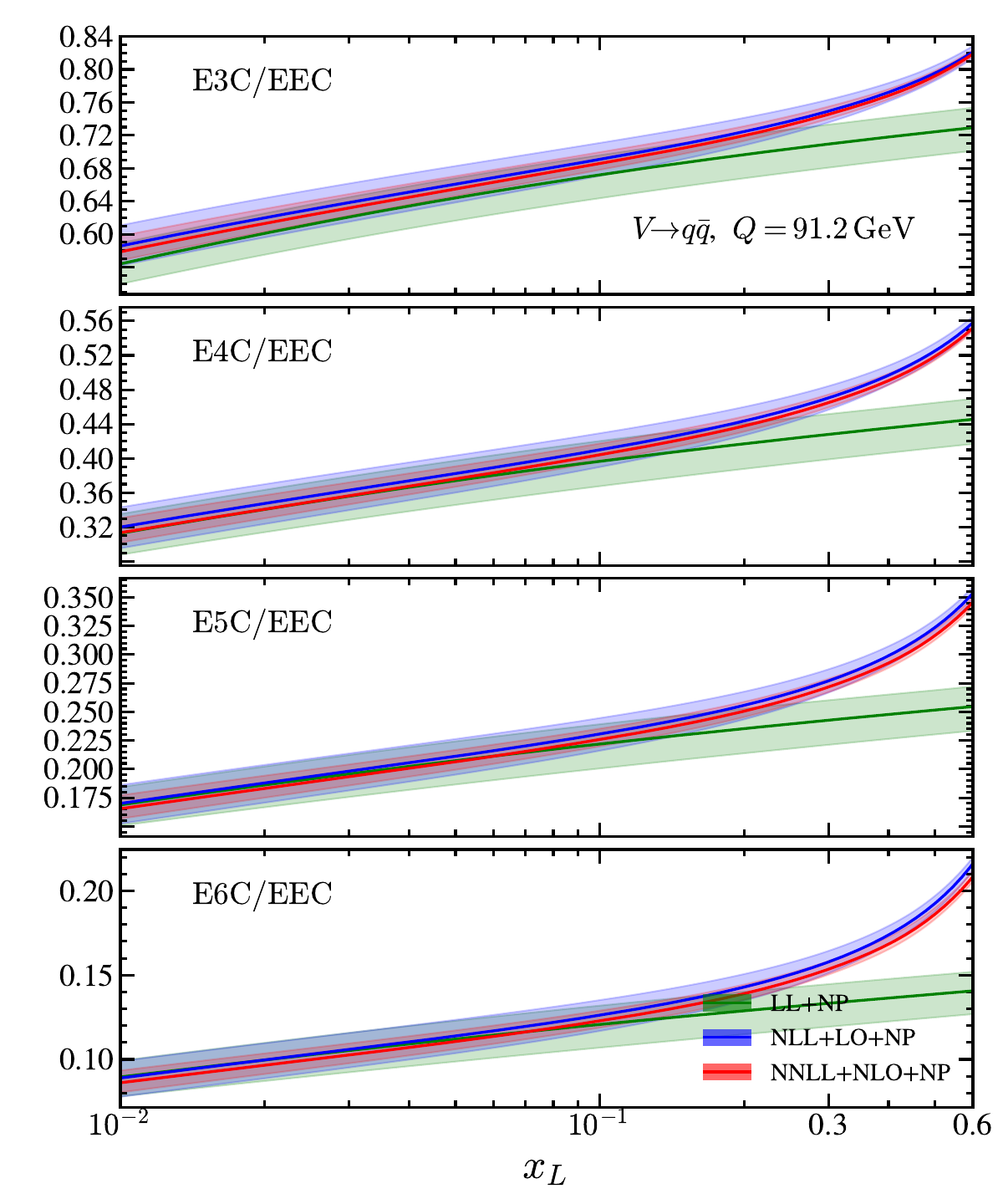}
    \includegraphics[width=0.46\linewidth]{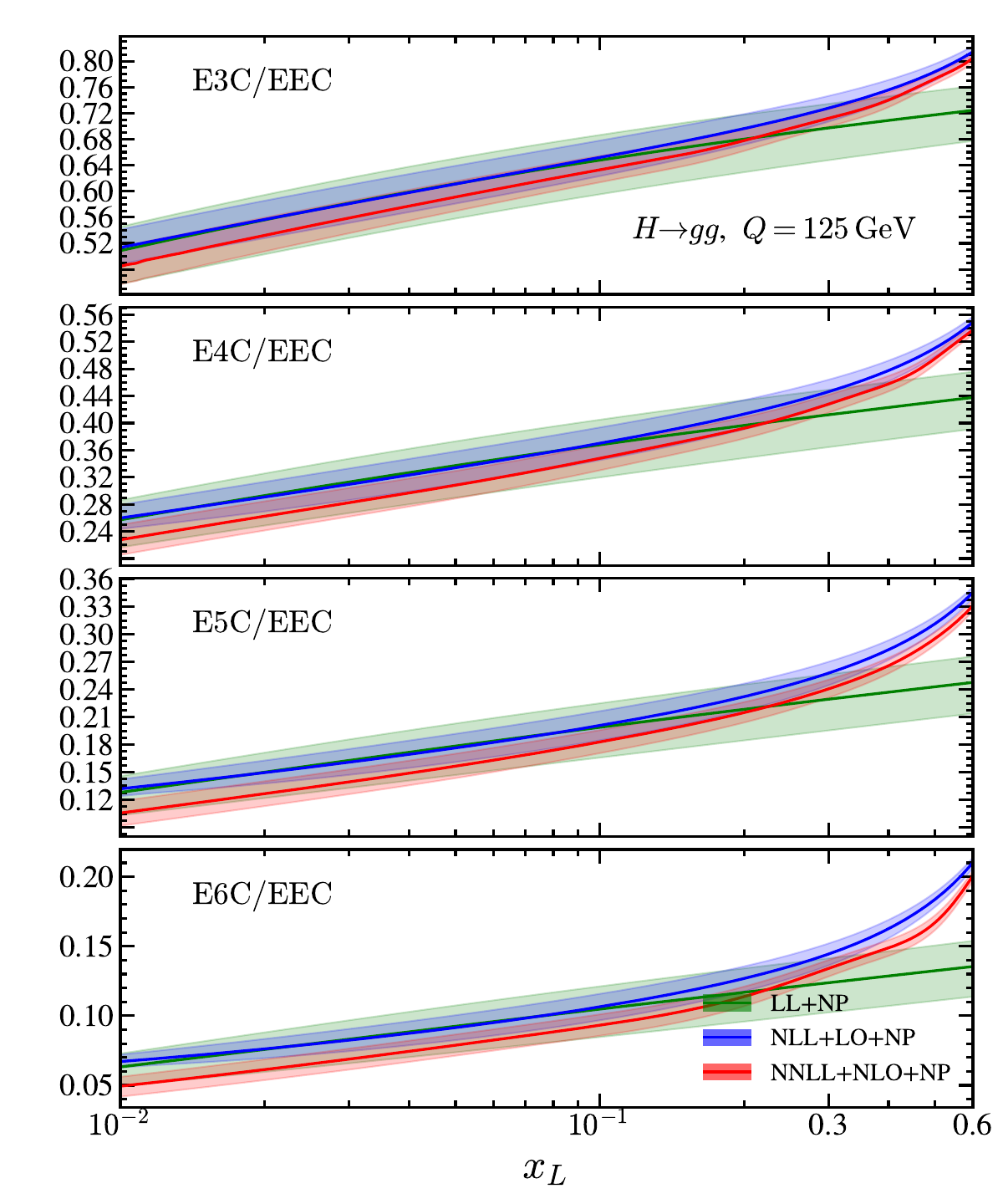}
    \caption{The same as Fig.~\ref{fig:ratio_pert_spectra}, but include leading non-perturbative correction.}
    \label{fig:ratio_pert+NP_spectra}
\end{figure}

\begin{figure}[!hbp]
    \centering
    \includegraphics[width=0.47\linewidth]{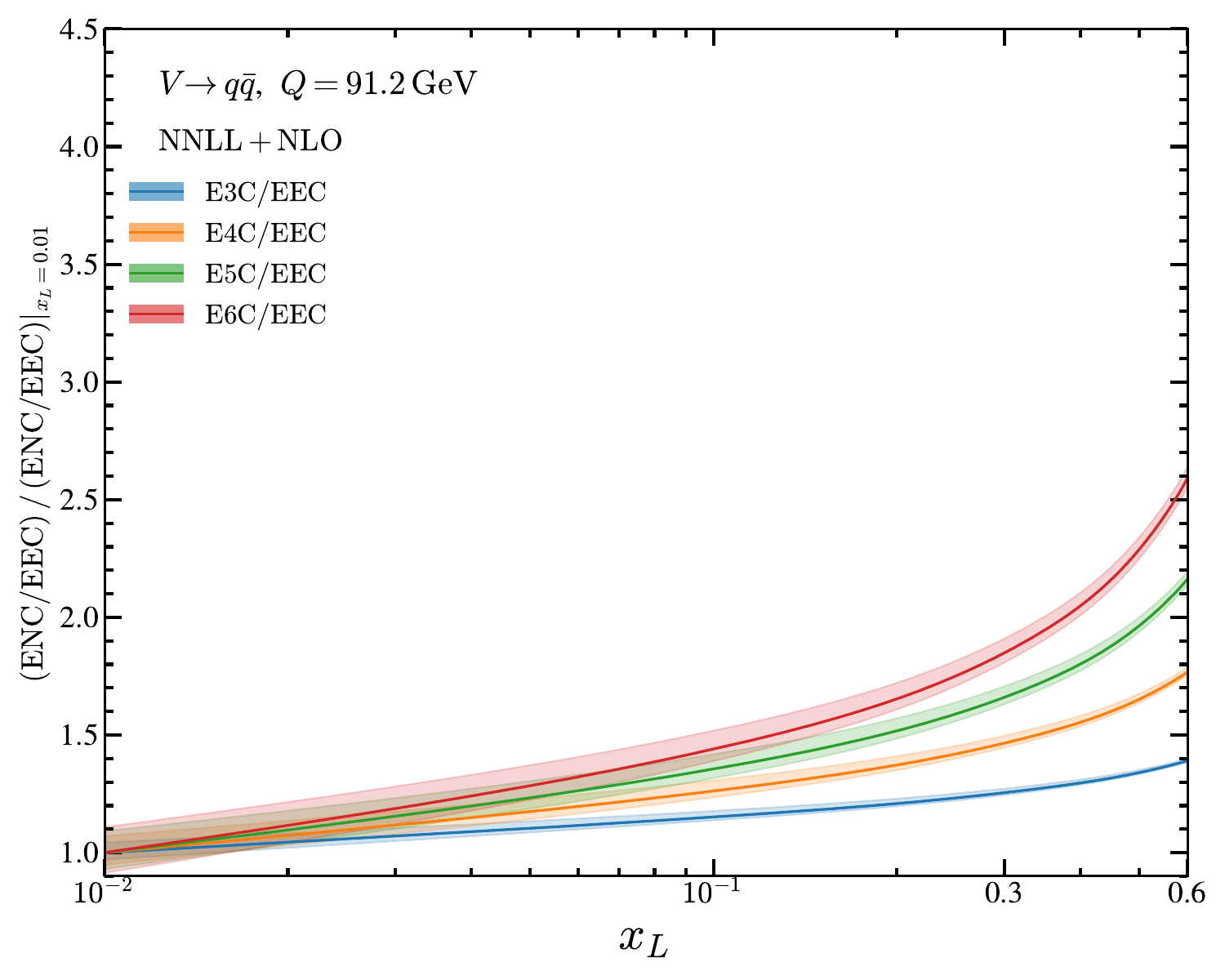}
    \includegraphics[width=0.47\linewidth]{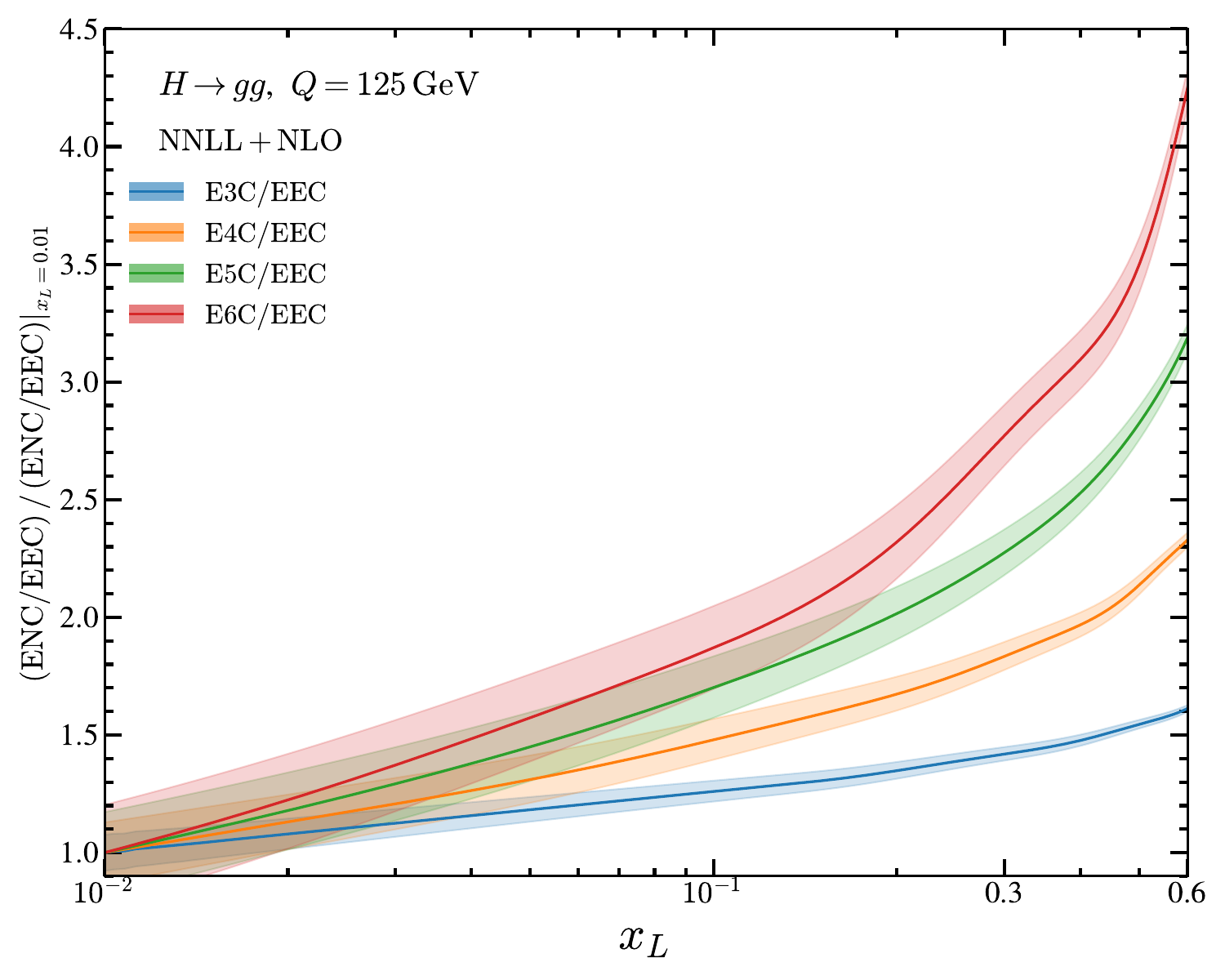}\\
    \includegraphics[width=0.47\linewidth]{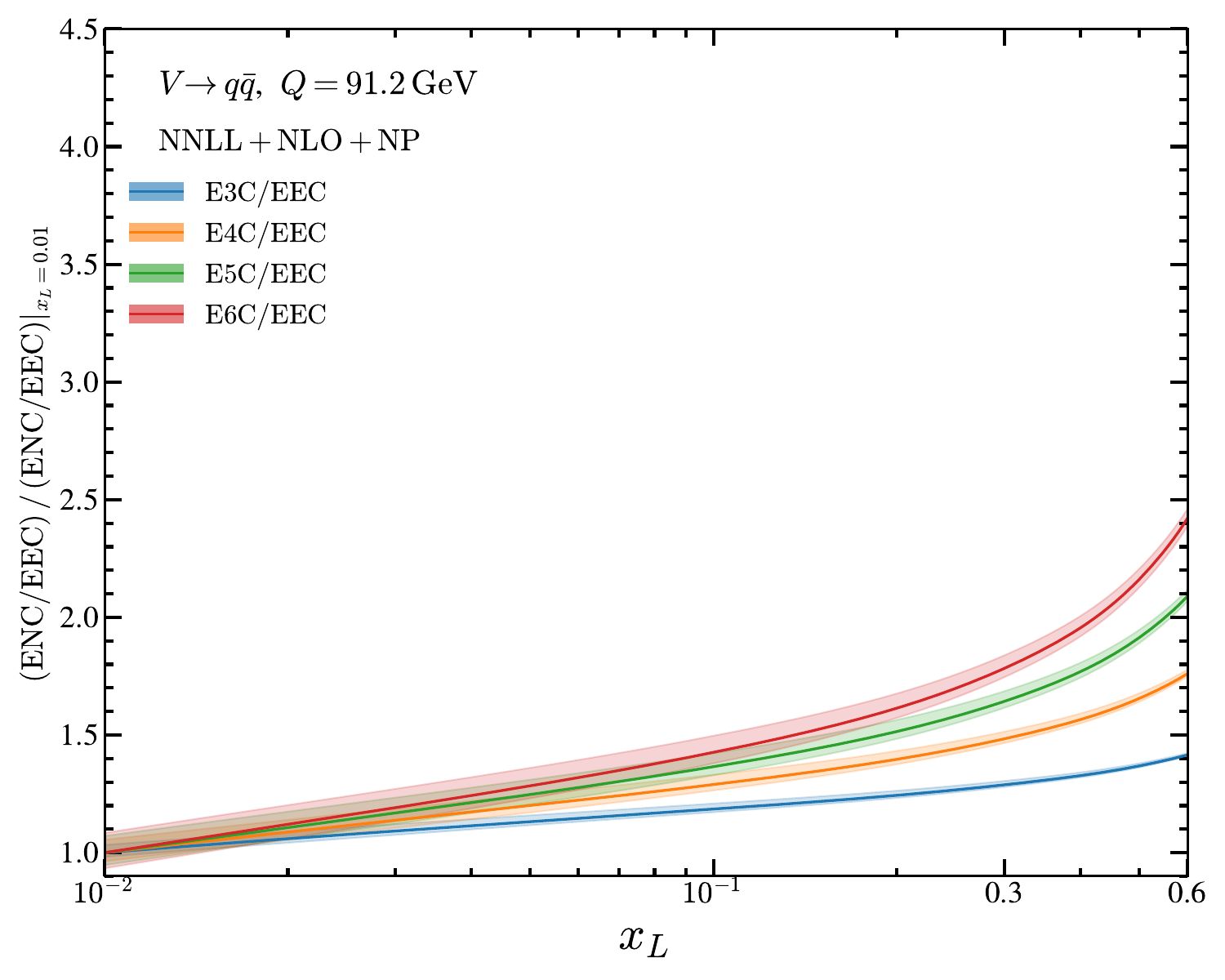}
    \includegraphics[width=0.47\linewidth]{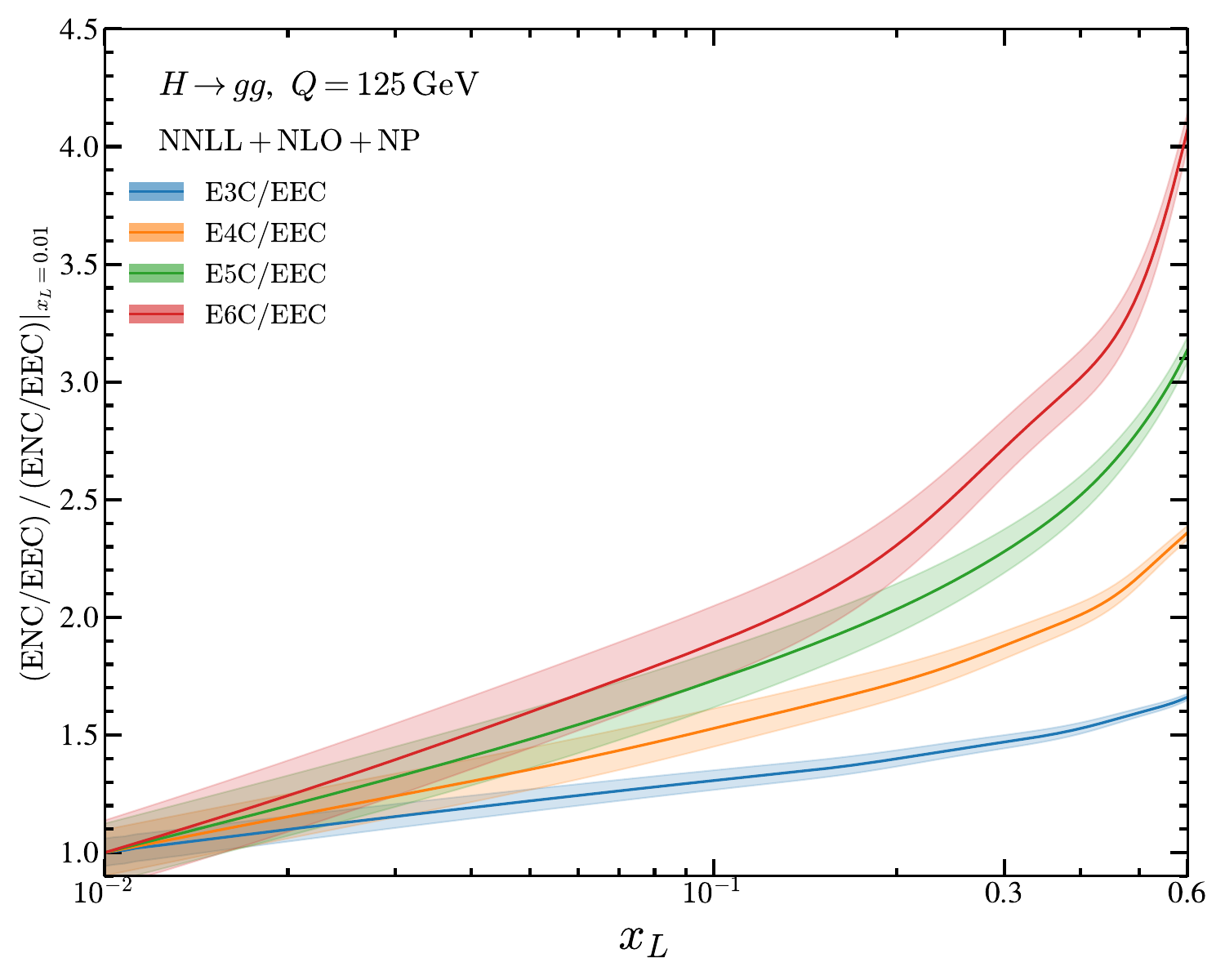}
    \caption{Ratios of projected $N$-point energy correlators to the EEC, normalized to unity at $x_L=0.01$, for $N=3,4,5,6$. The top row is the matched NNLL+NLO perturbative prediction and the bottom row additionally includes the leading non-perturbative correction. }
    \label{fig:ratioAnom}
\end{figure}

\subsection{Ratio of projected energy correlators}
Ratios of projected energy correlators provide a useful way to isolate the anomalous scaling. As discussed in the introduction, the anomalous scaling of the $N$-point projected correlator is governed by twist-two spin-$(N+1)$ operators. 
Ignoring the mixing of different partonic channels and the running of QCD coupling, the ENCs in the perturbative collinear limit obey
\begin{align}
\label{eq:scalingabs}
\frac{d\sigma^{[N]}}{dx_L} \sim x_L^{-1+\gamma(N+1)}\,,
\end{align} 
where $\gamma(N+1)$  denotes the anomalous dimension of the twist-two spin-$(N+1)$ operator.  Forming the ratio of the $N$-point correlator to the EEC thus cancels the classical $x_L^{-1}$ scaling and leaves the anomalous scaling
\begin{align}
\label{eq:scalingratio}
\frac{d\sigma^{[N]}/dx_L}{d\sigma^{[2]}/dx_L} \sim x_L^{\gamma(N+1)-\gamma(3)}\,.
\end{align}
The slope of the log-log plot of the ratio is therefore set by a difference of anomalous dimensions, and is consequently sensitive to $\alpha_s$. This naive expectation is of course modified by the running coupling and by non-perturbative corrections, but the schematic picture illustrates why ratios are useful observables. Note also that $\gamma(N+1)$ grows monotonically with $N$~\cite{Nachtmann:1973mr}, so the slope of the ratio is expected to steepen as $N$ increases.

In Fig.~\ref{fig:ratio_pert_spectra} and Fig.~\ref{fig:ratio_pert+NP_spectra}, we show the perturbative convergence of the ratios for both $e^+e^-$ and $H\to gg$ up to $N=6$, without and with the leading non-perturbative power corrections respectively. We use the canonical seven-point scale variations of Eq.~\eqref{eq:can_scale_var} for both numerator and denominator, treating the two as fully correlated. Since both inherit their anomalous scaling from twist-two operators, differing only in spin and accessed through different moments of the same QCD splitting functions, this correlated variation is a natural starting point. We emphasize, however, that for ratio observables a quantitative treatment of the correlation between numerator and denominator uncertainties is particularly important and deserves careful study~\cite{Tackmann:2024kci}. We find good perturbative convergence in both figures, especially once non-perturbative corrections are included and most clearly in the $e^+e^-$ channel. Ref.~\cite{Lee:2024esz} showed that non-perturbative corrections reduce the slope of the ratio, consistent with the $N$-scaling of Eq.~\eqref{eq:np_fixed}; we observe the same trend here in both $e^+e^-$ and $H\to gg$. The same reference also showed that omitting these corrections at $Q=1000$ GeV can shift the extracted $\alpha_s$ by about $10\%$. In Sec.~\ref{sec:pheno} we examine in detail the sensitivity of the ratio to $\alpha_s$ and $\overline{\Omega}_{1\kappa}$.

In Fig.~\ref{fig:ratioAnom}, we plot the ratios of $N$-point projected correlators to the EEC for $N=3,4,5,6$, normalized to $1$ at $x_L=0.01$. With this common normalization, the slope reads off the relative size of the anomalous scaling of Eq.~\eqref{eq:scalingratio} directly, making it easy to compare by eye how the scaling changes with $N$, with process, and with the inclusion of non-perturbative corrections. As expected, the slope steepens monotonically with $N$, reflecting the growth of $\gamma(N+1)$ with $N$. The slope is also visibly larger in $H \to gg$ decay than in $e^+e^-$ annihilation, consistent with the gluon twist-two anomalous dimensions being larger than the quark ones. Comparing the top and bottom rows, including the leading non-perturbative correction reduces the slope in both processes and at every $N$ as well.

\subsection{$Q$ dependence of projected energy correlators}
\begin{figure}[!htbp]
    \centering
    \includegraphics[width=0.48\linewidth]{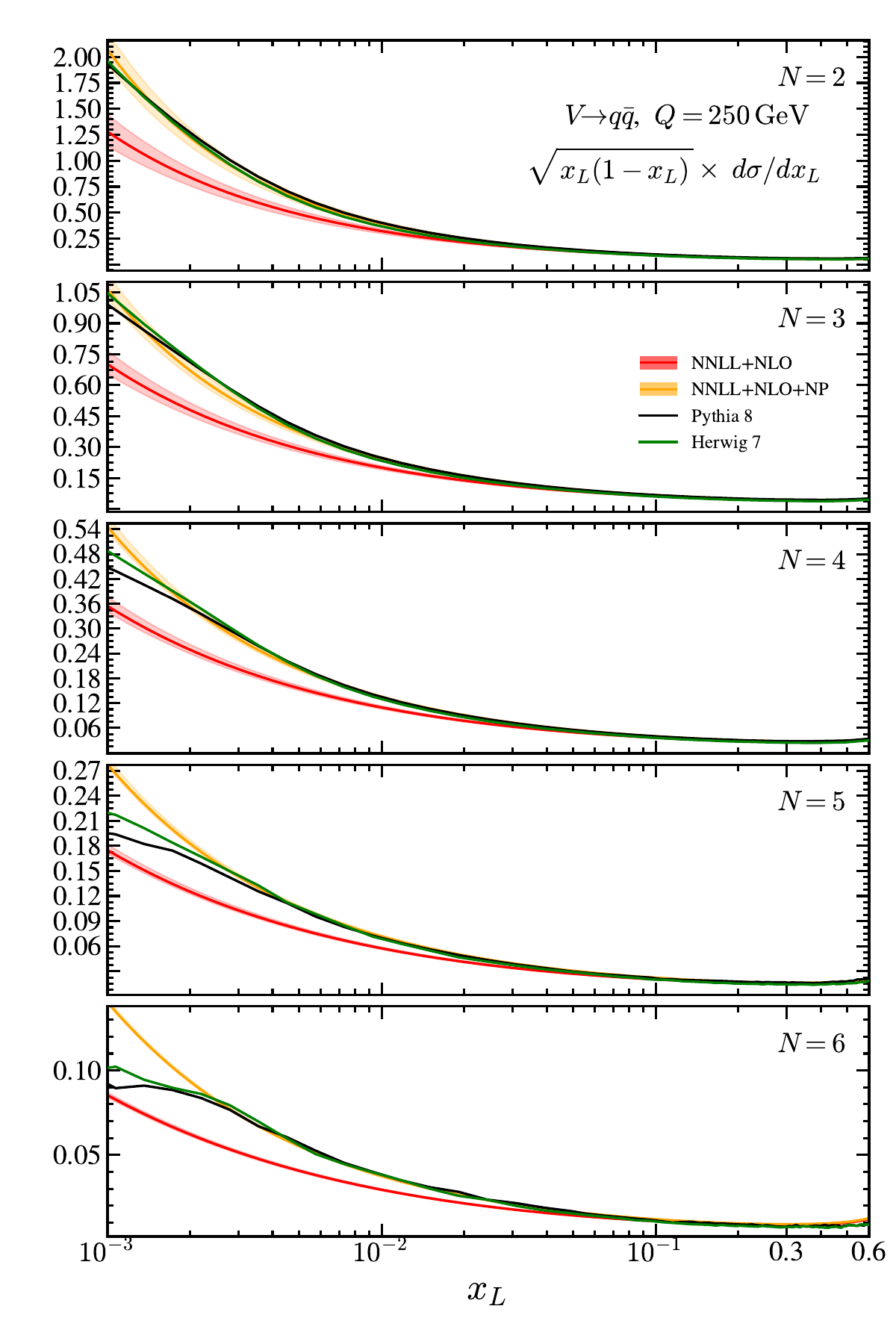}
    \includegraphics[width=0.48\linewidth]{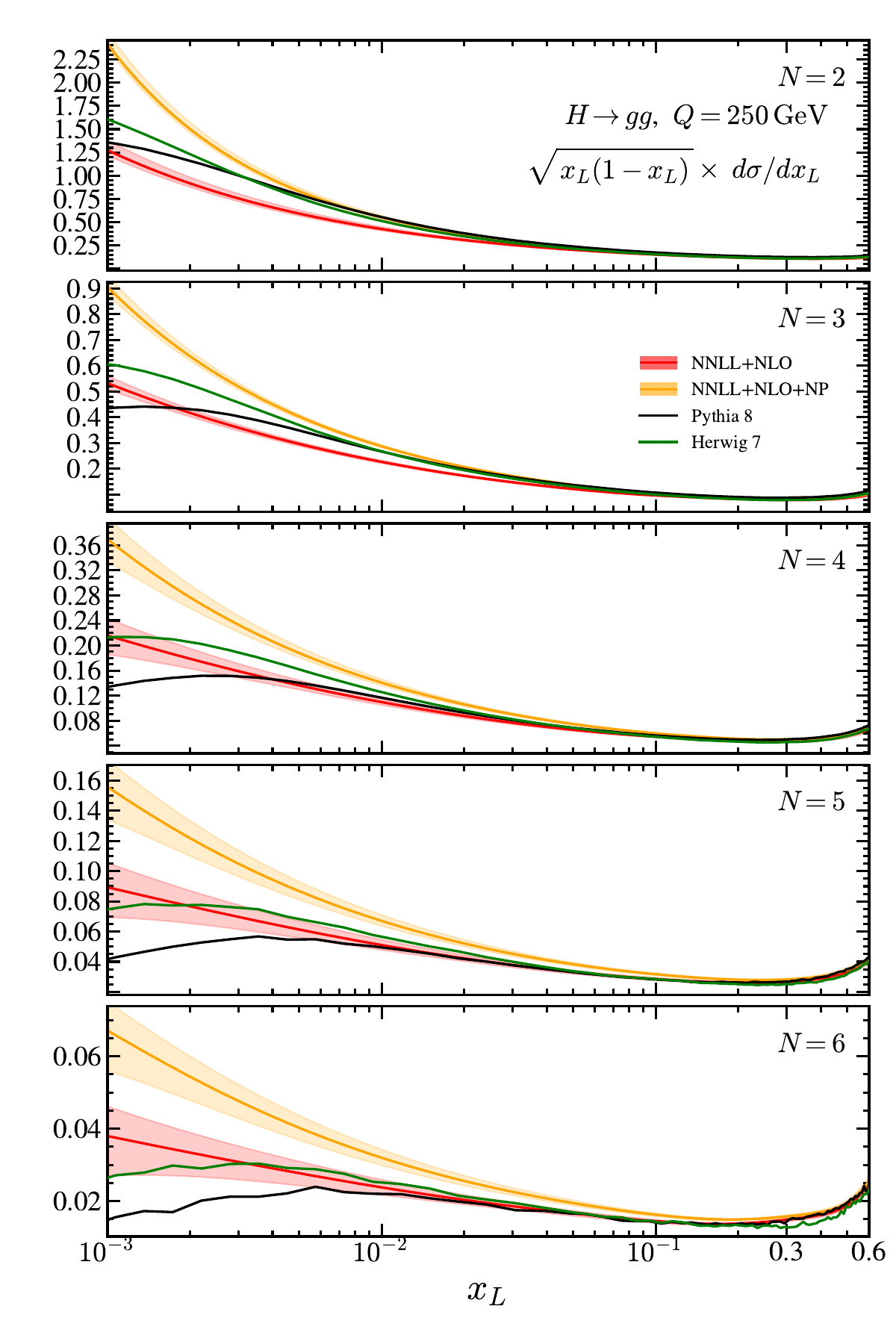}
    \caption{The same as Fig.~\ref{fig:matched_NP_spectra}, but at $Q=250~\rm{GeV}$ for both $e^+e^-\to q\bar q$ and $H\to gg$.}
    \label{fig:matched_NP_spectra_Q250}
\end{figure}

The previous subsections fixed the hard scale to $Q=m_Z$ for $e^+e^-\to q\bar q$ and to $Q=m_H$ for $H\to gg$. It is also useful to study the $Q$ dependence of the formalism, both as a check and for direct phenomenological reasons. For $e^+e^-$ annihilation, LEP2~\cite{ALEPH:2013dgf,Altarelli:1996ww}, the second phase of LEP, operated at center-of-mass energies above the $Z$ pole, reaching up to about $209$ GeV. Higher-energy runs are also envisioned at future Higgs factories such as FCC-ee and CEPC~\cite{FCC:2018evy,CEPCStudyGroup:2018ghi}. For $H\to gg$, while the Higgs mass itself is of course fixed in nature, the same formalism can be applied to any sufficiently narrow color-singlet scalar decaying to gluons, including BSM scalars with different masses. 

As a representative example, in Fig.~\ref{fig:matched_NP_spectra_Q250} we study the matched NNLL+NLO predictions at $Q=250$~GeV for both processes, with the same $\overline{\Omega}_{1q}=0.305$~GeV and $\overline{\Omega}_{1g}=(C_A/C_F)\,\overline{\Omega}_{1q} =0.686$~GeV used in Fig.~\ref{fig:matched_NP_spectra}. For $H\to gg$, we set the Higgs mass parameter in \textsc{Pythia8} and \textsc{Herwig7} to $m_H=250$~GeV. The qualitative pattern of agreement with the parton showers is consistent with the $Q=m_Z,m_H$ case discussed previously: the $e^+e^-$ channel shows good agreement once the leading non-perturbative correction is included, while in $H\to gg$ the deviations grow with $N$. 

Since both $e^+e^-$ and the Higgs decay are run at the same hard scale, we can more easily compare them as functions of the same angular scale $\sim Q\sqrt{x_L}$. As discussed above and in Ref.~\cite{Lee:2024esz}, the leading non-perturbative correction relative to the perturbative spectrum scales as $N\overline{\Omega}_{1\kappa}/(Q\sqrt{x_L})$, so the transition to hadronic degrees of freedom occurs around $N\overline{\Omega}_{1\kappa}\sim Q\sqrt{x_L}$. With $Q$ now identical between the two processes, the only difference is the representation of the soft matrix element. The figure shows clearly that the $H\to gg$ distributions transition into the non-perturbative regime at larger $x_L$ than $e^+e^-$, clearly indicating that $\overline{\Omega}_{1g}>\overline{\Omega}_{1q}$.

\begin{figure}[!hbp]
    \centering
    \includegraphics[width=0.45\linewidth]{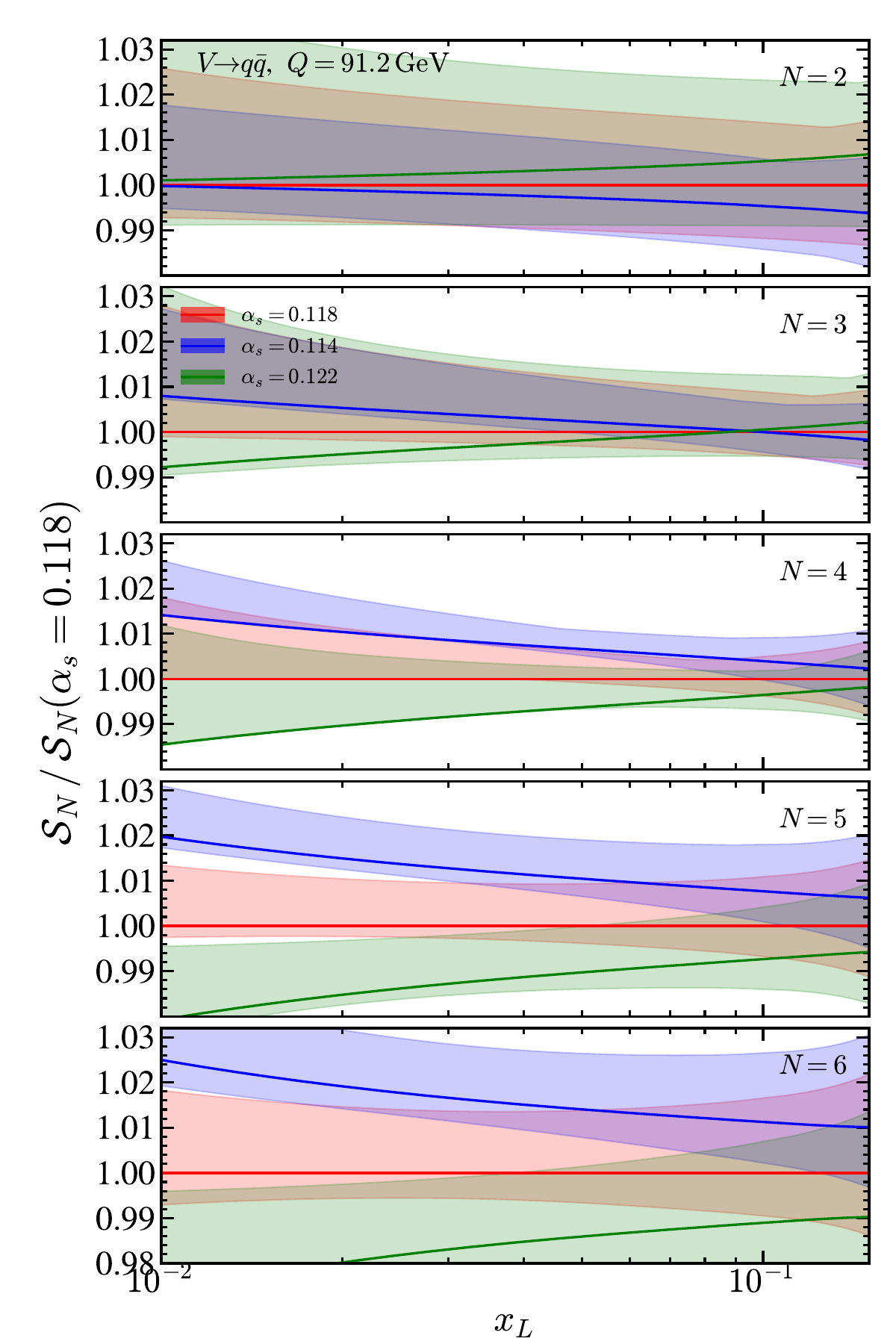}
    \includegraphics[width=0.45\linewidth]{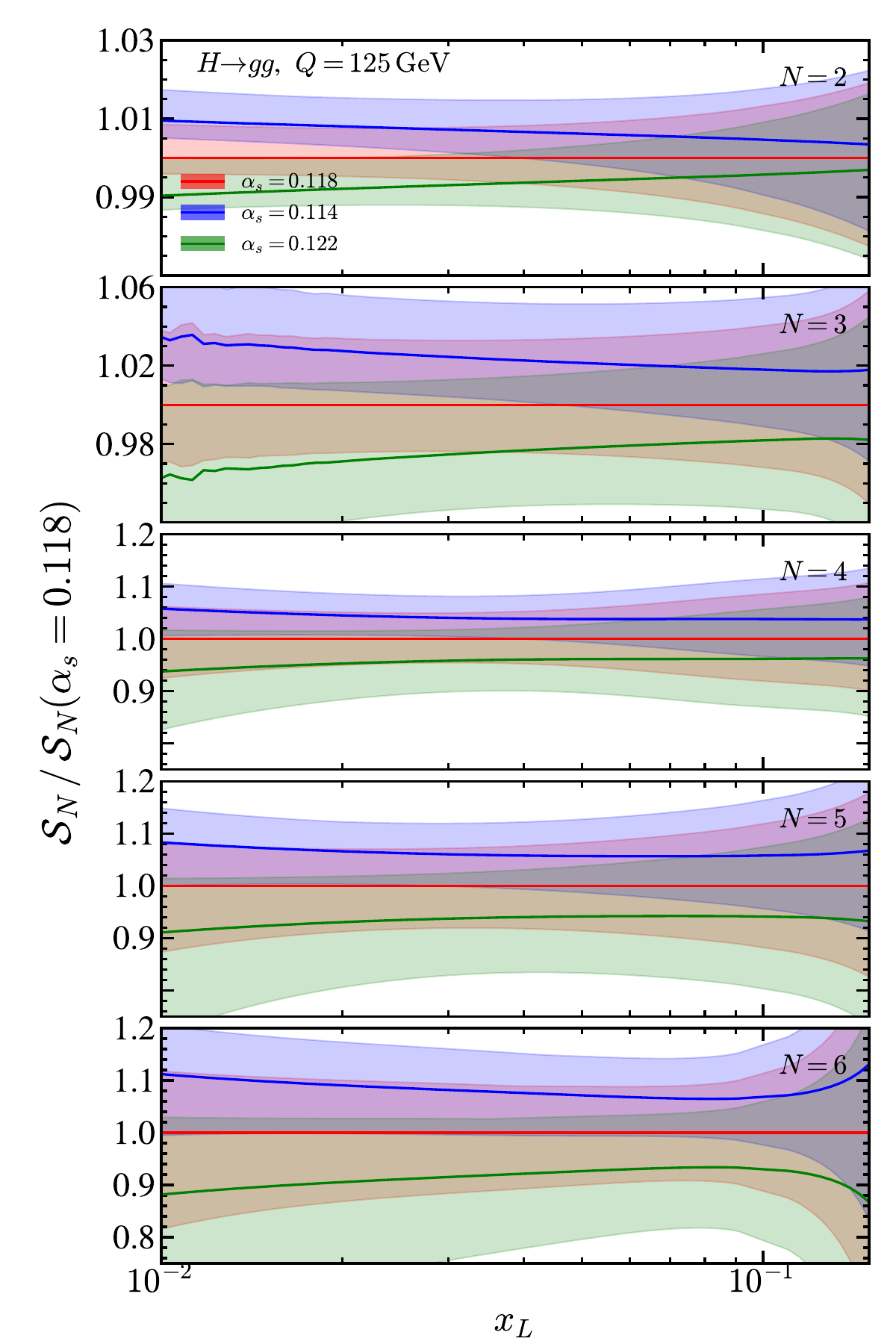}
    \caption{Change in the logarithmic slope $\mathcal{S}_N$ of the absolute projected energy correlator spectrum with respect to the central value $\alpha_s(m_Z)=0.118$, for $e^+e^-\to q\bar q$ (left) and $H\to gg$ (right). Each panel shows $N=2,\ldots,6$ from top to bottom.}
    \label{fig:slope_alphas_abs}
\end{figure}

\begin{figure}[!hbp]
    \centering
    \includegraphics[width=0.47\linewidth]{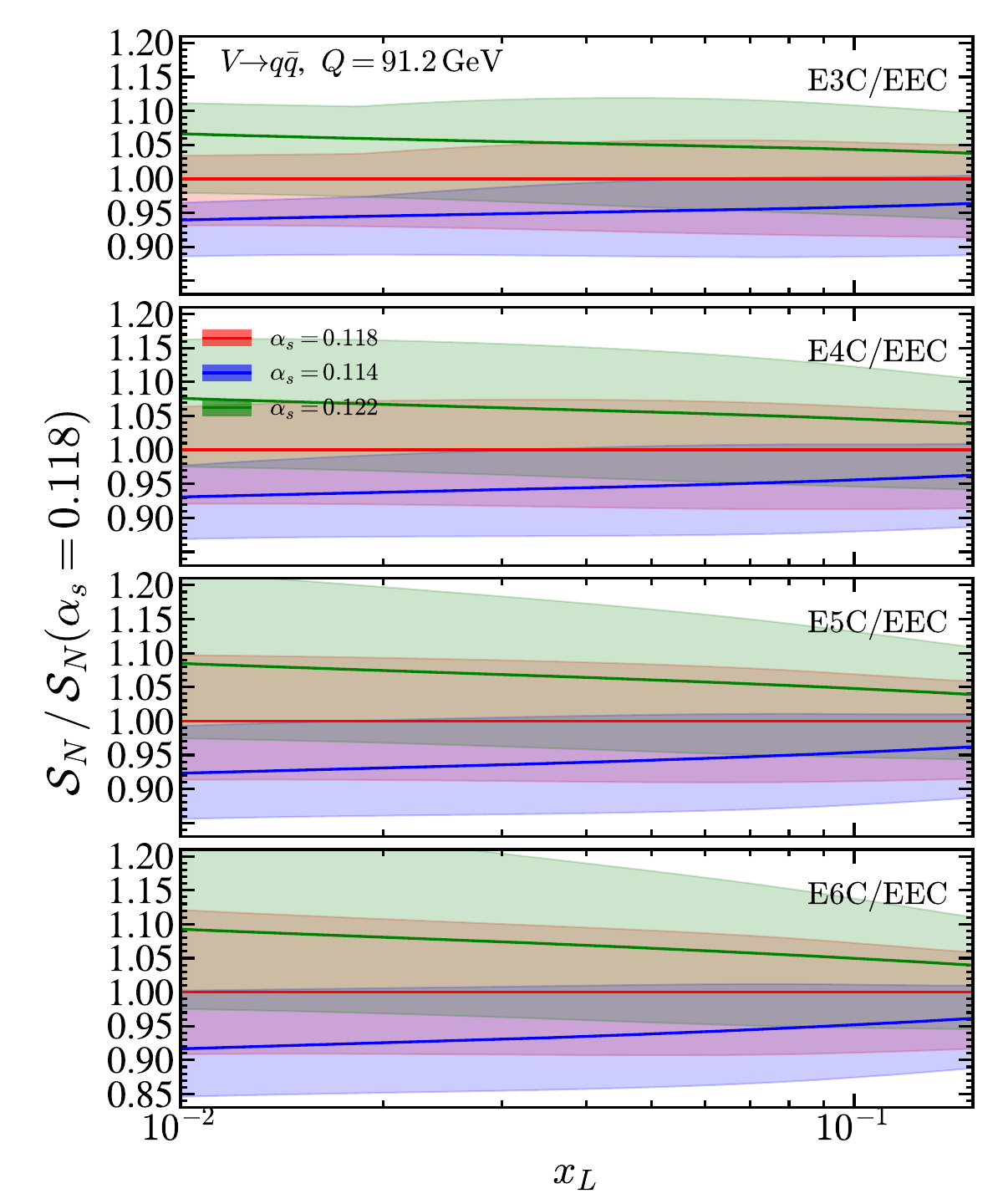}
    \includegraphics[width=0.47\linewidth]{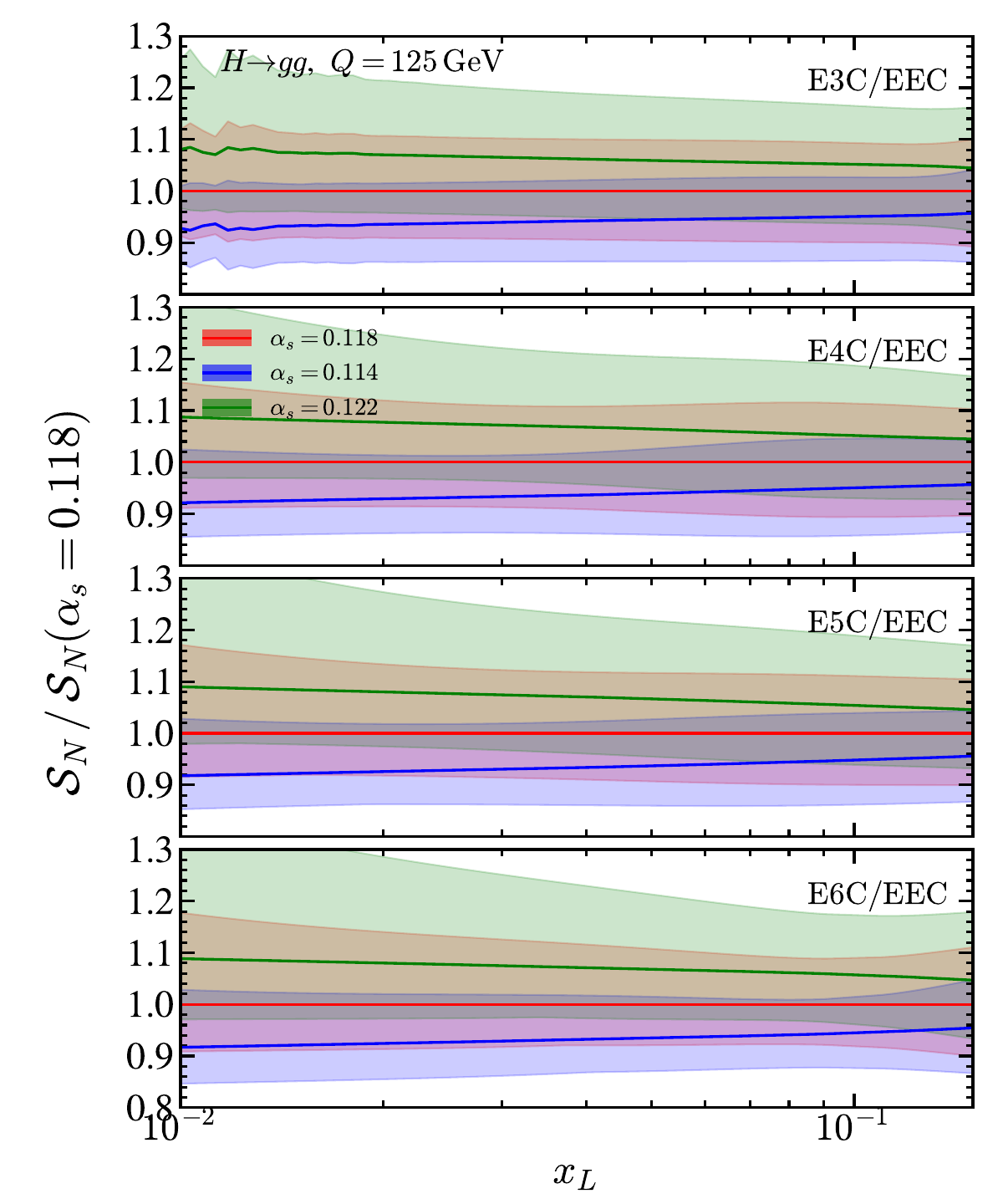}
    \caption{Change in the logarithmic slope $\mathcal{S}_N(\text{ENC/EEC})$ of the ratio of projected correlators to the EEC with respect to the central value $\alpha_s(m_Z)=0.118$, for $e^+e^-\to q\bar q$ (left) and $H\to gg$ (right). Each panel shows $N=3,\ldots,6$ from top to bottom.}
    \label{fig:slope_alphas_ratio}
\end{figure}
%%%%%%%%%%%%%%%%%%%%%%%%%%%%%%%%%%%%%%%%%%%%%%%%%%%%%%%%%%%%%%%%%%%%%%%%%%%%%%%%%%%%%%%%%%%%%%%%%%%%%%%%%%%%%%%%%%%%%%%%%%%%%%%%%%%%%%%%%%%%%%%%%%%%%%%%%%%%%%%%%%%%%%%%%%%%%%%%%%%%%%%%%%%%%%%%%%%%%%%%%%%%%%%
\section{Sensitivity to the strong coupling and non-perturbative corrections}\label{sec:pheno}

In this section, we discuss the potential phenomenological applications of projected $N$-point energy correlators to carry out precision studies. In particular, we want to study their sensitivity to extract the strong coupling constant $\alpha_s$ and the non-perturbative soft matrix elements $\overline{\Omega}_{1\kappa}$. As closely related observables, the family of projected energy correlators for different $N$ provides an ideal set to validate the extractions among different $N$-point observables and to study theory correlations and uncertainties. Being precise about the correlations between different $N$-point projected correlators is beyond the scope of this work and deserves further investigation; here we will carry out a more naive sensitivity study of the observables to $\alpha_s$ and $\overline{\Omega}_{1\kappa}$ with uncertainties and matching prescriptions described in Sec.~\ref{sec:nnll_resum}.  We study the sensitivity to $\alpha_s$ in Sec.~\ref{sec:alphas_sensitivity}, to $\overline{\Omega}_{1\kappa}$ as a whole in Sec.~\ref{sec:omega1k_sensitivity}, and to the less-constrained gluon parameter $\overline{\Omega}_{1g}$ in particular in Sec.~\ref{sec:omega1g_sensitivity}.

\subsection{Sensitivity to $\alpha_s$ extraction}
\label{sec:alphas_sensitivity}

To illustrate the sensitivity of energy correlators to different $\alpha_s(m_Z)$ values, we define the logarithmic slope of both the absolute spectrum and the ratio of higher $N$-point projected energy correlators to $N=2$:
\begin{equation}
\label{eq:slope_def}
\mathcal{S}_N(\text{ENC})\equiv
\frac{d}{d\ln x_L}\ln \frac{d\sigma^{[N]}}{dx_L}\,,\quad  \mathcal{S}_N(\text{ENC/EEC})\equiv
\frac{d}{d\ln x_L}\ln \frac{d\sigma^{[N]}/dx_L}{d\sigma^{[2]}/dx_L}
\end{equation}
where both of them are functions of $x_L$.
In Figs.~\ref{fig:slope_alphas_abs} and~\ref{fig:slope_alphas_ratio}, we plot the logarithmic slope of the absolute spectrum $d\sigma^{[N]}/dx_L$ and of the ratio $(d\sigma^{[N]}/dx_L)/(d\sigma^{[2]}/dx_L)$. As shown in Eqs.~\eqref{eq:scalingabs} and~\eqref{eq:scalingratio}, this slope gives the exponent of the scaling and is therefore directly sensitive to $\alpha_s$. We quantify the sensitivity by plotting the change in the logarithmic slope with respect to the central choice $\alpha_s=0.118$. The sensitivity grows with $N$ and is larger in $H\to gg$ than in $e^+e^-$ annihilation, consistent with the expectation that the twist-two anomalous dimensions $\gamma(N+1)$ grow with $N$ and that the gluon anomalous dimensions are numerically larger than the quark ones, as also observed in Ref.~\cite{Dixon:2019uzg}. This growth of the $\alpha_s$ sensitivity with $N$ provides a clear motivation for measurements of higher-point projected energy correlators.
 
We further observe higher sensitivity to $\alpha_s$ change in the ratio than in the absolute spectrum. This can be understood from the fact that Eqs.~\eqref{eq:scalingabs} and~\eqref{eq:scalingratio} also receive non-perturbative corrections proportional to $\overline{\Omega}_{1\kappa}$; in the ratio, the non-perturbative corrections in the numerator and denominator are related by the $N$-dependent prefactor as in Eq.~\eqref{eq:np_fixed} and they partially cancel, leaving the slope of the ratio more sharply sensitive to $\alpha_s$.

\begin{figure}[!hbp]
    \centering
    \includegraphics[width=0.45\linewidth]{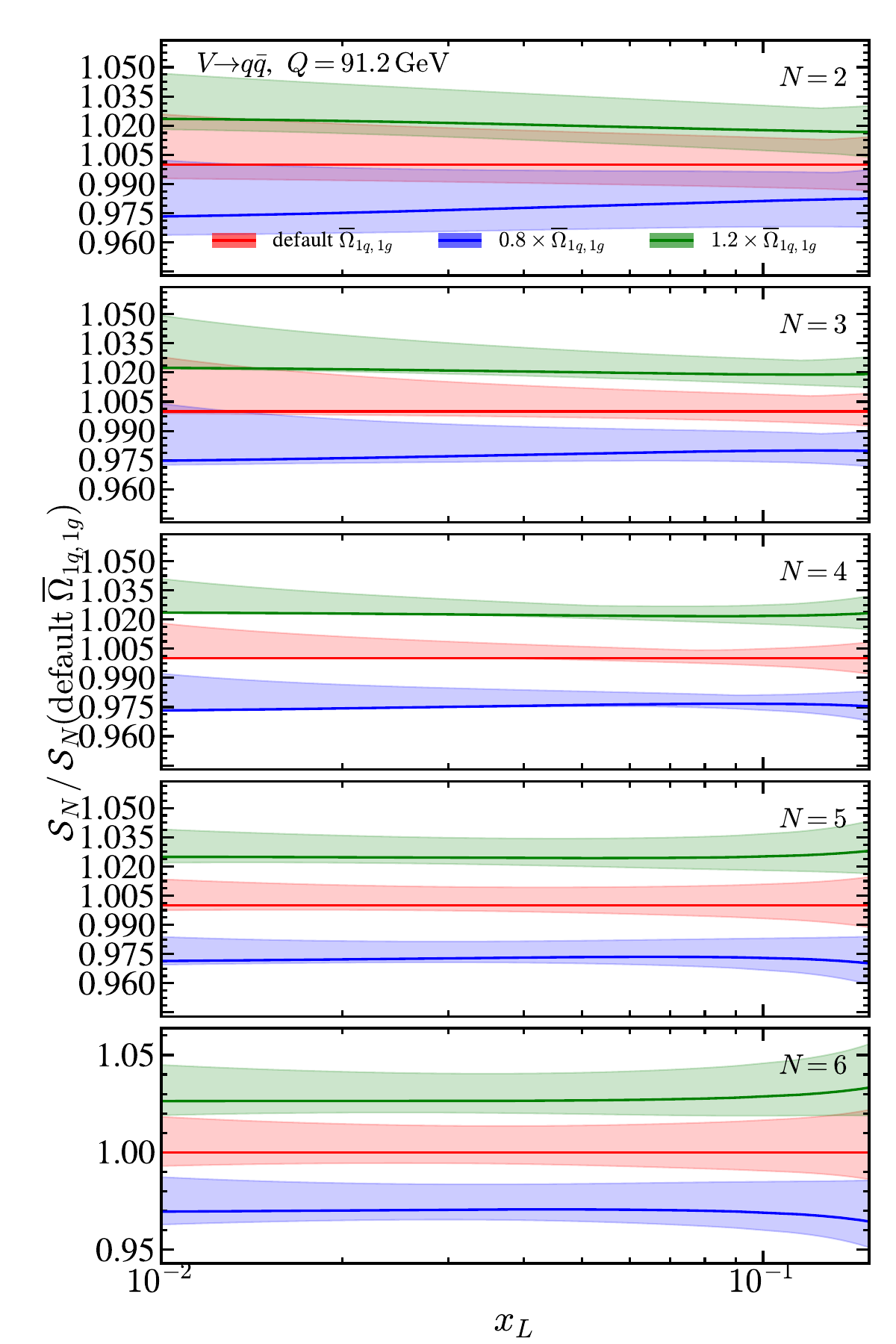}
    \includegraphics[width=0.45\linewidth]{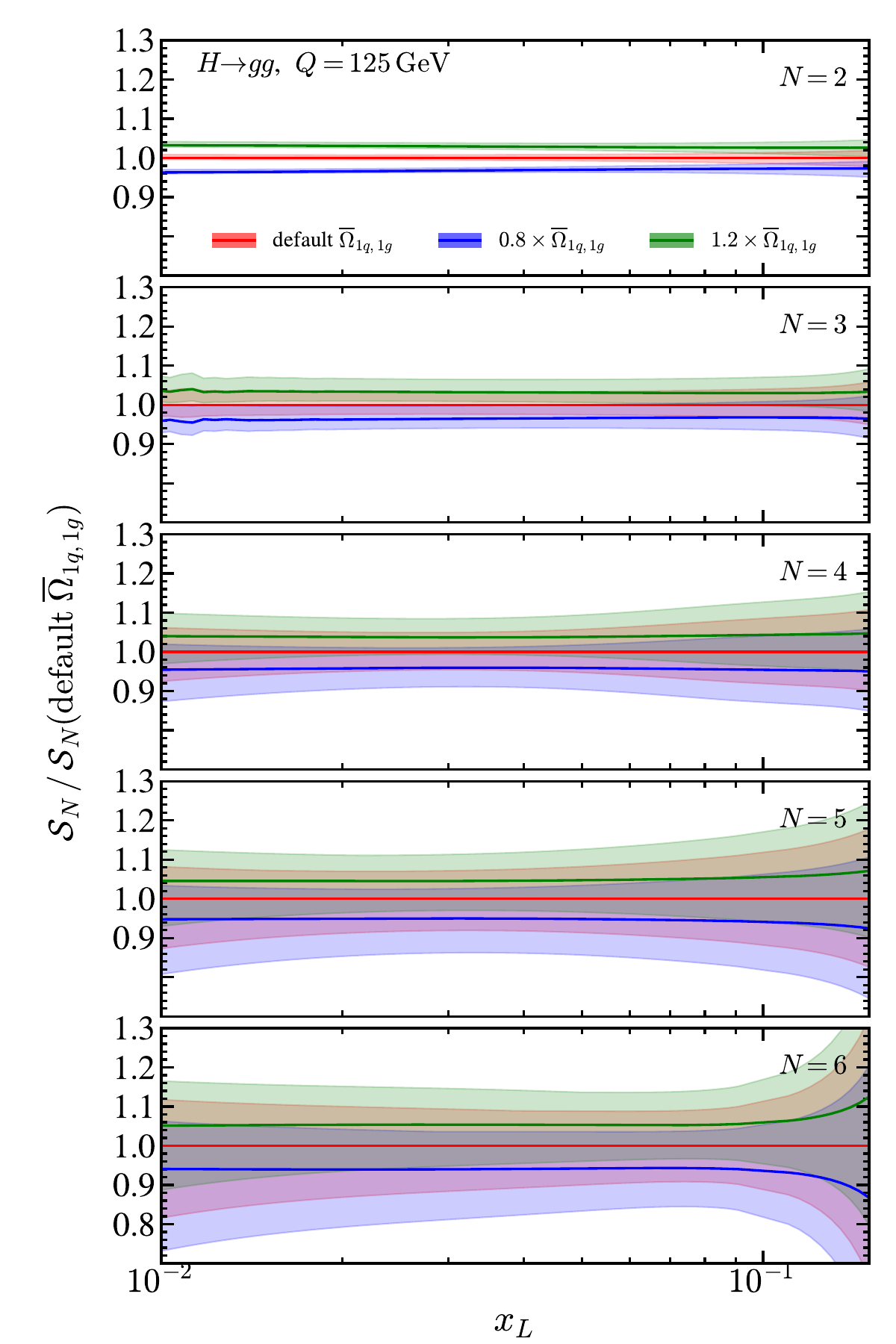}
    \caption{Change in the logarithmic slope $\mathcal{S}_N$ of the absolute projected energy correlator spectrum from different $\overline{\Omega}_{1q,1g}$ values. The distributions are normalized to the default choice $\overline{\Omega}_{1q}=0.305$ GeV and $\overline{\Omega}_{1g}=0.686$ GeV. We present $e^+e^-\to q\bar q$ (left) and $H\to gg$ (right). Each panel shows $N=2,\ldots,6$ from top to bottom.}
    \label{fig:slope_Omega1qg_abs}
\end{figure}

\begin{figure}[!hbp]
    \centering
    \includegraphics[width=0.45\linewidth]{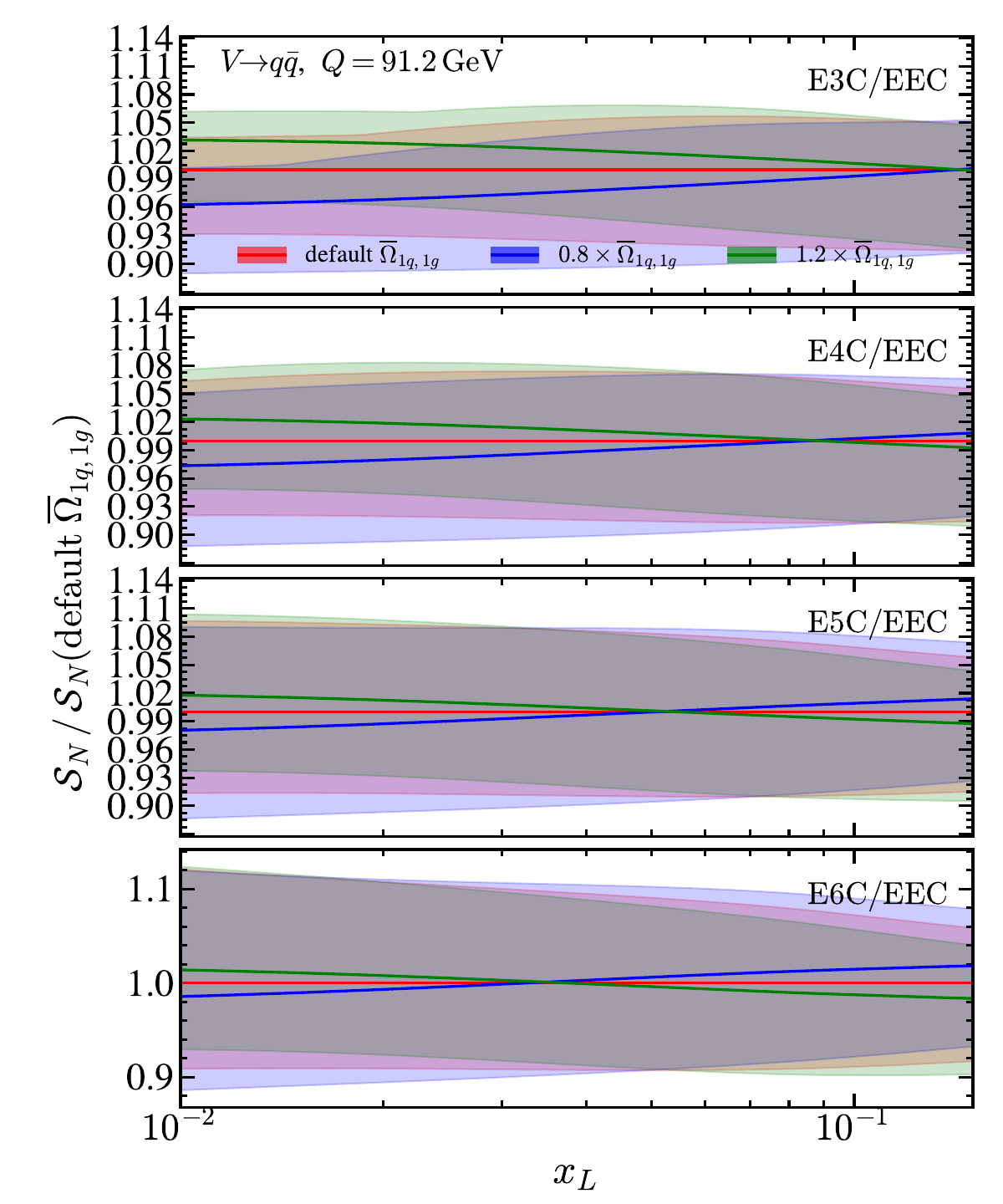}
    \includegraphics[width=0.45\linewidth]{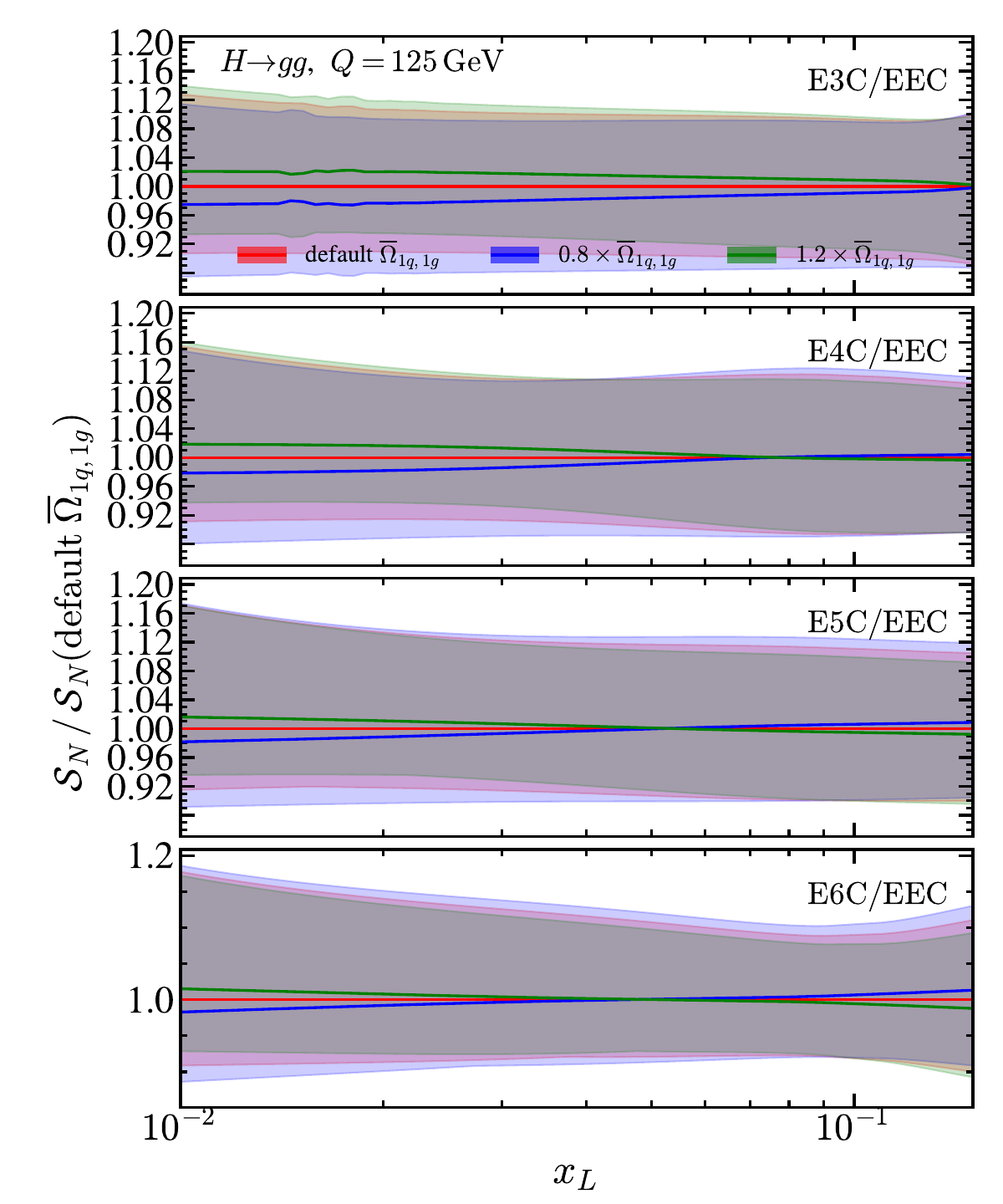}
    \caption{Change in the logarithmic slope $\mathcal{S}_N$ of ratio correlator spectrum from different $\overline{\Omega}_{1q,1g}$ values. The distributions are normalized to the default choice $\overline{\Omega}_{1q}=0.305$ GeV and $\overline{\Omega}_{1g}=0.686$ GeV. We present $e^+e^-\to q\bar q$ (left) and $H\to gg$ (right). Each panel shows $N=3,\ldots,6$ from top to bottom.}
    \label{fig:slope_Omega1qg_ratio}
\end{figure}

\subsection{Sensitivity to $\overline{\Omega}_{1\kappa}$ extraction}
\label{sec:omega1k_sensitivity}
In many event-shape extractions of $\alpha_s$, the strong coupling and the leading non-perturbative parameter $\overline{\Omega}_{1\kappa}$ are strongly correlated, so robust determinations require simultaneous fits of $\alpha_s$ and $\overline{\Omega}_{1\kappa}$~\cite{Abbate:2010xh,Benitez:2024nav,Benitez:2025vsp}. It is therefore important to also study the sensitivity of projected energy correlators to $\overline{\Omega}_{1\kappa}$.

In Figs.~\ref{fig:slope_Omega1qg_abs} and~\ref{fig:slope_Omega1qg_ratio}, we plot the sensitivity of the logarithmic slope of the absolute spectrum $d\sigma^{[N]}/dx_L$ and of the ratio $(d\sigma^{[N]}/dx_L)/(d\sigma^{[2]}/dx_L)$ to variations of $\overline{\Omega}_{1\kappa}$. For definiteness, we impose the naive Casimir relation $\overline{\Omega}_{1g}=(C_A/C_F),\overline{\Omega}_{1q}$ and vary $\overline{\Omega}_{1q}$ around its central value $\overline{\Omega}_{1q}=0.305$~GeV taken from the global thrust fit. The figures indicate that absolute spectrum has a higher sensitivity to variations of $\overline{\Omega}_{1\kappa}$ than the ratio, consistent with the partial cancellation of power corrections between numerator and denominator. This also shows that the ratio is the cleaner probe of $\alpha_s$, while the absolute spectrum is useful for extracting $\overline{\Omega}_{1\kappa}$.

\subsection{Sensitivity to $\overline{\Omega}_{1g}$ extraction}
\label{sec:omega1g_sensitivity}

\begin{figure}[!hbp]
    \centering
    \includegraphics[width=0.45\linewidth]{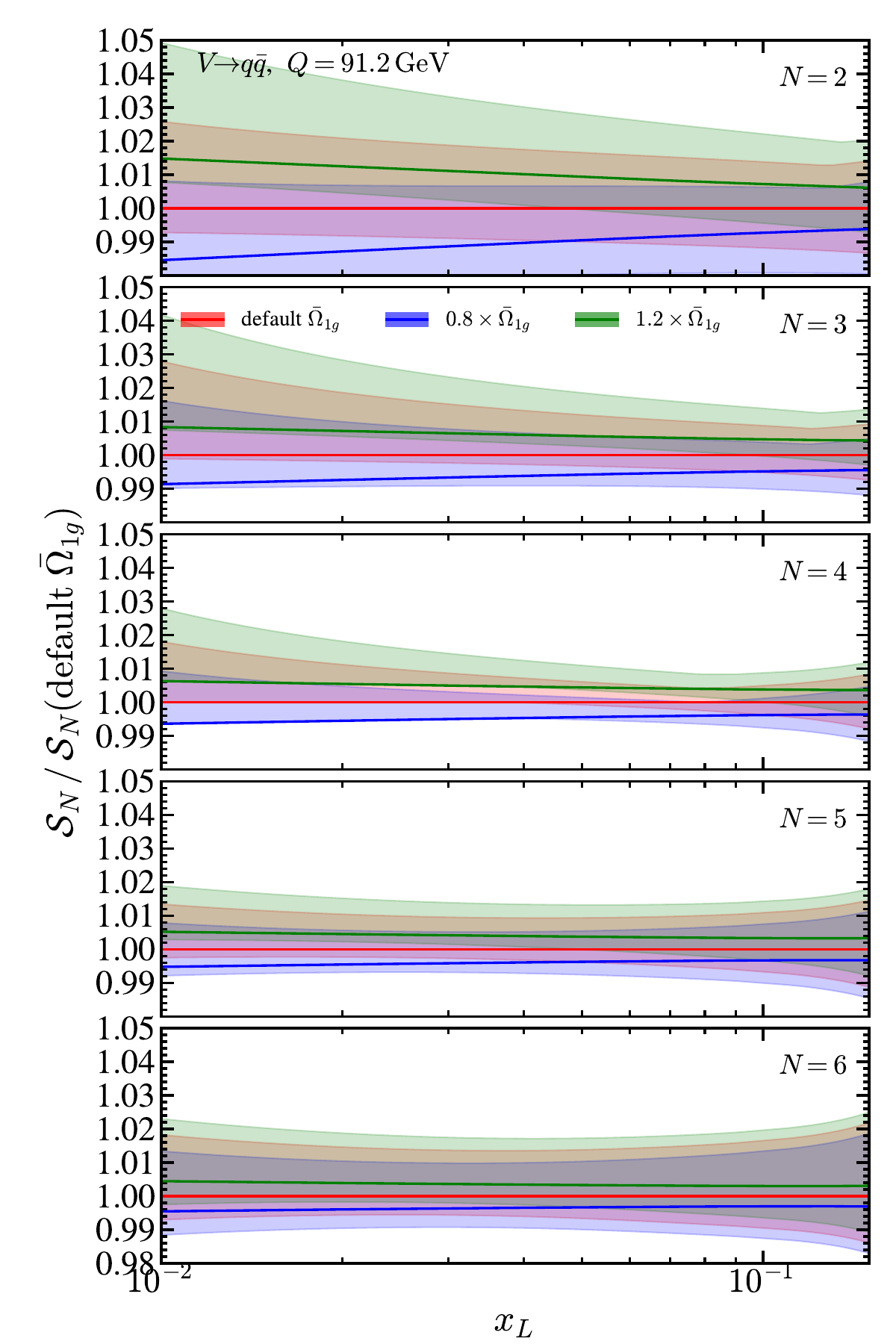}
    \includegraphics[width=0.45\linewidth]{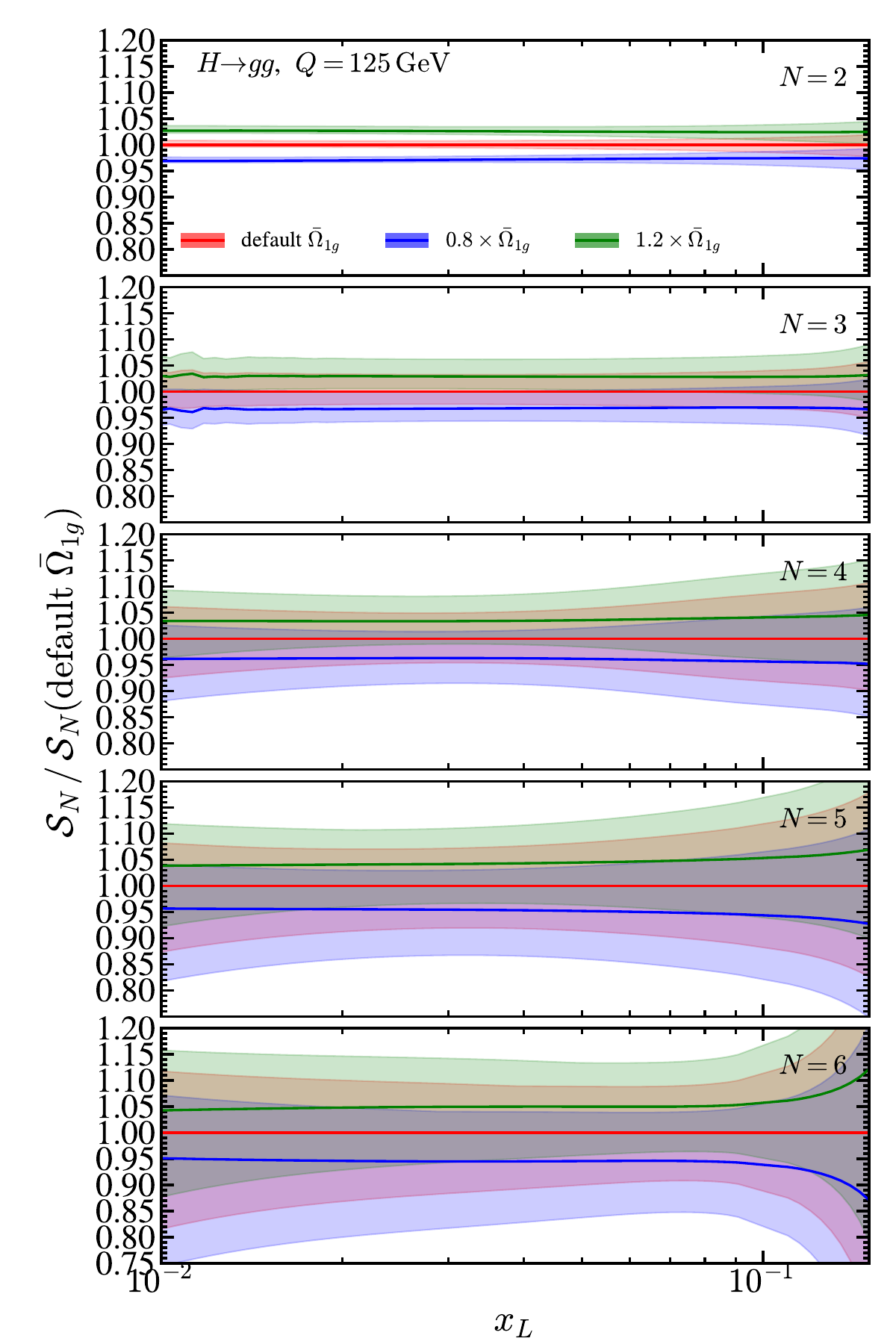}
    \caption{Change in the logarithmic slope $\mathcal{S}_N$ of the absolute projected energy correlator spectrum from different $\overline{\Omega}_{1g}$ values. The distributions are normalized to the default choice $\overline{\Omega}_{1g}=0.686$ GeV, and we choose $\overline{\Omega}_{1q}=0.305$ GeV for all curves. We present $e^+e^-\to q\bar q$ (left) and $H\to gg$ (right). Each panel shows $N=2,\ldots,6$ from top to bottom.}
    \label{fig:slope_Omega1g_abs}
\end{figure}

\begin{figure}[!htbp]
    \centering
    \includegraphics[width=0.45\linewidth]{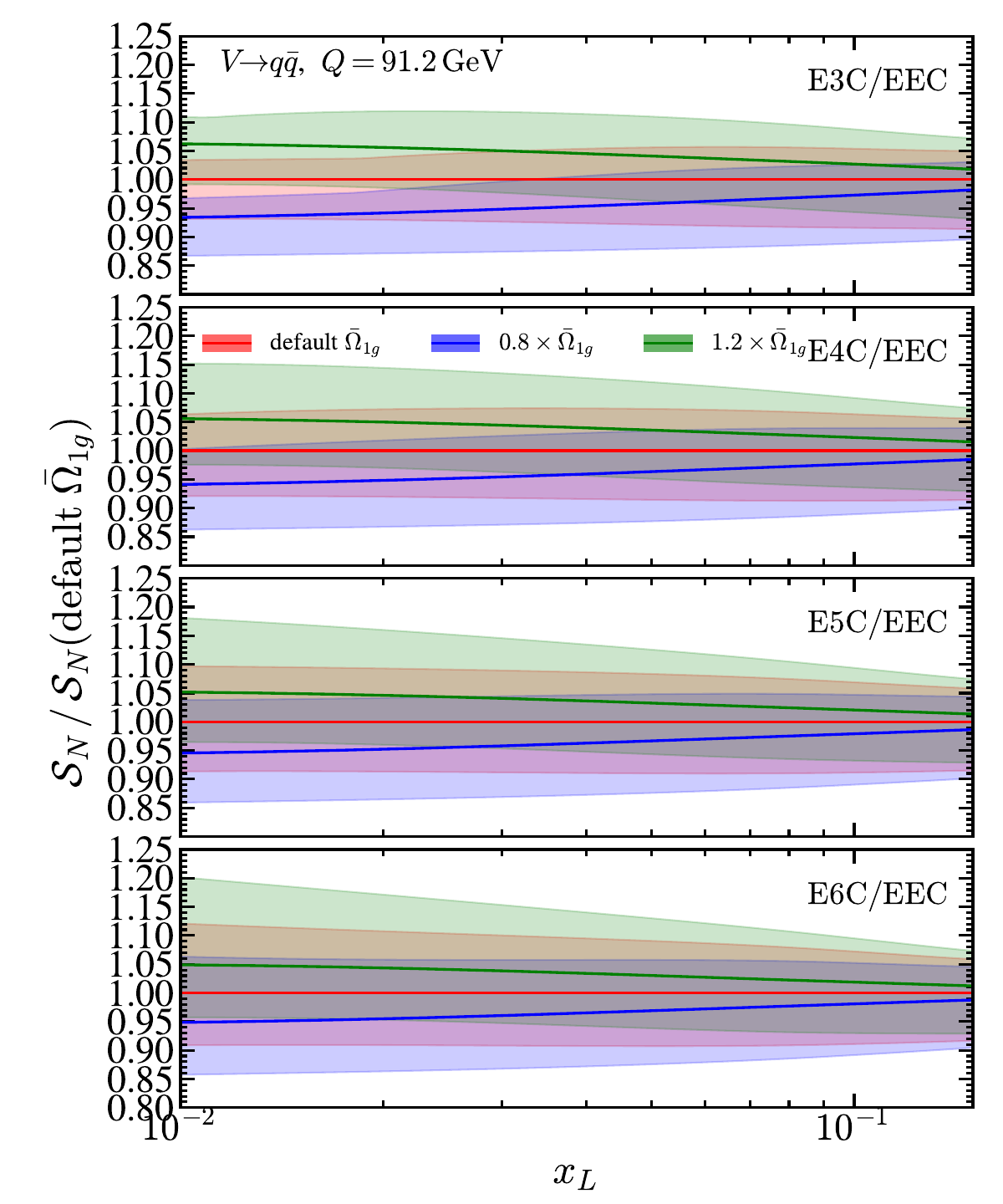}
    \includegraphics[width=0.45\linewidth]{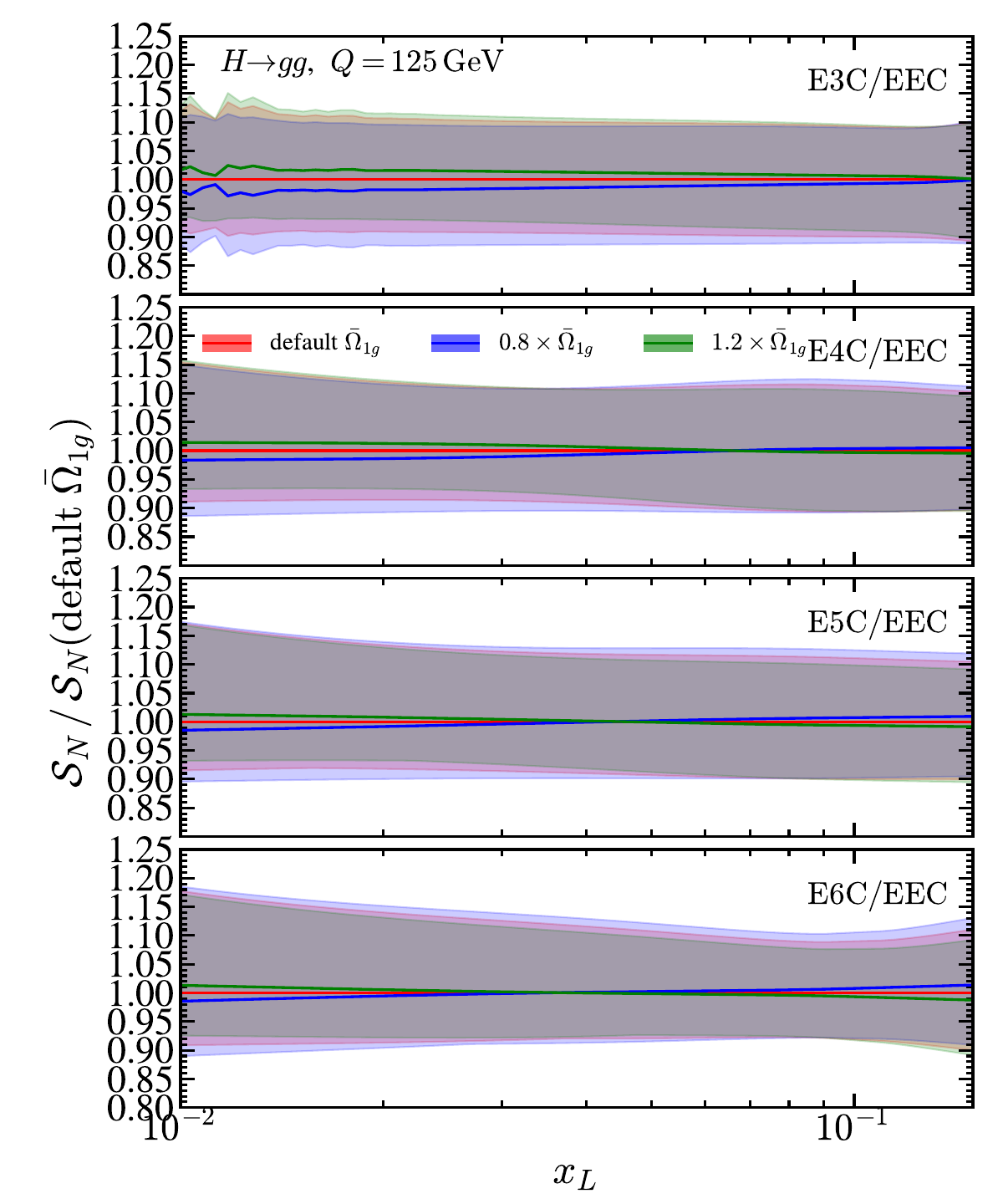}
    \caption{Change in the logarithmic slope $\mathcal{S}_N$ of ratio correlator spectrum from different $\overline{\Omega}_{1g}$ values. The distributions are normalized to the default choice $\overline{\Omega}_{1g}=0.686$ GeV, and we choose $\overline{\Omega}_{1q}=0.305$ GeV for all curves. We present $e^+e^-\to q\bar q$ (left) and $H\to gg$ (right). Each panel shows $N=3,\ldots,6$ from top to bottom.}
    \label{fig:slope_Omega1g_ratio}
\end{figure}

Reliable independent extractions of $\alpha_s$ and $\overline{\Omega}_{1q}$ from other event-shape observables will in turn enable the extraction of $\overline{\Omega}_{1g}$ from projected energy correlators, which is currently poorly constrained and has not been directly extracted from data. In Figs.~\ref{fig:slope_Omega1g_abs} and~\ref{fig:slope_Omega1g_ratio}, we plot the sensitivity of the logarithmic slope of the absolute spectrum $d\sigma^{[N]}/dx_L$ and of the ratio $(d\sigma^{[N]}/dx_L)/(d\sigma^{[2]}/dx_L)$ to variations of $\overline{\Omega}_{1g}$, for fixed $\alpha_s(m_Z)=0.118$ and $\overline{\Omega}_{1q}=0.305$~GeV.

The figures indicate that the absolute spectrum in $H\to gg$ has much higher sensitivity to variations of $\overline{\Omega}_{1g}$ than the corresponding $e^+e^-$ spectrum. This follows from Eq.~\eqref{eq:np_fixed}: the leading non-perturbative correction is controlled by $\overline{\Omega}_{1q}$ for $e^+e^-$ and by $\overline{\Omega}_{1g}$ for $H\to gg$. In the $e^+e^-$ case, $\overline{\Omega}_{1g}$ enters only indirectly, through DGLAP mixing of the quark and gluon jet functions in the resummation evolution.

More surprisingly, the $e^+e^-$ ratio retains substantial sensitivity to $\overline{\Omega}_{1g}$ in the resummation region, while the $H\to gg$ ratio shows much weaker sensitivity, as expected. The likely explanation is that the leading $\overline{\Omega}_{1q}$ correction largely cancels between numerator and denominator of the $e^+e^-$ ratio, leaving the $\overline{\Omega}_{1g}$ contribution from mixing comparatively more visible.

%%%%%%%%%%%%%%%%%%%%%%%%%%%%%%%%%%%%%%%%%%%%%%%%%%%%%%%%%%%%%%%%%%%%%%%%%%%%%%%%%%%%%%%%%%%%%%%%%%%%%%%%%%%%%%%%%%%%%%%%%%%%%%%%%%%%%%%%%%%%%%%%%%%%%%%%%%%%%%%%%%%%%%%%%%%%%%%%%%%%%%%%%%%%%%%%%%%%%%%%%%%%%%%
\section{Conclusion}\label{sec:conclusion}

In this paper, we extend the NNLL collinear resummation of projected energy correlators to four-, five-, and six-point correlators. To accomplish this precision, we have computed the two-loop jet function for projected $N$-point correlators with $N=4,5,6$, for both quark and gluon jets. We extracted these boundary constants by matching the singular $x_L\to0$ limit of the fixed-order distributions onto the collinear factorization theorem in Eq.~\eqref{eq:fac_nu}. The calculation was organized using the measurement decomposition of Eqs.~\eqref{eq:projected_exact_decomp}--\eqref{eq:W_definition}, which separates the two-loop problem into contact terms and resolved three-particle terms.  The contact terms were calculated using IBP reductions and differential equations, while the remaining three-particle contributions were obtained via Monte Carlo integration. We verified our calculation against the known $N=2,3$ results in the literature and the $x_L\to 0$ behavior from numerical programs.

With these ingredients, we obtained NNLL+NLO predictions for $N$-point projected correlators up to $N=6$ in both $e^+e^-$ annihilation and $H\to gg$ decay. In the perturbative collinear region, the matched distributions show relatively stable order-by-order behavior from NLL to NNLL. At sufficiently small $x_L$, the perturbative expansion becomes sensitive to the leading hadronization correction, whose parametric size is governed by $N\,\overline{\Omega}_{1\kappa}/(Q\sqrt{x_L})$.
This explains why the onset of non-perturbative effects moves to larger $x_L$ as $N$ increases. We incorporated these leading non-perturbative power corrections in the $\overline{\rm MS}$ scheme through the $(N-1)$-point evolution formula of Eq.~\eqref{eq:np_jet}, controlled by the two hadronic parameters $\overline{\Omega}_{1q}$ and $\overline{\Omega}_{1g}$.
We also compared the matched predictions with parton-shower simulations from \textsc{Pythia8} and \textsc{Herwig7}. In $e^+e^-$ annihilation, including the leading non-perturbative correction improves the qualitative agreement with the shower results in the small-angle region. In $H\to gg$ decay, and especially for higher-point correlators, we observe sizeable differences both between our predictions and the shower results and between the two showers themselves. These likely reflect the absence of precision $H\to gg$ data against which \textsc{Pythia8} and \textsc{Herwig7} could be tuned.

Beyond the absolute spectra, we studied ratios of projected correlators to the EEC. These ratios largely cancel the common classical $1/x_L$ behavior and isolate the anomalous scaling associated with twist-two operators of different spin. Their logarithmic slopes are therefore directly sensitive to the timelike anomalous dimensions and to the value of $\alpha_s$. We find that the sensitivity to $\alpha_s$ increases with $N$, consistent with the growth of the relevant anomalous dimensions, and is generally larger in the gluon channel than in the quark channel. The ratios are also less sensitive to the leading non-perturbative power correction than the absolute spectra, since the $N$-dependent prefactor of $\overline{\Omega}_{1\kappa}$ in Eq.~\eqref{eq:np_fixed} partially cancels between numerator and denominator. This partial cancellation sharpens the $\alpha_s$ sensitivity of the ratio relative to the absolute spectrum, and motivates the ratio as a particularly promising observable for future $\alpha_s$ determination. 

The framework developed here extends beyond $e^+e^-$ processes and color-singlet decays, and can be applied to projected energy correlators measured inside small-$R$ jets~\cite{Kang:2016mcy,Lee:2024icn,Generet:2025vth} at hadron colliders or in events with a high-$p_T$ recoil against a color-singlet boson such as a $Z$ or a photon. Indeed, the three-point projected energy correlator has already been measured inside jets by the CMS collaboration~\cite{CMS:2024mlf} to extract $\alpha_s$. A re-analysis of archival LEP data along the lines of Refs.~\cite{Bossi:2025xsi,Electron-PositronAlliance:2025fhk,Zhang:2025nlf} would also be especially valuable, enabling precision $\alpha_s$ extractions from a closely related family of observables and helping to disentangle the systematics that drive the longstanding tension between event-shape and lattice determinations of $\alpha_s$. Looking even further ahead, future Higgs factories such as FCC-ee and CEPC~\cite{FCC:2018evy,CEPCStudyGroup:2018ghi} will provide the first sample in which the $H\to gg$ predictions presented here can be confronted with precision data, and in which $\overline{\Omega}_{1g}$ can plausibly be extracted directly.

On the theoretical side, several extensions are also possible. Promoting the NNLL predictions for higher-point projected correlators to track-based observables using the track-function formalism~\cite{Chang:2013rca,Li:2021zcf,Chen:2022muj,Jaarsma:2023ell,Lee:2023npz,Lee:2023tkr} would make them directly comparable to high-precision measurements on tracks. It would also be interesting to extend the analytic continuation of the projected $\nu$-point correlator to two loops; the one-loop continuation in $\nu$ was carried out in Ref.~\cite{Chen:2020vvp}, and the two-loop jet constants computed here at integer $N=4,5,6$ may provide enough data to constrain or even bootstrap an analytic solution at general $\nu$.

\acknowledgments

We thank HuaXing Zhu for collaborating on the early stages of this project.
We also thank Hao Chen, Ian Moult, Iain Stewart, and Tong-Zhi Yang for useful discussions.
K.L is supported by the U.S. Department of Energy under contracts DE-AC02-06CH11357.
Y.L. is supported by funding from the European Research Council (ERC) under the European Union’s Horizon 2022 Research and Innovation Program (ERC Advanced Grant agreement No.101097780, EFT4jets). 
Z.X. is funded by the European Union (ERC, TOPMASS, 101165601).
Views and opinions expressed are however those of the author(s) only and do not necessarily reflect those of the European Union or the European Research Council.
Neither the European Union nor the granting authority can be held responsible for them.
XY.Z is supported by the MIT Pappalardo Fellowship.
The computations in this paper were run on the FASRC Cannon cluster at Harvard University and the theory cluster at DESY.

\clearpage
\appendix

\section{LO analytic results}
\label{app:lo_enc}

Below are the analytic expressions for LO projected energy correlators up to six-point, for both $e^+e^-\to q\bar q$ and $H\to gg$ processes. 

For $e^+e^-\to q\bar q$, the result for $N$-point is 
\begin{equation}
\frac{1}{\sigma_0}\frac{d\sigma_{e^+e^-}^{[2],\rm LO}}{dx_L}= \frac{\alpha_s C_F}{4\pi x_L^N (1-x_L) }\times \left[q_1^{[N]}(x_L)+\theta\left(x_L-\frac{3}{4}\right)q_2^{[N]}(x_L) \right] \, ,
\end{equation}
with
\begin{align}
q_1^{[2]}&=\left(\frac{36}{x_L}-\frac{48}{x_L^2}+\frac{18}{x_L^3}-8\right) \log \left(1-x_L\right)+18-\frac{39}{x_L}+\frac{18}{x_L^2}\,,\quad q_2^{[2]}=0\,,\notag\\
q_1^{[3]}&=\left(\frac{225}{2 x_L}-\frac{261}{2 x_L^2}+\frac{45}{x_L^3}-30\right) \log \left(1-x_L\right)+\frac{249}{4}-6 x_L-\frac{108}{x_L}+\frac{45}{x_L^2} \, ,\notag\\
q_2^{[3]}&=\left(-24 x_L-\frac{369}{x_L}+\frac{315}{x_L^2}-\frac{90}{x_L^3}+168\right) \log \left(4-4
   x_L\right)+96 x_L+\frac{783}{x_L}-\frac{4221}{8 x_L^2}+\frac{999}{8 x_L^3}-474\, ,\notag\\
q_1^{[4]}&=\left(6 x_L+\frac{555}{2 x_L}-\frac{282}{x_L^2}+\frac{90}{x_L^3}-95\right) \log
   \left(1-x_L\right)-\frac{3 x_L^2}{2}-\frac{111 x_L}{4}-\frac{237}{x_L}+\frac{90}{x_L^2}+\frac{333}{2}\, ,\notag\\
q_2^{[4]}&=\left(-48 x_L^2+336 x_L+\frac{630}{x_L}-\frac{180}{x_L^2}-738\right) \log \left(4-4 x_L\right)\notag\\
&\hspace{2cm}+192 x_L^2-948 x_L-\frac{4221}{4 x_L}+\frac{999}{4 x_L^2}+1566\, ,\notag\\
q_1^{[5]}&=\left(35 x_L+\frac{2565}{4 x_L}-\frac{4605}{8 x_L^2}+\frac{1365}{8
   x_L^3}-275\right) \log \left(1-x_L\right)\notag\\
   &\hspace{2cm}-\frac{13 x_L^3}{24}+\frac{47 x_L^2}{32}-\frac{3315 x_L}{32}-\frac{7845}{16 x_L}+\frac{1365}{8 x_L^2}+\frac{6565}{16}\, ,\notag\\
q_2^{[5]}&=\left(-80 x_L^3+690 x_L^2-2150 x_L-\frac{6285}{2
   x_L}+\frac{6405}{4 x_L^2}-\frac{1365}{4 x_L^3}+\frac{6845}{2}\right) \log \left(4-4 x_L\right)\notag\\
   &\hspace{1cm}+\frac{1004 x_L^3}{3}-2109 x_L^2+5285 x_L+\frac{370455}{64 x_L}-\frac{655701}{256 x_L^2}+\frac{121113}{256 x_L^3}-\frac{57635}{8}\, ,\notag\\
q_1^{[6]}&=\left(-5 x_L^2+\frac{1155 x_L}{8}+\frac{11571}{8
   x_L}-\frac{9273}{8 x_L^2}+\frac{1281}{4 x_L^3}-\frac{3003}{4}\right) \log \left(1-x_L\right)\notag\\
   &\hspace{2cm}-\frac{9 x_L^4}{40}+\frac{13 x_L^3}{160}+\frac{4063 x_L^2}{160}-\frac{2671 x_L}{8}-\frac{999}{x_L}+\frac{1281}{4 x_L^2}+\frac{15577}{16}\, ,\notag\\
q_2^{[6]}&=\left(-120 x_L^4+1230 x_L^3-4605 x_L^2+\frac{17385 x_L}{2}+\frac{19215}{4
   x_L}-\frac{4095}{4 x_L^2}-\frac{17955}{2}\right) \log \left(4-4 x_L\right)\notag\\
   &\hspace{1cm}+524 x_L^4-3957 x_L^3+11940 x_L^2-18975 x_L-\frac{1967103}{256 x_L}+\frac{363339}{256 x_L^2}+\frac{1071405}{64}\, .\notag\\
\end{align}

For $H\to gg$, the result for $N$-point is 
\begin{equation}
\frac{1}{\sigma_0}\frac{d\sigma_{H}^{[2],\rm LO}}{dx_L}= \frac{\alpha_s}{4\pi x_L^N (1-x_L) }\times \left[g_1^{[N]}(x_L)+\theta\left(x_L-\frac{3}{4}\right)g_2^{[N]}(x_L) \right]\, ,
\end{equation}
with
\begin{align}
    g_1^{[2]}&=C_A \left[+\left(\frac{28}{x_L}-\frac{102}{x_L^2}+\frac{148}{x_L^3}-\frac{72}{x_L^4}-4\right) \log
   \left(1-x_L\right)+\frac{25}{3}-\frac{52}{x_L}+\frac{112}{x_L^2}-\frac{72}{x_L^3}\right]\notag\\
   &+n_f
   \left[\left(-\frac{34}{x_L}+\frac{126}{x_L^2}-\frac{166}{x_L^3}+\frac{72}{x_L^4}+2\right) \log
   \left(1-x_L\right)+\frac{67}{x_L}-\frac{130}{x_L^2}+\frac{72}{x_L^3}-\frac{25}{3}\right]\, ,\notag\\
   g_2^{[2]}&=0\, ,\notag\\
   g_1^{[3]}&=C_A \left[\left(\frac{126}{x_L}-\frac{369}{x_L^2}+\frac{456}{x_L^3}-\frac{198}{x_L^4}-\frac{18 x_L}{5}-\frac{207}{x_L}+\frac{357}{x_L^2}-\frac{198}{x_L^3}-18\right)
   \log \left(1-x_L\right)+44\right]\notag\\
   &+n_f \left[\left(-\frac{297}{2 x_L}+\frac{873}{2
   x_L^2}-\frac{501}{x_L^3}+\frac{198}{x_L^4}+15\right) \log \left(1-x_L\right)+\frac{3 x_L}{5}+\frac{252}{x_L}-\frac{402}{x_L^2}+\frac{198}{x_L^3}-\frac{191}{4}\right]\, ,\notag\\
   g_2^{[3]}&=C_A \Bigg[\left(-12
   x_L-\frac{558}{x_L}+\frac{1182}{x_L^2}-\frac{1128}{x_L^3}+\frac{396}{x_L^4}+120\right) \log \left(4-4 x_L\right)+\frac{232 x_L}{5}\notag\\
   &+\frac{1524}{x_L}-\frac{5139}{2 x_L^2}+\frac{31365}{16 x_L^3}-\frac{43929}{80 x_L^4}-410\Bigg]+n_f \Bigg[\left(6
   x_L+\frac{675}{x_L}-\frac{1371}{x_L^2}+\frac{1218}{x_L^3}-\frac{396}{x_L^4}\right.\notag\\
   &\left.-132\right) \log \left(4-4 x_L\right)-\frac{172
   x_L}{5}-\frac{1830}{x_L}+\frac{23373}{8 x_L^2}-\frac{33363}{16 x_L^3}+\frac{43929}{80 x_L^4}+479\Bigg]\, ,\notag\\
   g_1^{[4]}&=C_A \Bigg[\left(3 x_L+\frac{813}{2
   x_L}-\frac{996}{x_L^2}+\frac{1078}{x_L^3}-\frac{424}{x_L^4}-71\right) \log \left(1-x_L\right)-\frac{16 x_L^2}{15}-\frac{301 x_L}{20}\notag\\
   &-\frac{1795}{3 x_L}+\frac{866}{x_L^2}-\frac{424}{x_L^3}+\frac{971}{6}\Bigg]+n_f \Bigg[\left(-\frac{3
   x_L}{2}-\frac{474}{x_L}+\frac{1149}{x_L^2}-\frac{1168}{x_L^3}+\frac{424}{x_L^4}+\frac{141}{2}\right) \notag\\
   &\log \left(1-x_L\right)+\frac{x_L^2}{15}+\frac{93
   x_L}{10}+\frac{2119}{3 x_L}-\frac{956}{x_L^2}+\frac{424}{x_L^3}-\frac{1097}{6}\Bigg]\, ,\notag\\
   g_2^{[4]}&=C_A \Bigg[\left(-24 x_L^2+240
   x_L+\frac{2364}{x_L}-\frac{2256}{x_L^2}+\frac{792}{x_L^3}-1116\right) \log \left(4-4 x_L\right)+\frac{464 x_L^2}{5}\notag\\
   &-820 x_L-\frac{5139}{x_L}+\frac{31365}{8 x_L^2}-\frac{43929}{40 x_L^3}+3048\Bigg]+n_f \Bigg[\left(12 x_L^2-264 x_L-\frac{2742}{x_L}+\frac{2436}{x_L^2}\right.\notag\\
   &\left.-\frac{792}{x_L^3}+1350\right) \log \left(4-4
   x_L\right)-\frac{344 x_L^2}{5}+958 x_L+\frac{23373}{4
   x_L}-\frac{33363}{8 x_L^2}+\frac{43929}{40 x_L^3}-3660\Bigg]\, ,\notag\\
   g_1^{[5]}&=C_A \Bigg[\left(20
   x_L+\frac{1165}{x_L}-\frac{4865}{2 x_L^2}+\frac{2340}{x_L^3}-\frac{3365}{4 x_L^4}-255\right) \log \left(1-x_L\right)-\frac{3 x_L^3}{7}+\frac{x_L^2}{2}\notag\\
   &-\frac{799 x_L}{12}-\frac{18515}{12 x_L}+\frac{15355}{8 x_L^2}-\frac{3365}{4 x_L^3}+\frac{8295}{16}\Bigg]+n_f
   \Bigg[\left(-15 x_L-\frac{5395}{4
   x_L}\right.\notag\\
   &\left.+\frac{22085}{8 x_L^2}-\frac{20085}{8 x_L^3}+\frac{3365}{4 x_L^4}+\frac{545}{2}\right) \log \left(1-x_L\right)+\frac{x_L^3}{84}-\frac{3 x_L^2}{32}+\frac{5657 x_L}{96}+\frac{85715}{48 x_L}\notag\\
   &-\frac{2090}{x_L^2}+\frac{3365}{4 x_L^3}-595\Bigg]\, ,\notag\\
   g_2^{[5]}&=C_A \Bigg[\left(-40 x_L^3+475 x_L^2-2675 x_L-\frac{12590}{x_L}+\frac{12235}{x_L^2}-\frac{6800}{x_L^3}+\frac{3365}{2 x_L^4}+\frac{15425}{2}\right)\notag\\
   &\log \left(4-4
   x_L\right)+\frac{3424 x_L^3}{21}-\frac{5080 x_L^2}{3}+\frac{47053 x_L}{6}+\frac{2577785}{96 x_L}-\frac{183369}{8 x_L^2}+\frac{355505}{32 x_L^3}\notag\\
   &-\frac{261237}{112
   x_L^4}-\frac{76055}{4}\Bigg]+n_f \Bigg[\left(20 x_L^3-\frac{995 x_L^2}{2}+\frac{6255 x_L}{2}+\frac{28345}{2 x_L}-\frac{53365}{4
   x_L^2}+\frac{28565}{4 x_L^3}\right.\notag\\
   &\left.-\frac{3365}{2 x_L^4}-8940\right) \log \left(4-4 x_L\right)-\frac{2444 x_L^3}{21}+\frac{5647 x_L^2}{3}-\frac{110261 x_L}{12}-\frac{1439125}{48 x_L}\notag\\
   &+\frac{6347793}{256
   x_L^2}-\frac{2965153}{256 x_L^3}+\frac{261237}{112 x_L^4}+\frac{87435}{4}\Bigg]\, ,\notag\\
   g_1^{[6]}&=C_A \Bigg[\left(-\frac{5 x_L^2}{2}+\frac{195 x_L}{2}+\frac{12607}{4 x_L}-\frac{22779}{4 x_L^2}+\frac{39357}{8 x_L^3}-\frac{6509}{4 x_L^4}-\frac{3393}{4}\right) \log
   \left(1-x_L\right)\notag\\
   &-\frac{213 x_L^4}{1120}-\frac{11 x_L^3}{280}+\frac{1039 x_L^2}{80}-\frac{42587 x_L}{160}-\frac{181313}{48 x_L}+\frac{4106}{x_L^2}-\frac{6509}{4
   x_L^3}+\frac{24599}{16}\Bigg]\notag\\
   &+n_f \Bigg[\left(\frac{5 x_L^2}{4}-\frac{705 x_L}{8}-\frac{14497}{4 x_L}+\frac{25545}{4 x_L^2}-\frac{41919}{8
   x_L^3}+\frac{6509}{4 x_L^4}+\frac{1875}{2}\right) \log \left(1-x_L\right)\notag\\
   &+\frac{3 x_L^4}{1120}+\frac{3 x_L^3}{112}-\frac{613 x_L^2}{80}+\frac{43137 x_L}{160}+\frac{206819}{48
   x_L}-\frac{17705}{4 x_L^2}+\frac{6509}{4 x_L^3}-\frac{28335}{16}\Bigg]\, ,\notag\\
   g_2^{[6]}&=C_A \Bigg[\left(-60
   x_L^4+825 x_L^3-5235 x_L^2+\frac{34455 x_L}{2}+\frac{34725}{x_L}-\frac{20400}{x_L^2}+\frac{10095}{2 x_L^3}-32130\right)\notag\\
   &\log \left(4-4
   x_L\right)+\frac{1800 x_L^4}{7}-3030 x_L^3+\frac{31813 x_L^2}{2}-\frac{176775 x_L}{4}-\frac{1056285}{16 x_L}+\frac{1066515}{32 x_L^2}\notag\\
   &-\frac{783711}{112 x_L^3}+\frac{2264135}{32}\Bigg]+n_f \Bigg[\left(30 x_L^4-\frac{1665 x_L^3}{2}+\frac{12015 x_L^2}{2}-19965 x_L-\frac{152175}{4 x_L}\right.\notag\\
   &\left.+\frac{85695}{4
   x_L^2}-\frac{10095}{2 x_L^3}+\frac{72855}{2}\right) \log \left(4-4 x_L\right)-\frac{1240 x_L^4}{7}+3252 x_L^3-\frac{73661 x_L^2}{4}\notag\\
   &+\frac{407745 x_L}{8}+\frac{18340515}{256
   x_L}-\frac{8895459}{256 x_L^2}+\frac{783711}{112 x_L^3}-\frac{636155}{8}\Bigg]\, .
\end{align}

These full expressions reproduce the known EEC and E3C limits and make clear that the resolved three-particle contribution always turns on through $\theta\left(x_L-3/4\right)$. We also provide the ancillary file for these expressions.

Note that for the Higgs decay, we do not include the matching coefficient $|C(m_t,\mu)|^2$. This matching coefficient cancels against the denominator, i.e. the Higgs to gluon decay width as discussed around Eq.~\eqref{eq:KHgg}.

\section{Jet anomalous dimensions}
\label{app:anomalous_dim}

The NNLL resummation requires the anomalous dimensions for solving the jet function RGE. As described in Eq.~\eqref{eq:jet_ana_dim}, they are Mellin moments of the time-like splitting kernel $\hat{P}(y)$. Note that due to the convolution structure in the jet function RGE, we also need the first two derivatives of jet anomalous dimensions with respect to $N$. Explicitly, if we define the splitting kernel
\begin{equation}
	P_{ij}(x)=\sum_{L=0}^{\infty}\left(\frac{\alpha_s}{4\pi}\right)^{L+1} P_{ij}^{(L)}(x) \,,
\end{equation}
and the required anomalous dimensions are
\begin{align}
	\gamma_{T,ij}^{(L)}&\equiv -\int_0^1 \df x \,x^N \,P_{ij}^{(L)}(x)\,,\nn\\
	\dot \gamma_{T,ij}^{(L)}&\equiv -\int_0^1 \df x\, \ln x \, x^N P_{ij}^{(L)}(x)\,,\nn\\
	\ddot \gamma_{T,ij}^{(L)}&\equiv -\int_0^1 \df x\, \ln^2 x \, x^N P_{ij}^{(L)}(x)\,.
\end{align}
Here $\{i,j\}=\{q,g\}$ and the anomalous dimension is a $2\times 2$ matrix. 
The $N=2$ anomalous dimensions are calculated in Ref.~\cite{Dixon:2019uzg}
and $N=3$ are summarized in Ref.~\cite{Chen:2023zlx}. For completeness, we include $N=3$ result and add the NNLO $N=4,5,6$ expressions below.

{\centering\subsection*{$N=3$ anomalous dimensions}}
At LO, we find
\begin{align}
	\gamma_{T,qq}^{(0)} &= \frac{157}{30} \, C_F\,, \qquad
	\gamma_{T,gq}^{(0)}= -\frac{11}{15} \, C_F \,, \qquad
	\gamma_{T,qg}^{(0)} = -\frac{11}{30} \, n_f \,, \qquad
	\gamma_{T,gg}^{(0)} = \frac{21}{5}\, C_A + \frac{2}{3} \, n_f \,,\nn\\
	\dot{\gamma}_{T,qq}^{(0)}&= \left( 4\zeta_2 - \frac{10169}{1800} \right) C_F \,, \quad
	\dot{\gamma}_{T,gq}^{(0)}= \frac{247}{900} \, C_F \,, \quad
	\dot{\gamma}_{T,qg}^{(0)} = \frac{137}{1800} \, n_f \,, \quad
	\dot{\gamma}_{T,gg}^{(0)} = \left( 4\zeta_2 - \frac{2453}{450} \right) C_A \,,   \nn\\
	\ddot{\gamma}_{T,qq}^{(0)}&= \left( - 8\zeta_3 + \frac{507103}{54000} \right) C_F \,, \quad
	\ddot{\gamma}_{T,gq}^{(0)}= - \frac{5489}{27000} \, C_F \,, \quad
	\ddot{\gamma}_{T,qg}^{(0)}= - \frac{1919}{54000} \, n_f  \,, \nn\\
	\ddot{\gamma}_{T,gg}^{(0)}&= \left( - 8\zeta_3 + \frac{124511}{13500} \right) C_A \,, 
\end{align}
and at NLO, we have
\begin{align}
	\gamma_{T,qq}^{(1)}&=
	\left( -\frac{628}{15}\zeta_2+\frac{2905763}{54000} \right) C_F^2
	+ \frac{16157}{675} C_A C_F
	- \frac{13427}{3000} \, C_F n_f \,, \nn \\
	\gamma_{T,gq}^{(1)} &= \left(\frac{88}{15}\zeta_2-\frac{104389}{27000}  \right) C_F^2
	-\frac{142591}{13500} C_A C_F  \,, \nn \\
	\gamma_{T,qg}^{(1)} &= \left(\frac{44}{15}\zeta_2 -\frac{60391}{27000} \right) C_A n_f
	- \frac{166729}{54000} \, C_F n_f - \frac{6}{25} \, n_f^2 \,, \nn \\
	\gamma_{T,gg}^{(1)} &=
	\left( -\frac{168}{5}\zeta_2+\frac{90047}{1500}\right) C_A^2
	+ \left(- \frac{16}{3}\zeta_2 +\frac{2273}{1350} \right) C_A n_f 
	+ \frac{57287}{27000} \, C_F n_f \,, \nn\\
	\dot{\gamma}_{T,qq}^{(1)}&= \left(-120\zeta_4+ \frac{422}{3}\zeta_3+\frac{10169}{150}\zeta_2-\frac{162656941}{1080000}\right) C_F^2\nn\\
	&+ \left(20\zeta_4-\frac{1181}{15}\zeta_3+\frac{268}{9}\zeta_2+\frac{992579}{36000} \right) C_F C_A  + \left(\frac{16}{3}\zeta_3-\frac{40}{9}\zeta_2-\frac{433757}{1620000} \right)  C_F n_f \,, \nn \\
	\dot{\gamma}_{T,gq}^{(1)}&= \left(-\frac{286}{15}\zeta_3+\frac{1034}{225}\zeta_2+\frac{15207541}{810000} \right) C_F C_A
	+ \left(\frac{44}{5}\zeta_3-\frac{71}{9}\zeta_2+\frac{235643}{540000} \right) C_F^2 \,, \nn \\
	\dot{\gamma}_{T,qg}^{(1)} &= \left(\frac{11}{5}\zeta_3-\frac{25}{18}\zeta_2-\frac{1490669}{1620000} \right) C_A n_f
	+ \left(-\frac{22}{3}\zeta_3+\frac{217}{225}\zeta_2+\frac{8521133}{1080000} \right) C_F n_f \nn\\ 
	&+ \left(-\frac{22}{45}\zeta_2+\frac{10121}{13500} \right) n_f^2 \,, \nn \\
	\dot{\gamma}_{T,gg}^{(1)} &= \left(-100\zeta_4+\frac{772}{15}\zeta_3+\frac{21418}{225}\zeta_2-\frac{42705619}{405000} \right) C_A^2
	+ \left(\frac{32}{3}\zeta_3-\frac{40}{9}\zeta_2-\frac{21958}{3375} \right) C_A n_f\nn\\
	&- \frac{59659}{540000} \, C_F n_f \,,\nn\\
	\ddot{\gamma}_{T,qq}^{(1)}&= \left(280\zeta_5+128\zeta_2\zeta_3-\frac{2357}{5}\zeta_4-\frac{20338}{75}\zeta_3-\frac{507103}{3375}\zeta_2+\frac{8749145327}{16200000}\right) C_F^2 \nn\\
	&+ \left(-140\zeta_5+\frac{2609}{10}\zeta_4+\frac{3707}{450}\zeta_3-\frac{1438459003}{9720000}\right) C_F C_A
	+ \left(-16\zeta_4+\frac{80}{9}\zeta_3+\frac{56957663}{8100000}\right) C_F n_f \,, \nn\\
	\ddot{\gamma}_{T,gq}^{(1)}&= \left(\frac{1001}{15}\zeta_4+\frac{127}{25}\zeta_3-\frac{42568}{3375}\zeta_2-\frac{180241183}{3037500}\right) C_F C_A \nn\\
	&+ \left(-\frac{506}{15}\zeta_4+\frac{24}{5}\zeta_3+\frac{53546}{3375}\zeta_2+\frac{46181009}{8100000}\right) C_F^2 \,, \nn\\
	\ddot{\gamma}_{T,qg}^{(1)}&= \left(-\frac{55}{6}\zeta_4+\frac{97}{150}\zeta_3+\frac{1862}{375}\zeta_2+\frac{28905223}{24300000}\right) C_A n_f \nn\\
	&+ \left(\frac{77}{3}\zeta_4+\frac{251}{225}\zeta_3-\frac{15524}{3375}\zeta_2-\frac{264257617}{12150000}\right) C_F n_f
	+ \left(\frac{44}{45}\zeta_3+\frac{137}{675}\zeta_2-\frac{892817}{607500}\right) n_f^2 \,, \nn\\
	\ddot{\gamma}_{T,gg}^{(1)}&= \left(140\zeta_5+128\zeta_2\zeta_3-178\zeta_4-\frac{57554}{225}\zeta_3-\frac{498044}{3375}\zeta_2+\frac{8340937507}{24300000}\right) C_A^2 \nn\\
	&+ \left(-32\zeta_4+\frac{80}{9}\zeta_3+\frac{19602857}{810000}\right) C_A n_f
	+ \frac{860411}{6075000} \, C_F n_f \,,
\end{align}
as well as NNLO:
\begin{align}
	\gamma_{T,qq}^{(2)}&=
	\left( \frac{1439}{75} \zeta_3+ \frac{136066373}{972000} \right) C_F C_A^2  \nn \\
	&+ \left( -\frac{628}{3}\zeta_4+ \frac{172466}{225} \zeta_3- \frac{113212}{225}\zeta_2-\frac{443247883}{9720000} \right) C_F^2 C_A \nn \\
	&+\left( 1256 \zeta_4-\frac{14936}{15}\zeta_3-\frac{2251148}{3375}\zeta_2+ \frac{47976425617}{48600000} \right) C_F^3 \nn \\
	&+\left(-\frac{2126}{45}\zeta_3 + \frac{8492}{3375}\zeta_2-\frac{57923471}{4050000} \right) C_A C_F n_f\nn\\
	&+\left(-\frac{2656}{225}\zeta_3+\frac{88163}{1125}\zeta_2-\frac{638186993}{8100000} \right) C_F^2 n_f
	-\frac{19711}{18000} \, C_F n_f^2 \,, \nn \\
	\gamma_{T,gq}^{(2)} &=
	\left(\frac{6448}{75}\zeta_3-\frac{10898}{375}\zeta_2-\frac{2010250477}{12150000} \right) C_F C_A^2 \nn\\ &+\left(\frac{88}{3}\zeta_4-\frac{31346}{225}\zeta_3+\frac{234407}{1125}\zeta_2-\frac{1694499413}{24300000} \right) C_F^2 C_A \nn \\
	&+ \left(-176 \zeta_4 +\frac{1796}{15}\zeta_3+\frac{79268}{3375}\zeta_2-\frac{1061823161}{24300000} \right) C_F^3  \nn \\
	&+ \left(\frac{704}{45}\zeta_3-\frac{3736}{675}\zeta_2+\frac{2334509}{405000} \right)C_A C_F n_f+ \left(-\frac{88}{45}\zeta_3+\frac{152}{225}\zeta_2-\frac{14837573}{4050000} \right) C_F^2 n_f\,, \nn \\
    %%%%%qg
	\gamma_{T,qg}^{(2)} &=
	\left(-\frac{220}{3}\zeta_4+\frac{1004}{225}\zeta_3+\frac{188462}{3375}\zeta_2-\frac{140682763}{6075000} \right) C_A^2 n_f \nn\\
	&+\left(\frac{6503}{225}\zeta_3+\frac{7003}{750}\zeta_2-\frac{509985949}{24300000} \right) C_A C_F n_f \nn \\
	&+\left(\frac{622}{225}\zeta_3+\frac{79361}{6750}\zeta_2-\frac{2412861131}{48600000}\right) C_F^2 n_f
	+\left(\frac{176}{45}\zeta_3-\frac{116}{135}\zeta_2-\frac{51449}{9000}\right) C_A n_f^2 \nn \\
	&+ \left(\frac{3454}{675}\zeta_2-\frac{915539}{300000}\right) C_F n_f^2
	- \frac{86}{375} \, n_f^3\,, \nn \\
	\gamma_{T,gg}^{(2)} &=
	\left(840 \zeta_4-\frac{3752}{25}\zeta_3-\frac{342578}{375}\zeta_2+\frac{1069405919}{1350000} \right)C_A^3 \nn \\
	&+\left(\frac{400}{3} \zeta_4-\frac{29534}{225}\zeta_3-\frac{30316}{675}\zeta_2+\frac{129284923}{2430000} \right) C_A^2 n_f \nn \\
	&+\left(\frac{2744}{45}\zeta_3-\frac{2158}{125}\zeta_2-\frac{188283293}{6075000} \right) C_A C_F n_f
	+\left(-\frac{352}{225}\zeta_3+\frac{4037}{3375}\zeta_2+\frac{27742123}{24300000} \right)C_F^2 n_f \nn \\
	&+\left(-\frac{64}{9}\zeta_3+\frac{160}{27}\zeta_2-\frac{71341}{27000} \right) C_A n_f^2+\left(-\frac{484}{675}\zeta_2-\frac{165553}{270000} \right) C_F n_f^2\,.
\end{align}

{\centering\subsection*{$N=4$ anomalous dimensions}}
At LO, we find
\begin{align}
	\gamma_{T,qq}^{(0)} &= \frac{91}{15} \, C_F\,, \qquad
	\gamma_{T,gq}^{(0)}= -\frac{8}{15} \, C_F \,, \qquad
	\gamma_{T,qg}^{(0)} = -\frac{32}{105} \, n_f \,, \qquad
	\gamma_{T,gg}^{(0)} = \frac{181}{35}\, C_A + \frac{2}{3} \, n_f \,,\nn\\
	\dot{\gamma}_{T,qq}^{(0)}&= \left(4\zeta_2-\frac{583}{100}\right) C_F \,, \quad
	\dot{\gamma}_{T,gq}^{(0)}= \frac{131}{900} \, C_F \,, \quad
	\dot{\gamma}_{T,qg}^{(0)} = \frac{557}{11025} \, n_f \,, \quad
	\dot{\gamma}_{T,gg}^{(0)} = \left(4\zeta_2-\frac{21076}{3675}\right) C_A \,,   \nn\\
	\ddot{\gamma}_{T,qq}^{(0)}&= \left(-8\zeta_3+\frac{255739}{27000}\right) C_F \,, \quad
	\ddot{\gamma}_{T,gq}^{(0)}= -\frac{2147}{27000} \, C_F \,, \quad
	\ddot{\gamma}_{T,qg}^{(0)}= -\frac{21169}{1157625} \, n_f  \,, \nn\\
	\ddot{\gamma}_{T,gg}^{(0)}&= \left(-8\zeta_3+\frac{10893926}{1157625}\right) C_A \,,
\end{align}
and at NLO, we have
\begin{align}
	\gamma_{T,qq}^{(1)}&=
	\left(-16\zeta_3+24\zeta_2-\frac{474221}{13500}\right) C_F^2
	+ \left(8\zeta_3-\frac{544}{15}\zeta_2+\frac{520837}{6750}\right) C_A C_F
	- \frac{604601}{110250} \, C_F n_f \,, \nn \\
	\gamma_{T,gq}^{(1)} &= \left(-\frac{32}{15}\zeta_2-\frac{2882863}{661500}\right) C_A C_F
	+ \left(\frac{64}{15}\zeta_2-\frac{9374}{3375}\right) C_F^2 \,, \nn \\
	\gamma_{T,qg}^{(1)} &= \left(\frac{128}{105}\zeta_2+\frac{1999}{18375}\right) C_A n_f
	- \frac{19792}{7875} \, C_F n_f - \frac{1024}{3675} \, n_f^2 \,, \nn \\
	\gamma_{T,gg}^{(1)} &=
	\left(-8\zeta_3-\frac{632}{105}\zeta_2+\frac{4706626}{165375}\right) C_A^2
	+ \left(-\frac{16}{3}\zeta_2+\frac{26399}{33075}\right) C_A n_f
	+ \frac{340066}{165375} \, C_F n_f \,, \nn\\
	\dot{\gamma}_{T,qq}^{(1)}&= \left(-56\zeta_4-\frac{904}{15}\zeta_3+\frac{583}{25}\zeta_2+\frac{25624327}{270000}\right) C_F^2\nn\\
	&+ \left(-12\zeta_4+\frac{376}{15}\zeta_3+\frac{11947}{225}\zeta_2-\frac{41156197}{405000}\right) C_F C_A
	+ \left(\frac{16}{3}\zeta_3-\frac{40}{9}\zeta_2+\frac{1383283}{69457500}\right) C_F n_f \,, \nn \\
	\dot{\gamma}_{T,gq}^{(1)}&= \left(-\frac{112}{15}\zeta_3+\frac{928}{225}\zeta_2+\frac{597670243}{138915000}\right) C_F C_A
	+ \left(\frac{32}{5}\zeta_3-\frac{238}{45}\zeta_2-\frac{165889}{810000}\right) C_F^2 \,, \nn \\
	\dot{\gamma}_{T,qg}^{(1)} &= \left(\frac{192}{35}\zeta_3-\frac{11432}{11025}\zeta_2-\frac{367614857}{69457500}\right) C_A n_f
	+ \left(-\frac{128}{21}\zeta_3+\frac{11456}{11025}\zeta_2+\frac{10112437}{1653750}\right) C_F n_f \nn\\
	&+ \left(-\frac{128}{315}\zeta_2+\frac{743629}{1157625}\right) n_f^2 \,, \nn \\
	\dot{\gamma}_{T,gg}^{(1)} &= \left(-68\zeta_4-\frac{4304}{105}\zeta_3+\frac{834124}{11025}\zeta_2+\frac{72821129}{34728750}\right) C_A^2
	+ \left(\frac{32}{3}\zeta_3-\frac{40}{9}\zeta_2-\frac{87547147}{13891500}\right) C_A n_f\nn\\
	&- \frac{215261}{5788125} \, C_F n_f \,,\nn\\
	\ddot{\gamma}_{T,qq}^{(1)}&= \left(104\zeta_5+64\zeta_2\zeta_3+\frac{1176}{5}\zeta_4-\frac{255739}{3375}\zeta_2-\frac{4425861547}{12150000}\right) C_F^2 \nn\\
	&+ \left(-52\zeta_5+32\zeta_2\zeta_3-\frac{512}{5}\zeta_4-\frac{29141}{225}\zeta_3-\frac{255739}{6750}\zeta_2+\frac{3875072533}{12150000}\right) C_F C_A \nn\\
	&+ \left(-16\zeta_4+\frac{80}{9}\zeta_3+\frac{199562516347}{29172150000}\right) C_F n_f \,, \nn\\
	\ddot{\gamma}_{T,gq}^{(1)}&= \left(\frac{392}{15}\zeta_4-\frac{677}{225}\zeta_3-\frac{5968}{675}\zeta_2-\frac{215094415271}{19448100000}\right) C_F C_A \nn\\
	&+ \left(-\frac{368}{15}\zeta_4+\frac{214}{45}\zeta_3+\frac{66121}{6750}\zeta_2+\frac{127516051}{24300000}\right) C_F^2 \,, \nn\\
	\ddot{\gamma}_{T,qg}^{(1)}&= \left(-\frac{2144}{105}\zeta_4-\frac{548}{3675}\zeta_3+\frac{662546}{165375}\zeta_2+\frac{76846745467}{4862025000}\right) C_A n_f \nn\\
	&+ \left(\frac{64}{3}\zeta_4-\frac{632}{11025}\zeta_3-\frac{4539754}{1157625}\zeta_2-\frac{4351226597}{260465625}\right) C_F n_f \nn\\
	&+ \left(\frac{256}{315}\zeta_3+\frac{4456}{33075}\zeta_2-\frac{859614167}{729303750}\right) n_f^2 \,, \nn\\
	\ddot{\gamma}_{T,gg}^{(1)}&= \left(52\zeta_5+96\zeta_2\zeta_3+\frac{5232}{35}\zeta_4-\frac{1415336}{11025}\zeta_3-\frac{43575704}{385875}\zeta_2-\frac{40304071316}{607753125}\right) C_A^2 \nn\\
	&+ \left(-32\zeta_4+\frac{80}{9}\zeta_3+\frac{70339456927}{2917215000}\right) C_A n_f
	+ \frac{21035824}{607753125} \, C_F n_f \,,
\end{align}
as well as NNLO:
\begin{align}
	\gamma_{T,qq}^{(2)}&=
	\left(112\zeta_5+48\zeta_2\zeta_3-\frac{12752}{15}\zeta_4+\frac{281558}{225}\zeta_3-\frac{730997}{3375}\zeta_2-\frac{4180165661}{16200000}\right) C_F C_A^2  \nn \\
	&+ \left(-432\zeta_5-208\zeta_2\zeta_3+\frac{51064}{15}\zeta_4-\frac{235402}{75}\zeta_3-\frac{1101667}{1125}\zeta_2+\frac{3623379121}{1518750}\right) C_F^2 C_A \nn \\
	&+ \left(416\zeta_5+224\zeta_2\zeta_3-\frac{7312}{3}\zeta_4+\frac{29756}{15}\zeta_3+\frac{973994}{1125}\zeta_2-\frac{161877491}{81000}\right) C_F^3 \nn \\
	&+ \left(\frac{68}{3}\zeta_4-\frac{257336}{1575}\zeta_3+\frac{2198426}{33075}\zeta_2-\frac{59504161907}{4167450000}\right) C_A C_F n_f\nn\\
	&+ \left(-\frac{136}{3}\zeta_4+\frac{337208}{1575}\zeta_3-\frac{5647384}{165375}\zeta_2-\frac{3829448611}{33075000}\right) C_F^2 n_f
	- \frac{19521281}{13891500} \, C_F n_f^2 \,, \nn \\
	\gamma_{T,gq}^{(2)} &=
	\left(\frac{448}{15}\zeta_4-\frac{3632}{1575}\zeta_3-\frac{8690054}{165375}\zeta_2-\frac{778603499}{22680000}\right) C_F C_A^2 \nn\\
	&+ \left(\frac{1168}{15}\zeta_4-\frac{4160}{21}\zeta_3+\frac{13794439}{165375}\zeta_2+\frac{4485362407}{34020000}\right) C_F^2 C_A \nn \\
	&+ \left(-\frac{704}{5}\zeta_4+\frac{3744}{25}\zeta_3+\frac{91562}{3375}\zeta_2-\frac{1414558213}{12150000}\right) C_F^3  \nn \\
	&+ \left(\frac{416}{45}\zeta_3-\frac{12}{5}\zeta_2+\frac{359997809}{92610000}\right) C_A C_F n_f
	+ \left(-\frac{64}{45}\zeta_3+\frac{196}{225}\zeta_2-\frac{79552063}{25725000}\right) C_F^2 n_f \,, \nn \\
	%%%%%qg
	\gamma_{T,qg}^{(2)} &=
	\left(-\frac{1152}{35}\zeta_4-\frac{7552}{735}\zeta_3+\frac{18545284}{1157625}\zeta_2+\frac{1901298533}{119070000}\right) C_A^2 n_f \nn\\
	&+ \left(-\frac{192}{35}\zeta_4+\frac{104064}{1225}\zeta_3+\frac{1961024}{165375}\zeta_2-\frac{89580154991}{1041862500}\right) C_A C_F n_f \nn \\
	&+ \left(\frac{2048}{105}\zeta_4-\frac{1073536}{11025}\zeta_3-\frac{154544}{33075}\zeta_2+\frac{2857982668}{37209375}\right) C_F^2 n_f \nn\\
	&+ \left(-\frac{128}{63}\zeta_3-\frac{7304}{2205}\zeta_2+\frac{32725682}{5788125}\right) C_A n_f^2
	+ \left(\frac{163456}{33075}\zeta_2-\frac{304184494}{86821875}\right) C_F n_f^2
	- \frac{34936}{128625} \, n_f^3\,, \nn \\
	\gamma_{T,gg}^{(2)} &=
	\left(96\zeta_5+64\zeta_2\zeta_3-\frac{2524}{105}\zeta_4+\frac{698828}{11025}\zeta_3-\frac{185369599}{1157625}\zeta_2+\frac{11996763263}{463050000}\right) C_A^3 \nn \\
	&+ \left(104\zeta_4+\frac{2336}{1575}\zeta_3-\frac{4654366}{55125}\zeta_2-\frac{88337469301}{4167450000}\right) C_A^2 n_f \nn \\
	&+ \left(\frac{12672}{175}\zeta_3-\frac{3258284}{165375}\zeta_2-\frac{13864219709}{347287500}\right) C_A C_F n_f \nn \\
	&+ \left(-\frac{256}{225}\zeta_3+\frac{112936}{55125}\zeta_2-\frac{110295583}{74418750}\right) C_F^2 n_f
	+ \left(-\frac{64}{9}\zeta_3+\frac{160}{27}\zeta_2-\frac{128595883}{41674500}\right) C_A n_f^2 \nn \\
	&+ \left(-\frac{2048}{4725}\zeta_2-\frac{43545391}{52093125}\right) C_F n_f^2 \,.
\end{align}

{\centering\subsection*{$N=5$ anomalous dimensions}}
At LO, we find
\begin{align}
	\gamma_{T,qq}^{(0)} &= \frac{709}{105} \, C_F\,, \qquad
	\gamma_{T,gq}^{(0)}= -\frac{44}{105} \, C_F \,, \qquad
	\gamma_{T,qg}^{(0)} = -\frac{11}{42} \, n_f \,, \qquad
	\gamma_{T,gg}^{(0)} = \frac{83}{14}\, C_A + \frac{2}{3} \, n_f \,,\nn\\
	\dot{\gamma}_{T,qq}^{(0)}&= \left(4\zeta_2-\frac{29159}{4900}\right) C_F \,, \qquad
	\dot{\gamma}_{T,gq}^{(0)}= \frac{989}{11025} \, C_F \,, \nn\\
	\dot{\gamma}_{T,qg}^{(0)} &= \frac{257}{7056} \, n_f \,, \qquad
	\dot{\gamma}_{T,gg}^{(0)} = \left(4\zeta_2-\frac{13871}{2352}\right) C_A \,, \nn\\
	\ddot{\gamma}_{T,qq}^{(0)}&= \left(-8\zeta_3+\frac{88122829}{9261000}\right) C_F \,, \qquad
	\ddot{\gamma}_{T,gq}^{(0)}= -\frac{44713}{1157625} \, C_F \,, \nn\\
	\ddot{\gamma}_{T,qg}^{(0)}&= -\frac{6413}{592704} \, n_f  \,, \qquad
	\ddot{\gamma}_{T,gg}^{(0)}= \left(-8\zeta_3+\frac{5623381}{592704}\right) C_A \,,
\end{align}
and at NLO, we have
\begin{align}
	\gamma_{T,qq}^{(1)}&=
	\left(-\frac{5672}{105}\zeta_2+\frac{12724717}{171500}\right) C_F^2
	+ \frac{157415}{5292} \, C_A C_F
	- \frac{484268}{77175} \, C_F n_f \,, \nn \\
	\gamma_{T,gq}^{(1)} &= -\frac{948127}{154350} \, C_A C_F
	+ \left(\frac{352}{105}\zeta_2-\frac{2626061}{1157625}\right) C_F^2 \,, \nn \\
	\gamma_{T,qg}^{(1)} &= \left(\frac{44}{21}\zeta_2-\frac{1249361}{740880}\right) C_A n_f
	- \frac{427303}{205800} \, C_F n_f - \frac{218}{735} \, n_f^2 \,, \nn \\
	\gamma_{T,gg}^{(1)} &=
	\left(-\frac{332}{7}\zeta_2+\frac{13375435}{148176}\right) C_A^2
	+ \left(-\frac{16}{3}\zeta_2+\frac{2071}{26460}\right) C_A n_f
	+ \frac{940633}{463050} \, C_F n_f \,, \nn\\
	\dot{\gamma}_{T,qq}^{(1)}&= \left(-120\zeta_4+\frac{3592}{21}\zeta_3+\frac{87477}{1225}\zeta_2-\frac{375290653573}{1944810000}\right) C_F^2 \nn\\
	&+ \left(20\zeta_4-\frac{9224}{105}\zeta_3+\frac{268}{9}\zeta_2+\frac{138472247}{3704400}\right) C_F C_A
	+ \left(\frac{16}{3}\zeta_3-\frac{40}{9}\zeta_2+\frac{35972869}{194481000}\right) C_F n_f \,, \nn \\
	\dot{\gamma}_{T,gq}^{(1)}&= \left(-\frac{1144}{105}\zeta_3+\frac{32552}{11025}\zeta_2+\frac{1875831403}{194481000}\right) C_F C_A
	+ \left(\frac{176}{35}\zeta_3-\frac{8884}{2205}\zeta_2-\frac{113414743}{486202500}\right) C_F^2 \,, \nn \\
	\dot{\gamma}_{T,qg}^{(1)} &= \left(\frac{11}{7}\zeta_3-\frac{10163}{8820}\zeta_2-\frac{7873753}{25930800}\right) C_A n_f
	+ \left(-\frac{110}{21}\zeta_3+\frac{2347}{2205}\zeta_2+\frac{53256353}{10804500}\right) C_F n_f \nn\\
	&+ \left(-\frac{22}{63}\zeta_2+\frac{2085599}{3704400}\right) n_f^2 \,, \nn \\
	\dot{\gamma}_{T,gg}^{(1)} &= \left(-100\zeta_4+\frac{227}{3}\zeta_3+\frac{177367}{1764}\zeta_2-\frac{112901605}{777924}\right) C_A^2
	+ \left(\frac{32}{3}\zeta_3-\frac{40}{9}\zeta_2-\frac{17138021}{2778300}\right) C_A n_f\nn\\
	&- \frac{44197}{2701125} \, C_F n_f \,,\nn\\
	\ddot{\gamma}_{T,qq}^{(1)}&= \left(280\zeta_5+128\zeta_2\zeta_3-\frac{20008}{35}\zeta_4-\frac{349908}{1225}\zeta_3-\frac{176245658}{1157625}\zeta_2+\frac{136627878615667}{204205050000}\right) C_F^2 \nn\\
	&+ \left(-140\zeta_5+\frac{1464}{5}\zeta_4+\frac{130693}{11025}\zeta_3-\frac{72487456823}{388962000}\right) C_F C_A \nn\\
	&+ \left(-16\zeta_4+\frac{80}{9}\zeta_3+\frac{1104878422199}{163364040000}\right) C_F n_f \,, \nn\\
	\ddot{\gamma}_{T,gq}^{(1)}&= \left(\frac{572}{15}\zeta_4-\frac{13676}{11025}\zeta_3-\frac{7458088}{1157625}\zeta_2-\frac{4853397718847}{163364040000}\right) C_F C_A \nn\\
	&+ \left(-\frac{2024}{105}\zeta_4+\frac{1408}{315}\zeta_3+\frac{8173496}{1157625}\zeta_2+\frac{425881880509}{102102525000}\right) C_F^2 \,, \nn\\
	\ddot{\gamma}_{T,qg}^{(1)}&= \left(-\frac{275}{42}\zeta_4+\frac{1259}{980}\zeta_3+\frac{720833}{205800}\zeta_2-\frac{37355534579}{261382464000}\right) C_A n_f \nn\\
	&+ \left(\frac{55}{3}\zeta_4-\frac{2963}{4410}\zeta_3-\frac{6346747}{1852200}\zeta_2-\frac{8817222855829}{653456160000}\right) C_F n_f \nn\\
	&+ \left(\frac{44}{63}\zeta_3+\frac{257}{2646}\zeta_2-\frac{2307600973}{2333772000}\right) n_f^2 \,, \nn\\
	\ddot{\gamma}_{T,gg}^{(1)}&= \left(140\zeta_5+128\zeta_2\zeta_3-\frac{3581}{14}\zeta_4-\frac{479573}{1764}\zeta_3-\frac{5623381}{37044}\zeta_2+\frac{950440411537}{2091059712}\right) C_A^2 \nn\\
	&+ \left(-32\zeta_4+\frac{80}{9}\zeta_3+\frac{224594970149}{9335088000}\right) C_A n_f
	+ \frac{648731477}{54454680000} \, C_F n_f \,,
\end{align}
as well as NNLO:
\begin{align}
	\gamma_{T,qq}^{(2)}&=
	\left(\frac{69862}{3675}\zeta_3+\frac{115237918583}{666792000}\right) C_F C_A^2  \nn \\
	&+ \left(-\frac{5672}{21}\zeta_4+\frac{12450874}{11025}\zeta_3-\frac{4234318}{6615}\zeta_2-\frac{210810601457}{1166886000}\right) C_F^2 C_A \nn \\
	&+ \left(\frac{11344}{7}\zeta_4-\frac{269924}{175}\zeta_3-\frac{117695764}{128625}\zeta_2+\frac{14023453930387}{8508543750}\right) C_F^3 \nn \\
	&+ \left(-\frac{123952}{2205}\zeta_3+\frac{102388}{77175}\zeta_2-\frac{78937316743}{4084101000}\right) C_A C_F n_f\nn\\
	&+ \left(-\frac{37768}{2205}\zeta_3+\frac{557588}{5145}\zeta_2-\frac{2475762140309}{20420505000}\right) C_F^2 n_f
	- \frac{7960262}{4862025} \, C_F n_f^2 \,, \nn \\
	\gamma_{T,gq}^{(2)} &=
	\left(\frac{84856}{1225}\zeta_3-\frac{283272}{8575}\zeta_2-\frac{48038223017}{453789000}\right) C_F C_A^2 \nn\\
	&+ \left(\frac{352}{21}\zeta_4-\frac{1021184}{11025}\zeta_3+\frac{35199932}{231525}\zeta_2-\frac{586547975651}{10210252500}\right) C_F^2 C_A \nn \\
	&+ \left(-\frac{704}{7}\zeta_4+\frac{13584}{175}\zeta_3-\frac{1660784}{1157625}\zeta_2-\frac{1086821243677}{51051262500}\right) C_F^3  \nn \\
	&+ \left(\frac{2816}{315}\zeta_3-\frac{47696}{11025}\zeta_2+\frac{1388619349}{291721500}\right) C_A C_F n_f \nn\\
	&+ \left(-\frac{352}{315}\zeta_3+\frac{10736}{11025}\zeta_2-\frac{768418561}{283618125}\right) C_F^2 n_f \,, \nn \\
	%%%%%qg
	\gamma_{T,qg}^{(2)} &=
	\left(-\frac{1100}{21}\zeta_4+\frac{2843}{2205}\zeta_3+\frac{8024111}{185220}\zeta_2-\frac{48371790659}{2178187200}\right) C_A^2 n_f \nn\\
	&+ \left(\frac{73564}{2205}\zeta_3+\frac{97154}{46305}\zeta_2-\frac{225197009129}{16336404000}\right) C_A C_F n_f \nn \\
	&+ \left(\frac{3002}{2205}\zeta_3+\frac{131792}{15435}\zeta_2-\frac{3406359780601}{81682020000}\right) C_F^2 n_f \nn\\
	&+ \left(\frac{176}{63}\zeta_3+\frac{844}{1323}\zeta_2-\frac{3014751143}{466754400}\right) C_A n_f^2
	+ \left(\frac{31196}{6615}\zeta_2-\frac{5077590437}{1361367000}\right) C_F n_f^2
	- \frac{22906}{77175} \, n_f^3\,, \nn \\
	\gamma_{T,gg}^{(2)} &=
	\left(\frac{8300}{7}\zeta_4-\frac{49219}{147}\zeta_3-\frac{50191979}{37044}\zeta_2+\frac{611826686767}{435637440}\right) C_A^3 \nn \\
	&+ \left(\frac{400}{3}\zeta_4-\frac{391588}{2205}\zeta_3-\frac{126247}{6615}\zeta_2+\frac{7117482587}{155584800}\right) C_A^2 n_f \nn \\
	&+ \left(\frac{167896}{2205}\zeta_3-\frac{258578}{15435}\zeta_2-\frac{24824403614}{510512625}\right) C_A C_F n_f \nn\\
	&+ \left(-\frac{1672}{2205}\zeta_3+\frac{15488}{15435}\zeta_2-\frac{6007293823}{10210252500}\right) C_F^2 n_f
	+ \left(-\frac{64}{9}\zeta_3+\frac{160}{27}\zeta_2-\frac{56262253}{16669800}\right) C_A n_f^2 \nn\\
	&+ \left(-\frac{1936}{6615}\zeta_2-\frac{39577991}{41674500}\right) C_F n_f^2 \,.
\end{align}

{\centering\subsection*{$N=6$ anomalous dimensions}}
At LO, we find
\begin{align}
	\gamma_{T,qq}^{(0)} &= \frac{1027}{140} \, C_F\,, \qquad
	\gamma_{T,gq}^{(0)}= -\frac{29}{84} \, C_F \,, \qquad
	\gamma_{T,qg}^{(0)} = -\frac{29}{126} \, n_f \,, \qquad
	\gamma_{T,gg}^{(0)} = \frac{4129}{630}\, C_A + \frac{2}{3} \, n_f \,,\nn\\
	\dot{\gamma}_{T,qq}^{(0)}&= \left(4\zeta_2-\frac{2130073}{352800}\right) C_F \,, \qquad
	\dot{\gamma}_{T,gq}^{(0)}= \frac{857}{14112} \, C_F \,, \nn\\
	\dot{\gamma}_{T,qg}^{(0)} &= \frac{1759}{63504} \, n_f \,, \qquad
	\dot{\gamma}_{T,gg}^{(0)} = \left(4\zeta_2-\frac{9532891}{1587600}\right) C_A \,, \nn\\
	\ddot{\gamma}_{T,qq}^{(0)}&= \left(-8\zeta_3+\frac{157096321}{16464000}\right) C_F \,, \qquad
	\ddot{\gamma}_{T,gq}^{(0)}= -\frac{25517}{1185408} \, C_F \,, \nn\\
	\ddot{\gamma}_{T,qg}^{(0)}&= -\frac{112193}{16003008} \, n_f  \,, \qquad
	\ddot{\gamma}_{T,gg}^{(0)}= \left(-8\zeta_3+\frac{19058167189}{2000376000}\right) C_A \,,
\end{align}
and at NLO, we have
\begin{align}
	\gamma_{T,qq}^{(1)}&=
	\left(-16\zeta_3+24\zeta_2-\frac{571080347}{16464000}\right) C_F^2
	+ \left(8\zeta_3-\frac{1447}{35}\zeta_2+\frac{3347931437}{37044000}\right) C_A C_F \nn\\
	&- \frac{924429029}{133358400} \, C_F n_f \,, \nn \\
	\gamma_{T,gq}^{(1)} &= \left(-\frac{29}{21}\zeta_2-\frac{358501999}{133358400}\right) C_A C_F
	+ \left(\frac{58}{21}\zeta_2-\frac{19635271}{9878400}\right) C_F^2 \,, \nn \\
	\gamma_{T,qg}^{(1)} &= \left(\frac{58}{63}\zeta_2-\frac{674773}{22226400}\right) C_A n_f
	- \frac{77139049}{44452800} \, C_F n_f - \frac{18079}{59535} \, n_f^2 \,, \nn \\
	\gamma_{T,gg}^{(1)} &=
	\left(-8\zeta_3-\frac{3638}{315}\zeta_2+\frac{2907487777}{66679200}\right) C_A^2
	+ \left(-\frac{16}{3}\zeta_2-\frac{1262143}{2381400}\right) C_A n_f
	+ \frac{10772855}{5334336} \, C_F n_f \,, \nn\\
	\dot{\gamma}_{T,qq}^{(1)}&= \left(-56\zeta_4-\frac{2287}{35}\zeta_3+\frac{2130073}{88200}\zeta_2+\frac{4134151961719}{41489280000}\right) C_F^2 \nn\\
	&+ \left(-12\zeta_4+\frac{6863}{210}\zeta_3+\frac{528497}{9800}\zeta_2-\frac{878408535427}{7779240000}\right) C_F C_A \nn\\
	&+ \left(\frac{16}{3}\zeta_3-\frac{40}{9}\zeta_2+\frac{98858394157}{336063168000}\right) C_F n_f \,, \nn \\
	\dot{\gamma}_{T,gq}^{(1)}&= \left(-\frac{29}{6}\zeta_3+\frac{49577}{17640}\zeta_2+\frac{733074793007}{336063168000}\right) C_F C_A
	+ \left(\frac{29}{7}\zeta_3-\frac{58147}{17640}\zeta_2-\frac{3518955577}{24893568000}\right) C_F^2 \,, \nn \\
	\dot{\gamma}_{T,qg}^{(1)} &= \left(\frac{29}{7}\zeta_3-\frac{38863}{39690}\zeta_2-\frac{22499184329}{6223392000}\right) C_A n_f
	+ \left(-\frac{290}{63}\zeta_3+\frac{21124}{19845}\zeta_2+\frac{457639515377}{112021056000}\right) C_F n_f \nn\\
	&+ \left(-\frac{58}{189}\zeta_2+\frac{150415297}{300056400}\right) n_f^2 \,, \nn \\
	\dot{\gamma}_{T,gg}^{(1)} &= \left(-68\zeta_4-\frac{12041}{315}\zeta_3+\frac{15442291}{198450}\zeta_2-\frac{996806891321}{168031584000}\right) C_A^2 \nn\\
	&+ \left(\frac{32}{3}\zeta_3-\frac{40}{9}\zeta_2-\frac{18229482653}{3000564000}\right) C_A n_f
	- \frac{22751783}{2688505344} \, C_F n_f \,,\nn\\
	\ddot{\gamma}_{T,qq}^{(1)}&= \left(104\zeta_5+64\zeta_2\zeta_3+\frac{18063}{70}\zeta_4-\frac{157096321}{2058000}\zeta_2-\frac{10144876691716111}{26138246400000}\right) C_F^2 \nn\\
	&+ \left(-52\zeta_5+32\zeta_2\zeta_3-\frac{2581}{20}\zeta_4-\frac{11643019}{88200}\zeta_3-\frac{157096321}{4116000}\zeta_2+\frac{4593861844067443}{13069123200000}\right) C_F C_A \nn\\
	&+ \left(-16\zeta_4+\frac{80}{9}\zeta_3+\frac{474491371031743}{70573265280000}\right) C_F n_f \,, \nn\\
	\ddot{\gamma}_{T,gq}^{(1)}&= \left(\frac{203}{12}\zeta_4-\frac{60589}{17640}\zeta_3-\frac{39045151}{7408800}\zeta_2-\frac{411461493638437}{70573265280000}\right) C_F C_A \nn\\
	&+ \left(-\frac{667}{42}\zeta_4+\frac{2623}{630}\zeta_3+\frac{20479463}{3704400}\zeta_2+\frac{17102202998017}{5227649280000}\right) C_F^2 \,, \nn\\
	\ddot{\gamma}_{T,qg}^{(1)}&= \left(-\frac{1943}{126}\zeta_4+\frac{6523}{8820}\zeta_3+\frac{43650167}{14288400}\zeta_2+\frac{191073910750489}{17643316320000}\right) C_A n_f \nn\\
	&+ \left(\frac{145}{9}\zeta_4-\frac{13507}{13230}\zeta_3-\frac{152262247}{50009400}\zeta_2-\frac{265119996478997}{23524421760000}\right) C_F n_f \nn\\
	&+ \left(\frac{116}{189}\zeta_3+\frac{1759}{23814}\zeta_2-\frac{17925977441}{21003948000}\right) n_f^2 \,, \nn\\
	\ddot{\gamma}_{T,gg}^{(1)}&= \left(52\zeta_5+96\zeta_2\zeta_3+\frac{30521}{210}\zeta_4-\frac{17412091}{132300}\zeta_3-\frac{19058167189}{166698000}\zeta_2-\frac{1946750538175069}{35286632640000}\right) C_A^2 \nn\\
	&+ \left(-32\zeta_4+\frac{80}{9}\zeta_3+\frac{60563152132537}{2520473760000}\right) C_A n_f
	+ \frac{63636521}{12546358272} \, C_F n_f \,,
\end{align}
as well as NNLO:
\begin{align}
	\gamma_{T,qq}^{(2)}&=
	\left(112\zeta_5+48\zeta_2\zeta_3-\frac{200381}{210}\zeta_4+\frac{134198969}{88200}\zeta_3-\frac{17764211749}{74088000}\zeta_2-\frac{21141062718937129}{52276492800000}\right) C_F C_A^2  \nn \\
	&+ \left(-432\zeta_5-208\zeta_2\zeta_3+\frac{403286}{105}\zeta_4-\frac{12323051}{3150}\zeta_3-\frac{125635543}{102900}\zeta_2+\frac{565988843066047}{174254976000}\right) C_F^2 C_A \nn \\
	&+ \left(416\zeta_5+224\zeta_2\zeta_3-\frac{18838}{7}\zeta_4+\frac{11460541}{4410}\zeta_3+\frac{6200882381}{6174000}\zeta_2-\frac{23498573204731811}{8712748800000}\right) C_F^3 \nn \\
	&+ \left(\frac{68}{3}\zeta_4-\frac{4944871}{26460}\zeta_3+\frac{5101643923}{66679200}\zeta_2-\frac{2412733463680139}{141146530560000}\right) C_A C_F n_f\nn\\
	&+ \left(-\frac{136}{3}\zeta_4+\frac{1568579}{6615}\zeta_3-\frac{106402739}{3333960}\zeta_2-\frac{69188615691221}{470488435200}\right) C_F^2 n_f \nn\\
	&- \frac{917070648731}{504094752000} \, C_F n_f^2 \,, \nn \\
	\gamma_{T,gq}^{(2)} &=
	\left(\frac{58}{3}\zeta_4+\frac{115181}{52920}\zeta_3-\frac{5196678647}{133358400}\zeta_2-\frac{7235729013003301}{282293061120000}\right) C_F C_A^2 \nn\\
	&+ \left(\frac{2117}{42}\zeta_4-\frac{1929847}{13230}\zeta_3+\frac{584536927}{9525600}\zeta_2+\frac{22593692881537}{210039480000}\right) C_F^2 C_A \nn \\
	&+ \left(-\frac{638}{7}\zeta_4+\frac{941989}{8820}\zeta_3+\frac{11984773}{740880}\zeta_2-\frac{40048690625761}{435637440000}\right) C_F^3  \nn \\
	&+ \left(\frac{377}{63}\zeta_3-\frac{2193}{980}\zeta_2+\frac{3929616013}{1260236880}\right) C_A C_F n_f
	+ \left(-\frac{58}{63}\zeta_3+\frac{2252}{2205}\zeta_2-\frac{34054426196017}{14114653056000}\right) C_F^2 n_f \,, \nn \\
	%%%%%qg
	\gamma_{T,qg}^{(2)} &=
	\left(-\frac{174}{7}\zeta_4-\frac{1163237}{79380}\zeta_3+\frac{2303401993}{200037600}\zeta_2+\frac{2695491344802233}{141146530560000}\right) C_A^2 n_f \nn\\
	&+ \left(-\frac{29}{7}\zeta_4+\frac{64543}{810}\zeta_3+\frac{105089071}{11113200}\zeta_2-\frac{682173379142099}{8821658160000}\right) C_A C_F n_f \nn \\
	&+ \left(\frac{928}{63}\zeta_4-\frac{673261}{7938}\zeta_3-\frac{4295}{1134}\zeta_2+\frac{131783952793697}{1960368480000}\right) C_F^2 n_f \nn\\
	&+ \left(-\frac{290}{189}\zeta_3-\frac{140062}{59535}\zeta_2+\frac{482452191023}{126023688000}\right) C_A n_f^2 \nn\\
	&+ \left(\frac{268192}{59535}\zeta_2-\frac{27218712341927}{7057326528000}\right) C_F n_f^2
	- \frac{5858554}{18753525} \, n_f^3\,, \nn \\
	\gamma_{T,gg}^{(2)} &=
	\bigg[96\zeta_5+64\zeta_2\zeta_3+\frac{19429}{315}\zeta_4+\frac{36942781}{198450}\zeta_3 -\frac{74091536963}{250047000}\zeta_2+\frac{71458615798919}{1680315840000}\bigg] C_A^3 \nn\\
	&+ \bigg[104\zeta_4-\frac{348137}{26460}\zeta_3-\frac{1003304173}{13335840}\zeta_2 -\frac{996178580420963}{28229306112000}\bigg] C_A^2 n_f \nn \\
	&+ \left(\frac{364499}{4410}\zeta_3-\frac{60695777}{3333960}\zeta_2-\frac{94455666451589}{1764331632000}\right) C_A C_F n_f \nn\\
	&+ \left(-\frac{29}{49}\zeta_3+\frac{4168811}{3333960}\zeta_2-\frac{1343905087691}{940976870400}\right) C_F^2 n_f \nn\\
	&+ \left(-\frac{64}{9}\zeta_3+\frac{160}{27}\zeta_2-\frac{32292832169}{9001692000}\right) C_A n_f^2 + \left(-\frac{841}{3969}\zeta_2-\frac{59798107}{58786560}\right) C_F n_f^2 \,.
\end{align}

\section{Singular expansions at NNLO}
\label{app:singularNNLO}

In this subsection, we list the $x_L\to 0$ singular expansion of projected $N$-point correlators to NNLO.
For $e^+e^-\to q\bar{q}$, we have
\begin{align}
    \frac{1}{\sigma_0} \frac{d\sigma^{[N]}_{e^+e^-}}{dx_L}&=\left(\frac{\alpha_s}{4\pi}\right)e_{1,0}^{[N]}\left(\frac{1}{x_L}\right)_{+}+\left(\frac{\alpha_s}{4\pi}\right)^2\left[e_{2,1}^{[N]}\left(\frac{\ln x_L}{x_L}\right)_{+}+e_{2,0}^{[N]}\left(\frac{1}{x_L}\right)_{+}\right]\nn\\
    &+\left(\frac{\alpha_s}{4\pi}\right)^3\left[e_{3,2}^{[N]}\left(\frac{\ln^2 x_L}{x_L}\right)_{+}+e_{3,1}^{[N]}\left(\frac{\ln x_L}{x_L}\right)_{+}+e_{3,0}^{[N]}\left(\frac{1}{x_L}\right)_{+}\right] \, ,
\end{align}
with coefficients $e_{i,j}^{[N]}$. Below we show the analytic expressions for all these coefficients from $N=2$ to $N=6$. For $N=2$, we have
\begin{align}
    e_{1,0}^{[2]}&=\frac{3}{2}C_F  \, ,\nn\\
    e_{2,1}^{[2]}&=-\frac{107 C_A C_F}{15}+\frac{53 C_F n_f}{60}+\frac{25 C_F^2}{4} \, ,\nn\\
    e_{2,0}^{[2]}&=\left(4 \zeta _3-\frac{25 \pi ^2}{9}+\frac{35366}{675}\right) C_A C_F-\frac{4913 C_F n_f}{900}+\left(-8 \zeta _3+\frac{43 \pi ^2}{9}-\frac{8263}{216}\right) C_F^2 \, ,\nn\\
    e_{3,2}^{[2]}&=-\frac{16259 C_A C_F n_f}{1800}-\frac{340}{9} C_A C_F^2+\frac{8059}{300} C_A^2 C_F+\frac{4619}{720} C_F^2 n_f+\frac{23}{45} C_F n_f^2+\frac{625 C_F^3}{48} \, ,\nn\\
    e_{3,1}^{[2]}&=\left(\frac{16 \zeta _3}{3}-\frac{18 \pi ^2}{5}+\frac{6644267}{54000}\right) C_A C_F n_f+\left(\frac{262 \zeta _3}{3}-\frac{550 \pi ^2}{9}+\frac{105425}{144}\right) C_A
   C_F^2\nn\\
   &+\left(-\frac{74 \zeta _3}{3}+\frac{503 \pi ^2}{30}-\frac{2916859}{6750}\right) C_A^2 C_F+\left(-\frac{32 \zeta _3}{3}+\frac{208 \pi
   ^2}{27}-\frac{6760183}{64800}\right) C_F^2 n_f\nn\\
   &-\frac{8867 C_F n_f^2}{1350}+\left(-\frac{172 \zeta _3}{3}+\frac{1849 \pi ^2}{54}-\frac{723533}{2592}\right) C_F^3 \, ,\nn\\
   e_{3,0}^{[2]}&=\left(-\frac{589 \zeta _3}{10}+\frac{17 \pi ^4}{135}+\frac{144097 \pi ^2}{32400}-\frac{10584871}{3240000}\right) C_A C_F n_f\nn\\
   &+\left(-12 \pi ^2 \zeta _3-\frac{24436 \zeta
   _3}{15}-216 \zeta _5+\frac{1367 \pi ^4}{108}+\frac{21463 \pi ^2}{1296}+\frac{12974003}{8640}\right) C_A C_F^2\nn\\
   &+\left(4 \pi ^2 \zeta _3+\frac{3385 \zeta _3}{9}+56 \zeta
   _5-\frac{629 \pi ^4}{180}-\frac{31252 \pi ^2}{2025}+\frac{55444211}{3240000}\right) C_A^2 C_F\nn\\
   &+\left(\frac{6451 \zeta _3}{45}-\frac{34
   \pi ^4}{135}-\frac{1387 \pi ^2}{144}-\frac{93368917}{518400}\right) C_F^2 n_f+\left(\frac{132467}{108000}-\frac{7 \pi ^2}{135}\right) C_F n_f^2\nn\\
   &+\left(8 \pi ^2 \zeta
   _3+\frac{6287 \zeta _3}{6}+208 \zeta _5-\frac{262 \pi ^4}{27}+\frac{28895 \pi ^2}{1296}-\frac{21234923}{20736}\right) C_F^3\nn\\
   &+C_F \left(\frac{25
   j_2^{q,[2]}}{12}-\frac{7 j_2^{g,[2]}}{12}\right)+j_2^{q,[2]}\left(-\frac{11}{3}  C_A+\frac{2}{3}  n_f\right) \, .
\end{align}

For $N=3$, we have
\begin{align}
    e_{1,0}^{[3]}&=\frac{9 C_F}{8} \, ,\nn\\
    e_{2,1}^{[3]}&=-\frac{979 C_A C_F}{200}+\frac{139 C_F n_f}{200}+\frac{471 C_F^2}{80} \, ,\nn\\
    e_{2,0}^{[3]}&=\frac{66769 C_A C_F}{3000}-\frac{24863 C_F n_f}{6000}-\frac{21 C_F^2}{10} \, ,\nn\\
    e_{3,2}^{[3]}&=-\frac{19019 C_A C_F n_f}{3000}-\frac{412753 C_A C_F^2}{12000}+\frac{17743 C_A^2 C_F}{1000}+\frac{35369 C_F^2 n_f}{6000}+\frac{32}{75} C_F n_f^2+\frac{24649 C_F^3}{1600} \, ,\nn\\
    e_{3,1}^{[3]}&=\frac{3055907 C_A C_F n_f}{45000}+\left(\frac{34399441}{120000}-\frac{11 \pi ^2}{2}\right) C_A C_F^2-\frac{4559891 C_A^2 C_F}{22500}\nn\\
    &
    +\left(\pi^2-\frac{1026851}{20000}\right) C_F^2 n_f
    -\frac{11747 C_F n_f^2}{2250}-\frac{814823 C_F^3}{48000} \, ,\nn\\
    e_{3,0}^{[3]}&=\left(-\frac{1037 \zeta _3}{100}-\frac{2167 \pi ^2}{9000}-\frac{24958553}{7200000}\right) C_A C_F n_f\nn\\
    &+\left(-\frac{42321 \zeta _3}{200}+\frac{284797 \pi
   ^2}{36000}+\frac{4941457181}{7200000}\right) C_A C_F^2
   \nn\\
   &
   +\left(-\frac{829 \zeta _3}{100}+\frac{4433 \pi ^2}{2250}+\frac{363491521}{5400000}\right) C_A^2 C_F
   +\left(\frac{1827 \zeta _3}{50}-\frac{3877 \pi ^2}{6000}-\frac{3239027203}{21600000}\right) C_F^2 n_f
   \nn\\
   &
   +\left(\frac{106027}{216000}-\frac{11 \pi ^2}{450}\right)
   C_F n_f^2
   +\left(\frac{3267 \zeta _3}{20}-\frac{111313 \pi ^2}{14400}-\frac{6031520921}{17280000}\right) C_F^3
   \nn\\
   &+C_F \left(\frac{157 j_2^{q,[3]}}{120}-\frac{11
   j_2^{g,[3]}}{60}\right)+j_2^{q,[3]}\left(\frac{1}{3}  n_f-\frac{11}{6}C_A\right) \, .
\end{align}

For $N=4$, we have
\begin{align}
    e_{1,0}^{[4]}&=\frac{83 C_F}{120} \, ,\nn\\
    e_{2,1}^{[4]}&=-\frac{36299 C_A C_F}{12600}+\frac{2753 C_F n_f}{6300}+\frac{7553 C_F^2}{1800} \, ,\nn\\
    e_{2,0}^{[4]}&=\left(\zeta _3-\frac{4 \pi ^2}{5}+\frac{108536621}{5292000}\right) C_A C_F-\frac{6863089 C_F n_f}{2646000}+\left(-2 \zeta _3+\frac{68 \pi ^2}{45}-\frac{26407}{1800}\right) C_F^2 \, ,\nn\\
    e_{3,2}^{[4]}&=-\frac{1247209 C_A C_F n_f}{330750}-\frac{434239 C_A C_F^2}{18000}+\frac{13631939 C_A^2 C_F}{1323000}+\frac{789857 C_F^2 n_f}{189000}\nn\\
    &+\frac{289 C_F n_f^2}{1050}+\frac{687323
   C_F^3}{54000} \, ,\nn\\
    e_{3,1}^{[4]}&=\left(\frac{4 \zeta _3}{3}-\frac{1684 \pi ^2}{1575}+\frac{14035265711}{277830000}\right) C_A C_F n_f+\left(\frac{394 \zeta _3}{15}-\frac{31721 \pi
   ^2}{1350}+\frac{30148033}{81000}\right) C_A C_F^2\nn\\
   &+\left(-\frac{34 \zeta _3}{5}+\frac{24778 \pi ^2}{4725}-\frac{30424279061}{185220000}\right) C_A^2 C_F+\left(-\frac{8 \zeta
   _3}{3}+\frac{71 \pi ^2}{27}-\frac{2135412109}{39690000}\right) C_F^2 n_f\nn\\
   &-\frac{6610307 C_F n_f^2}{1984500}+\left(-\frac{116 \zeta _3}{5}+\frac{3944 \pi
   ^2}{225}-\frac{52560749}{324000}\right) C_F^3 \, ,\nn\\
    e_{3,0}^{[4]}&=\left(-\frac{10616 \zeta _3}{525}+\frac{17 \pi ^4}{540}+\frac{5275043 \pi ^2}{3969000}-\frac{51350203409}{11113200000}\right) C_A C_F n_f\nn\\
    &+\left(-3 \pi ^2 \zeta _3-\frac{3903307
   \zeta _3}{6300}-54 \zeta _5+\frac{862 \pi ^4}{225}+\frac{703943 \pi ^2}{162000}+\frac{2474360976011}{3175200000}\right) C_A C_F^2\nn\\
   &+\left(\pi ^2 \zeta _3+\frac{60307 \zeta
   _3}{420}+14 \zeta _5-\frac{769 \pi ^4}{675}-\frac{34386871 \pi ^2}{7938000}+\frac{536813881}{102900000}\right) C_A^2 C_F\nn\\
   &+\left(\frac{26786
   \zeta _3}{525}-\frac{17 \pi ^4}{270}-\frac{30173 \pi ^2}{10125}-\frac{386812080629}{3704400000}\right) C_F^2 n_f+\left(\frac{60205297}{333396000}-\frac{152 \pi
   ^2}{14175}\right) C_F n_f^2\nn\\
   &+\left(2 \pi ^2 \zeta _3+\frac{204547 \zeta _3}{450}+52 \zeta _5-\frac{394 \pi ^4}{135}+\frac{58801 \pi
   ^2}{6000}-\frac{123996342479}{194400000}\right) C_F^3\nn\\
   &+C_F \left(\frac{91 j_2^{q,[4]}}{120}-\frac{j_2^{g,[4]}}{15}\right)+j_2^{q,[4]}\left(\frac{1}{6}  n_f-\frac{11}{12} C_A\right) \, .
\end{align}

For $N=5$, we have
\begin{align}
    e_{1,0}^{[5]}&=\frac{19 C_F}{48} \, ,\nn\\
    e_{2,1}^{[5]}&=-\frac{56683 C_A C_F}{35280}+\frac{1117 C_F n_f}{4410}+\frac{13471 C_F^2}{5040} \, ,\nn\\
    e_{2,0}^{[5]}&=\frac{118638559 C_A C_F}{14817600}-\frac{11192729 C_F n_f}{7408800}-\frac{31979 C_F^2}{18900} \, ,\nn\\
    e_{3,2}^{[5]}&=-\frac{3134417 C_A C_F n_f}{1481760}-\frac{12532993 C_A C_F^2}{823200}+\frac{8468977 C_A^2 C_F}{1481760}+\frac{9849067 C_F^2 n_f}{3704400}\nn\\
    &+\frac{1427 C_F
   n_f^2}{8820}+\frac{9550939 C_F^3}{1058400} \, ,\nn\\
    e_{3,1}^{[5]}&=\frac{464804299 C_A C_F n_f}{19448100}+\left(\frac{205068546709}{1555848000}-\frac{209 \pi ^2}{108}\right) C_A C_F^2-\frac{10620789397 C_A^2 C_F}{155584800}\nn\\
    &+\left(\frac{19 \pi
   ^2}{54}-\frac{9339476177}{388962000}\right) C_F^2 n_f-\frac{10889647 C_F n_f^2}{5556600}-\frac{3082024613 C_F^3}{222264000} \, ,\nn\\
    e_{3,0}^{[5]}&=\left(-\frac{433 \zeta _3}{126}-\frac{1159213 \pi ^2}{22226400}-\frac{834556171943}{261382464000}\right) C_A C_F n_f\nn\\
    &+\left(-\frac{473251 \zeta _3}{5880}+\frac{15298127 \pi
   ^2}{5556600}+\frac{130173460675817}{435637440000}\right) C_A C_F^2\nn\\
   &+\left(-\frac{3019 \zeta _3}{1960}+\frac{2670173 \pi ^2}{7408800}+\frac{14752927607993}{522764928000}\right)
   C_A^2 C_F\nn\\
   &+\left(\frac{1879 \zeta _3}{147}-\frac{962293 \pi ^2}{11113200}-\frac{42132393867671}{653456160000}\right) C_F^2
   n_f+\left(\frac{7334497}{116688600}-\frac{187 \pi ^2}{39690}\right) C_F n_f^2\nn\\
   &+\left(\frac{94901 \zeta _3}{1260}-\frac{10082689 \pi
   ^2}{3175200}-\frac{39129021455567}{186701760000}\right) C_F^3\nn\\
   &+C_F \left(\frac{709 j_2^{q,[5]}}{1680}-\frac{11 j_2^{g,[5]}}{420}\right)+j_2^{q,[5]}\left(\frac{1}{12}  n_f-\frac{11}{24}  C_A\right) \, .
\end{align}

For $N=6$, we have
\begin{align}
    e_{1,0}^{[6]}&=\frac{367 C_F}{1680} \, ,\nn\\
    e_{2,1}^{[6]}&=-\frac{1476173 C_A C_F}{1693440}+\frac{238649 C_F n_f}{1693440}+\frac{376909 C_F^2}{235200} \, ,\nn\\
    e_{2,0}^{[6]}&=\left(\frac{\zeta _3}{4}-\frac{2243 \pi ^2}{10080}+\frac{67974834583}{10668672000}\right) C_A C_F-\frac{904034737 C_F n_f}{1066867200}\nn\\
    &+\left(-\frac{\zeta _3}{2}+\frac{1447 \pi
   ^2}{3360}-\frac{5708227013}{1185408000}\right) C_F^2 \, ,\nn\\
    e_{3,2}^{[6]}&=-\frac{491526533 C_A C_F n_f}{426746880}-\frac{4302140999 C_A C_F^2}{474163200}+\frac{6602110381 C_A^2 C_F}{2133734400}+\frac{2267320537 C_F^2 n_f}{1422489600}\nn\\
    &+\frac{115337 C_F
   n_f^2}{1270080}+\frac{387085543 C_F^3}{65856000} \, ,\nn\\
    e_{3,1}^{[6]}&=\left(\frac{\zeta _3}{3}-\frac{378709 \pi ^2}{1270080}+\frac{42568368526933}{2688505344000}\right) C_A C_F n_f\nn\\
    &+\left(\frac{4059 \zeta _3}{560}-\frac{31174991 \pi
   ^2}{4233600}+\frac{44207341016831}{331914240000}\right) C_A C_F^2\nn\\
   &+\left(-\frac{587 \zeta _3}{336}+\frac{29989 \pi ^2}{19845}-\frac{27085781886323}{537701068800}\right) C_A^2
   C_F\nn\\
   &+\left(-\frac{2 \zeta _3}{3}+\frac{5809 \pi ^2}{7560}-\frac{6980683406971}{358467379200}\right) C_F^2 n_f-\frac{1769887759 C_F n_f^2}{1600300800}\nn\\
   &+\left(-\frac{6017 \zeta
   _3}{840}+\frac{8706599 \pi ^2}{1411200}-\frac{1147915867097}{18439680000}\right) C_F^3 \, ,\nn\\
    e_{3,0}^{[6]}&=\left(-\frac{4944931 \zeta _3}{846720}+\frac{17 \pi ^4}{2160}+\frac{233488393 \pi ^2}{609638400}-\frac{31331898211717}{15122842560000}\right) C_A C_F n_f\nn\\
    &+\left(-\frac{3}{4} \pi
   ^2 \zeta _3-\frac{828552661 \zeta _3}{4233600}-\frac{27 \zeta _5}{2}+\frac{216487 \pi ^4}{201600}+\frac{32473103 \pi
   ^2}{36288000}\right.\nn\\
   &\left.+\frac{4265592909571504753}{15055629926400000}\right) C_A C_F^2+\left(\frac{\pi ^2 \zeta _3}{4}+\frac{190574591 \zeta _3}{4233600}+\frac{7 \zeta
   _5}{2}-\frac{196321 \pi ^4}{604800}\right.\nn\\
   &\left.-\frac{4266737131 \pi ^2}{3556224000}-\frac{22183030097197607}{67750334668800000}\right) C_A^2 C_F\nn\\
   &+\left(\frac{1273973 \zeta _3}{84672}-\frac{17 \pi ^4}{1080}-\frac{1157609269 \pi ^2}{1422489600}-\frac{348772205186542007}{9033377955840000}\right) C_F^2
   n_f\nn\\
   &+\left(\frac{322291557109}{16131032064000}-\frac{1595 \pi ^2}{762048}\right) C_F n_f^2\nn\\
   &+\left(\frac{\pi ^2 \zeta _3}{2}+\frac{10389913 \zeta _3}{67200}+13 \zeta
   _5-\frac{61921 \pi ^4}{75600}+\frac{13002155137 \pi ^2}{3556224000}-\frac{220788639245910637}{836423884800000}\right) C_F^3\nn\\
   &+C_F \left(\frac{1027 j_2^{q,[6]}}{4480}-\frac{29
   j_2^{g,[6]}}{2688}\right)+j_2^{q,[6]}\left(\frac{1}{24} n_f-\frac{11}{48}
   C_A\right) \, .
\end{align}

Similarly, we can write the singular expansion of $H\to gg$ with the coefficients $f_{i,j}^{[N]}$:
\begin{align}
    \frac{1}{\sigma_0} \frac{d\sigma^{[N]}_{H\to gg}}{dx_L}&=\left(\frac{\alpha_s}{4\pi}\right)f_{1,0}^{[N]}\left(\frac{1}{x_L}\right)_{+}+\left(\frac{\alpha_s}{4\pi}\right)^2\left[f_{2,1}^{[N]}\left(\frac{\ln x_L}{x_L}\right)_{+}+f_{2,0}^{[N]}\left(\frac{1}{x_L}\right)_{+}\right]\nn\\
    &+\left(\frac{\alpha_s}{4\pi}\right)^3\left[f_{3,2}^{[N]}\left(\frac{\ln^2 x_L}{x_L}\right)_{+}+f_{3,1}^{[N]}\left(\frac{\ln x_L}{x_L}\right)_{+}+f_{3,0}^{[N]}\left(\frac{1}{x_L}\right)_{+}\right] \, .
\end{align}

Now for $N=2$, we have
\begin{align}
    f_{1,0}^{[2]}&=\frac{7 C_A}{5}+\frac{n_f}{10} \, ,\nn\\
    f_{2,1}^{[2]}&=\frac{89 C_A n_f}{50}-\frac{91 C_A^2}{75}-\frac{7 C_F n_f}{10}+\frac{2 n_f^2}{15} \, ,\nn\\
    f_{2,0}^{[2]}&=\left(-\frac{201371}{27000}-\frac{7 \pi ^2}{45}\right) C_A n_f+\left(-4 \zeta _3+\frac{97 \pi ^2}{45}+\frac{138427}{3375}\right) C_A^2+\frac{9 C_F n_f}{100}-\frac{43 n_f^2}{60} \, ,\nn\\
    f_{3,2}^{[2]}&=\frac{1463}{450} C_A C_F n_f-\frac{656}{125} C_A^2 n_f+\frac{133}{90} C_A n_f^2+\frac{3094 C_A^3}{1125}-\frac{1631 C_F n_f^2}{1800}-\frac{35}{24} C_F^2 n_f+\frac{2 n_f^3}{15} \, ,\nn\\
    f_{3,1}^{[2]}&=\left(-\frac{28 \zeta _3}{15}+\frac{112 \pi ^2}{135}-\frac{3140863}{81000}\right) C_A C_F n_f+\left(-\frac{44 \zeta _3}{5}+\frac{4187 \pi ^2}{675}+\frac{29866}{405}\right) C_A^2
   n_f\nn\\
   &+\left(-\frac{176981}{13500}-\frac{2 \pi ^2}{9}\right) C_A n_f^2+\left(\frac{104 \zeta _3}{15}-\frac{7142 \pi ^2}{675}-\frac{22091}{2025}\right) C_A^3\nn\\
   &+\left(\frac{56 \zeta
   _3}{15}-\frac{301 \pi ^2}{135}+\frac{97169}{6480}\right) C_F^2 n_f+\frac{149657 C_F n_f^2}{18000}-\frac{283 n_f^3}{225} \, ,\nn\\
    f_{3,0}^{[2]}&=\left(\frac{23878 \zeta _3}{225}-\frac{7 \pi ^4}{150}+\frac{94417 \pi ^2}{16200}-\frac{4143529}{16875}\right) C_A C_F n_f\nn\\
    &+\left(\frac{14126 \zeta _3}{225}-\frac{13 \pi
   ^4}{135}-\frac{175381 \pi ^2}{16200}-\frac{7964118803}{16200000}\right) C_A^2 n_f+\left(-\frac{68 \zeta _3}{45}+\frac{71 \pi ^2}{675}-\frac{23974453}{4860000}\right) C_A
   n_f^2\nn\\
   &+\left(-\frac{26047 \zeta _3}{75}+48 \zeta _5-\frac{931 \pi ^4}{1350}+\frac{21359 \pi ^2}{450}+\frac{12171537439}{8100000}\right) C_A^3\nn\\
   &+\left(-\frac{16613 \zeta _3}{225}+\frac{112 \pi ^4}{675}-\frac{46297 \pi ^2}{5400}+\frac{132484699}{777600}\right) C_F^2 n_f-\frac{2}{45} \pi ^2 n_f^3+\frac{86509 n_f^3}{40500}\nn\\
   &+\left(\frac{12 \zeta _3}{5}-\frac{7 \pi
   ^2}{24}+\frac{15337613}{1296000}\right) C_F n_f^2+j_2^{g,[2]}\left( n_f-\frac{34}{15} C_A\right)-\frac{7}{30} j_2^{q,[2]} n_f \, .
\end{align}

For $N=3$, we have
\begin{align}
    f_{1,0}^{[3]}&=\frac{21 C_A}{20}+\frac{3 n_f}{40} \, ,\nn\\
    f_{2,1}^{[3]}&=\frac{36 C_A n_f}{25}+\frac{14 C_A^2}{25}-\frac{33 C_F n_f}{80}+\frac{n_f^2}{10} \, ,\nn\\
    f_{2,0}^{[3]}&=-\frac{14127 C_A n_f}{2000}+\frac{124153 C_A^2}{3000}+\frac{7 C_F n_f}{80}-\frac{131 n_f^2}{240} \, ,\nn\\
    f_{3,2}^{[3]}&=\frac{4631 C_A C_F n_f}{3000}-\frac{212}{125} C_A^2 n_f+\frac{77}{60} C_A n_f^2-\frac{329 C_A^3}{375}-\frac{6479 C_F n_f^2}{12000}-\frac{1727 C_F^2 n_f}{1600}+\frac{n_f^3}{10} \, ,\nn\\
    f_{3,1}^{[3]}&=-\frac{3150047 C_A C_F n_f}{180000}+\left(\frac{400933}{5625}+\frac{17 \pi ^2}{30}\right) C_A^2 n_f+\left(\frac{\pi ^2}{15}-\frac{24983}{1800}\right) C_A
   n_f^2\nn\\
   &+\left(\frac{1413553}{22500}-\frac{77 \pi ^2}{15}\right) C_A^3+\frac{1817029 C_F n_f^2}{360000}-\frac{80131 C_F^2 n_f}{48000}-\frac{859 n_f^3}{900} \, ,\nn\\
    f_{3,0}^{[3]}&=\left(\frac{5097 \zeta _3}{100}+\frac{979 \pi ^2}{6000}-\frac{4481199733}{43200000}\right) C_A C_F n_f+\left(\frac{23 \zeta _3}{100}+\frac{15743 \pi
   ^2}{4500}-\frac{1124680027}{2400000}\right) C_A^2 n_f\nn\\
   &+\left(\frac{19 \zeta _3}{10}+\frac{\pi ^2}{180}-\frac{13714631}{2160000}\right) C_A n_f^2+\left(-\frac{1197 \zeta
   _3}{10}-\frac{19291 \pi ^2}{1500}+\frac{553753307}{400000}\right) C_A^3\nn\\
   &+\left(-\frac{81 \zeta _3}{5}+\frac{7799 \pi
   ^2}{14400}+\frac{450474103}{17280000}\right) C_F^2 n_f+\left(\frac{9 \zeta _3}{5}-\frac{6479 \pi ^2}{36000}+\frac{309593693}{43200000}\right) C_F n_f^2\nn\\
   &+n_f
   \left(\frac{j_2^{g,[3]}}{2}-\frac{11 j_2^{q,[3]}}{120}\right)-\frac{47}{60} j_2^{g,[3]} C_A+\left(\frac{91219}{54000}-\frac{\pi ^2}{30}\right) n_f^3 \, .
\end{align}

For $N=4$, we have
\begin{align}
    f_{1,0}^{[4]}&=\frac{181 C_A}{280}+\frac{19 n_f}{420} \, ,\nn\\
    f_{2,1}^{[4]}&=\frac{10253 C_A n_f}{11025}+\frac{14299 C_A^2}{14700}-\frac{332 C_F n_f}{1575}+\frac{19 n_f^2}{315} \, ,\nn\\
    f_{2,0}^{[4]}&=\left(-\frac{1329997}{308700}-\frac{8 \pi ^2}{315}\right) C_A n_f+\left(-\zeta _3+\frac{232 \pi ^2}{315}+\frac{27645409}{1481760}\right) C_A^2+\frac{14767 C_F
   n_f}{220500}-\frac{3539 n_f^2}{10584} \, ,\nn\\
    f_{3,2}^{[4]}&=\frac{7352 C_A C_F n_f}{11025}-\frac{150677 C_A^2 n_f}{4630500}+\frac{11441 C_A n_f^2}{13230}-\frac{3245873 C_A^3}{3087000}-\frac{45872 C_F n_f^2}{165375}\nn\\
    &-\frac{2158 C_F^2
   n_f}{3375}+\frac{19 n_f^3}{315} \, ,\nn\\
    f_{3,1}^{[4]}&=\left(-\frac{32 \zeta _3}{105}+\frac{488 \pi ^2}{4725}-\frac{1236752053}{138915000}\right) C_A C_F n_f+\left(-\frac{248 \zeta _3}{105}+\frac{71042 \pi
   ^2}{33075}+\frac{879657958}{40516875}\right) C_A^2 n_f\nn\\
   &+\left(-\frac{30886598}{3472875}-\frac{2 \pi ^2}{189}\right) C_A n_f^2+\left(-\frac{316 \zeta _3}{105}-\frac{62431 \pi
   ^2}{66150}+\frac{112080210239}{1944810000}\right) C_A^3\nn\\
   &+\left(\frac{64 \zeta _3}{105}-\frac{2176 \pi ^2}{4725}+\frac{147638}{50625}\right) C_F^2 n_f+\frac{1717766 C_F
   n_f^2}{643125}-\frac{16531 n_f^3}{28350} \, ,\nn\\
    f_{3,0}^{[4]}&=\left(\frac{769 \zeta _3}{21}-\frac{4 \pi ^4}{525}+\frac{3135287 \pi ^2}{1984500}-\frac{4722787369477}{58344300000}\right) C_A C_F n_f\nn\\
    &+\left(\frac{7247 \zeta _3}{350}-\frac{37
   \pi ^4}{1890}-\frac{63221999 \pi ^2}{27783000}-\frac{234523005612047}{816820200000}\right) C_A^2 n_f\nn\\
   &+\left(\frac{17 \zeta _3}{35}+\frac{712 \pi
   ^2}{33075}-\frac{31230478007}{11668860000}\right) C_A n_f^2+\left(\frac{1098499}{1029000}-\frac{19 \pi ^2}{945}\right) n_f^3\nn\\
   &+\left(-\frac{834658 \zeta _3}{11025}+12 \zeta _5-\frac{4877 \pi ^4}{18900}+\frac{5996201 \pi
   ^2}{411600}+\frac{1010512263758903}{1633640400000}\right) C_A^3\nn\\
   &+\left(-\frac{39056 \zeta _3}{2205}+\frac{128 \pi ^4}{4725}-\frac{84439 \pi
   ^2}{36750}+\frac{149994187}{3037500}\right) C_F^2 n_f\nn\\
   &+\left(\frac{38 \zeta _3}{35}-\frac{6536 \pi ^2}{70875}+\frac{53544544673}{14586075000}\right) C_F n_f^2+n_f
   \left(\frac{j_2^{g,[4]}}{4}-\frac{4 j_2^{q,[4]}}{105}\right)-\frac{227}{840} j_2^{g,[4]} C_A \, .
\end{align}

For $N=5$, we have
\begin{align}
    f_{1,0}^{[5]}&=\frac{83 C_A}{224}+\frac{17 n_f}{672} \, ,\nn\\
    f_{2,1}^{[5]}&=\frac{15559 C_A n_f}{28224}+\frac{7885 C_A^2}{9408}-\frac{209 C_F n_f}{2016}+\frac{17 n_f^2}{504} \, ,\nn\\
    f_{2,0}^{[5]}&=-\frac{64588933 C_A n_f}{23708160}+\frac{327604399 C_A^2}{23708160}+\frac{3587 C_F n_f}{80640}-\frac{32119 n_f^2}{169344} \, ,\nn\\
    f_{3,2}^{[5]}&=\frac{559537 C_A C_F n_f}{1975680}+\frac{1069039 C_A^2 n_f}{2370816}+\frac{44671 C_A n_f^2}{84672}-\frac{465215 C_A^3}{790272}-\frac{202763 C_F n_f^2}{1481760}\nn\\
    &-\frac{148181 C_F^2
   n_f}{423360}+\frac{17 n_f^3}{504} \, ,\nn\\
    f_{3,1}^{[5]}&=-\frac{9359137051 C_A C_F n_f}{2489356800}+\left(\frac{5802589423}{331914240}+\frac{311 \pi ^2}{1512}\right) C_A^2 n_f+\left(\frac{17 \pi ^2}{756}-\frac{41130125}{7112448}\right)
   C_A n_f^2\nn\\
   &+\left(\frac{11861917541}{199148544}-\frac{913 \pi ^2}{504}\right) C_A^3+\frac{10445973 C_F n_f^2}{7683200}-\frac{26440613 C_F^2 n_f}{59270400}-\frac{29941
   n_f^3}{90720} \, ,\nn\\
    f_{3,0}^{[5]}&=\left(\frac{3247 \zeta _3}{196}-\frac{30019 \pi ^2}{17781120}-\frac{65439560835163}{2091059712000}\right) C_A C_F n_f\nn\\
    &+\left(\frac{341 \zeta _3}{196}+\frac{5985731 \pi
   ^2}{3951360}-\frac{53396538441073}{278807961600}\right) C_A^2 n_f\nn\\
   &+\left(\frac{323 \zeta _3}{504}+\frac{1397 \pi ^2}{84672}-\frac{48602161849}{29872281600}\right) C_A
   n_f^2\nn\\
   &+\left(-\frac{11537 \zeta _3}{392}-\frac{21262013 \pi ^2}{3951360}+\frac{376022366587657}{836423884800}\right) C_A^3+\left(\frac{110548211}{177811200}-\frac{17
   \pi ^2}{1512}\right) n_f^3\nn\\
   &+\left(\frac{17 \zeta _3}{28}-\frac{202763 \pi
   ^2}{4445280}+\frac{936794568809}{522764928000}\right) C_F n_f^2\nn\\
   &+\left(-\frac{2071
   \zeta _3}{504}+\frac{156431 \pi ^2}{1270080}+\frac{426649977431}{49787136000}\right) C_F^2 n_f
   \nn\\
   &+n_f \left(\frac{j_2^{g,[5]}}{8}-\frac{11 j_2^{q,[5]}}{672}\right)-\frac{59}{672} j_2^{g,[5]} C_A \, .
\end{align}

For $N=6$, we have
\begin{align}
    f_{1,0}^{[6]}&=\frac{4129 C_A}{20160}+\frac{55 n_f}{4032} \, ,\nn\\
    f_{2,1}^{[6]}&=\frac{793717 C_A n_f}{2540160}+\frac{7510651 C_A^2}{12700800}-\frac{10643 C_F n_f}{211680}+\frac{55 n_f^2}{3024} \, ,\nn\\
    f_{2,0}^{[6]}&=\left(-\frac{807461891}{533433600}-\frac{29 \pi ^2}{6048}\right) C_A n_f+\left(-\frac{\zeta _3}{4}+\frac{6439 \pi ^2}{30240}+\frac{30227541313}{5334336000}\right)
   C_A^2\nn\\
   &+\frac{28983739 C_F n_f}{1066867200}-\frac{1578863 n_f^2}{15240960} \, ,\nn\\
    f_{3,2}^{[6]}&=\frac{255850789 C_A C_F n_f}{2133734400}+\frac{300593801 C_A^2 n_f}{640120320}+\frac{2327141 C_A n_f^2}{7620480}-\frac{3687729641 C_A^3}{16003008000}\nn\\
    &-\frac{28377109 C_F
   n_f^2}{426746880}-\frac{10930361 C_F^2 n_f}{59270400}+\frac{55 n_f^3}{3024} \, ,\nn\\
    f_{3,1}^{[6]}&=\left(-\frac{29 \zeta _3}{504}+\frac{4495 \pi ^2}{254016}-\frac{178564669351}{99574272000}\right) C_A C_F n_f\nn\\
    &+\left(-\frac{307 \zeta _3}{504}+\frac{303823 \pi
   ^2}{476280}+\frac{14276979570313}{4032758016000}\right) C_A^2 n_f+\left(\frac{23 \pi ^2}{9072}-\frac{31553863549}{9601804800}\right) C_A n_f^2\nn\\
   &+\left(-\frac{1819 \zeta
   _3}{1260}+\frac{2174551 \pi ^2}{9525600}+\frac{119645044280357}{4032758016000}\right) C_A^3\nn\\
   &+\left(\frac{29 \zeta _3}{252}-\frac{41963 \pi
   ^2}{423360}+\frac{93494123567}{149361408000}\right) C_F^2 n_f+\frac{122088294301 C_F n_f^2}{179233689600}-\frac{293821 n_f^3}{1632960} \, ,\nn\\
    f_{3,0}^{[6]}&=\left(\frac{1880183 \zeta _3}{181440}-\frac{29 \pi ^4}{20160}+\frac{1220346811 \pi ^2}{3200601600}-\frac{42313152704732657}{1935723847680000}\right) C_A C_F
   n_f\nn\\
   &+\left(\frac{1358407 \zeta _3}{211680}-\frac{397 \pi ^4}{90720}-\frac{8455304941 \pi ^2}{19203609600}-\frac{118046440117353577}{1129172244480000}\right) C_A^2
   n_f\nn\\
   &+\left(\frac{167 \zeta _3}{756}+\frac{87463 \pi ^2}{11430720}-\frac{24825645395783}{48393096192000}\right) C_A n_f^2\nn\\
   &+\left(-\frac{20085281 \zeta _3}{1587600}+3 \zeta
   _5-\frac{69847 \pi ^4}{907200}+\frac{41429974547 \pi ^2}{9601804800}+\frac{209608878647475173}{1254635827200000}\right) C_A^3\nn\\
   &+\left(-\frac{5106361 \zeta _3}{1270080}+\frac{29 \pi ^4}{5670}-\frac{149999107 \pi ^2}{266716800}+\frac{9512383086610937}{752781496320000}\right) C_F^2
   n_f\nn\\
   &+\left(\frac{55 \zeta _3}{168}-\frac{9453623 \pi ^2}{426746880}+\frac{2316694647929087}{2710013386752000}\right) C_F n_f^2\nn\\
   &+n_f \left(\frac{j_2^{g,[6]}}{16}-\frac{29
   j_2^{q,[6]}}{4032}\right)-\frac{491 j_2^{g,[6]}
   C_A}{20160}+\left(\frac{10006753259}{28805414400}-\frac{55 \pi ^2}{9072}\right) n_f^3 \, .
\end{align}

\bibliography{ENCR_Ref}{}

\providecommand{\href}[2]{#2}\begingroup\raggedright\begin{thebibliography}{100}

\bibitem{Sterman:1975xv}
G.~F. Sterman, {\it {Jet Structure in e+ e- Annihilation with Massless
  Hadrons}},  {\em ILL-TH-75-32} (1975).

\bibitem{Sveshnikov:1995vi}
N.~A. Sveshnikov and F.~V. Tkachov, {\it {Jets and quantum field theory}},
  {\em Phys. Lett. B} {\bf 382} (1996) 403--408,
  [\href{http://arxiv.org/abs/hep-ph/9512370}{{\tt hep-ph/9512370}}].

\bibitem{Tkachov:1995kk}
F.~V. Tkachov, {\it {Measuring multi - jet structure of hadronic energy flow or
  What is a jet?}},  {\em Int. J. Mod. Phys. A} {\bf 12} (1997) 5411--5529,
  [\href{http://arxiv.org/abs/hep-ph/9601308}{{\tt hep-ph/9601308}}].

\bibitem{Korchemsky:1999kt}
G.~P. Korchemsky and G.~F. Sterman, {\it {Power corrections to event shapes and
  factorization}},  {\em Nucl. Phys. B} {\bf 555} (1999) 335--351,
  [\href{http://arxiv.org/abs/hep-ph/9902341}{{\tt hep-ph/9902341}}].

\bibitem{Hofman:2008ar}
D.~M. Hofman and J.~Maldacena, {\it {Conformal collider physics: Energy and
  charge correlations}},  {\em JHEP} {\bf 05} (2008) 012,
  [\href{http://arxiv.org/abs/0803.1467}{{\tt arXiv:0803.1467}}].

\bibitem{Kologlu:2019mfz}
M.~Kologlu, P.~Kravchuk, D.~Simmons-Duffin, and A.~Zhiboedov, {\it {The
  light-ray OPE and conformal colliders}},  {\em JHEP} {\bf 01} (2021) 128,
  [\href{http://arxiv.org/abs/1905.01311}{{\tt arXiv:1905.01311}}].

\bibitem{Moult:2025nhu}
I.~Moult and H.~X. Zhu, {\it {Energy Correlators: A Journey From Theory to
  Experiment}},  \href{http://arxiv.org/abs/2506.09119}{{\tt
  arXiv:2506.09119}}.

\bibitem{Basham:1978bw}
C.~L. Basham, L.~S. Brown, S.~D. Ellis, and S.~T. Love, {\it {Energy
  Correlations in electron - Positron Annihilation: Testing QCD}},  {\em Phys.
  Rev. Lett.} {\bf 41} (1978) 1585.

\bibitem{Basham:1978zq}
C.~L. Basham, L.~S. Brown, S.~D. Ellis, and S.~T. Love, {\it {Energy
  Correlations in electron-Positron Annihilation in Quantum Chromodynamics:
  Asymptotically Free Perturbation Theory}},  {\em Phys. Rev. D} {\bf 19}
  (1979) 2018.

\bibitem{Basham:1979gh}
C.~L. Basham, L.~S. Brown, S.~D. Ellis, and S.~T. Love, {\it {Energy
  Correlations in Perturbative Quantum Chromodynamics: A Conjecture for All
  Orders}},  {\em Phys. Lett. B} {\bf 85} (1979) 297--299.

\bibitem{Basham:1977iq}
C.~L. Basham, L.~S. Brown, S.~D. Ellis, and S.~T. Love, {\it {Electron -
  Positron Annihilation Energy Pattern in Quantum Chromodynamics:
  Asymptotically Free Perturbation Theory}},  {\em Phys. Rev. D} {\bf 17}
  (1978) 2298.

\bibitem{Belitsky:2013ofa}
A.~V. Belitsky, S.~Hohenegger, G.~P. Korchemsky, E.~Sokatchev, and
  A.~Zhiboedov, {\it {Energy-Energy Correlations in N=4 Supersymmetric
  Yang-Mills Theory}},  {\em Phys. Rev. Lett.} {\bf 112} (2014), no.~7 071601,
  [\href{http://arxiv.org/abs/1311.6800}{{\tt arXiv:1311.6800}}].

\bibitem{Henn:2019gkr}
J.~M. Henn, E.~Sokatchev, K.~Yan, and A.~Zhiboedov, {\it {Energy-energy
  correlation in $N$=4 super Yang-Mills theory at next-to-next-to-leading
  order}},  {\em Phys. Rev. D} {\bf 100} (2019), no.~3 036010,
  [\href{http://arxiv.org/abs/1903.05314}{{\tt arXiv:1903.05314}}].

\bibitem{Dixon:2018qgp}
L.~J. Dixon, M.-X. Luo, V.~Shtabovenko, T.-Z. Yang, and H.~X. Zhu, {\it
  {Analytical Computation of Energy-Energy Correlation at Next-to-Leading Order
  in QCD}},  {\em Phys. Rev. Lett.} {\bf 120} (2018), no.~10 102001,
  [\href{http://arxiv.org/abs/1801.03219}{{\tt arXiv:1801.03219}}].

\bibitem{Luo:2019nig}
M.-X. Luo, V.~Shtabovenko, T.-Z. Yang, and H.~X. Zhu, {\it {Analytic
  Next-To-Leading Order Calculation of Energy-Energy Correlation in
  Gluon-Initiated Higgs Decays}},  {\em JHEP} {\bf 06} (2019) 037,
  [\href{http://arxiv.org/abs/1903.07277}{{\tt arXiv:1903.07277}}].

\bibitem{Gao:2020vyx}
J.~Gao, V.~Shtabovenko, and T.-Z. Yang, {\it {Energy-energy correlation in
  hadronic Higgs decays: analytic results and phenomenology at NLO}},  {\em
  JHEP} {\bf 02} (2021) 210, [\href{http://arxiv.org/abs/2012.14188}{{\tt
  arXiv:2012.14188}}].

\bibitem{ALEPH:1990vew}
{\bf ALEPH} Collaboration, D.~Decamp et~al., {\it {Measurement of alpha-s from
  the structure of particle clusters produced in hadronic Z decays}},  {\em
  Phys. Lett. B} {\bf 257} (1991) 479--491.

\bibitem{DELPHI:1990sof}
{\bf DELPHI} Collaboration, P.~Abreu et~al., {\it {Energy-energy correlations
  in hadronic final states from Z0 decays}},  {\em Phys. Lett. B} {\bf 252}
  (1990) 149--158.

\bibitem{Bossi:2024qeu}
H.~Bossi, A.~Baty, Y.~Chen, Y.-C. Chen, G.-M. Innocenti, M.~Maggi, C.~McGinn,
  and Y.-J. Lee, {\it {Measurement of the energy-energy correlator in the
  back-to-back limit using the archived ALEPH e+e- data at 91.2 GeV}},  {\em
  PoS} {\bf LHCP2024} (2025) 228, [\href{http://arxiv.org/abs/2501.01968}{{\tt
  arXiv:2501.01968}}].

\bibitem{Bossi:2025xsi}
H.~Bossi, Y.-C. Chen, Y.~Chen, J.~Zhang, G.~M. Innocenti, A.~Badea, A.~Baty,
  M.~Maggi, C.~McGinn, and Y.-J. Lee, {\it {Analysis note: measurement of
  energy-energy correlator in $e^{+}e^{-}$ collisions at $91$ GeV with archived
  ALEPH data}},  \href{http://arxiv.org/abs/2505.11828}{{\tt
  arXiv:2505.11828}}.

\bibitem{Electron-PositronAlliance:2025fhk}
{\bf Electron-Positron Alliance} Collaboration, H.~Bossi et~al., {\it {Energy
  Correlators from Partons to Hadrons: Unveiling the Dynamics of the Strong
  Interactions with Archival ALEPH Data}},
  \href{http://arxiv.org/abs/2511.00149}{{\tt arXiv:2511.00149}}.

\bibitem{Zhang:2025nlf}
J.~Zhang, T.-A. Sheng, Y.-C. Chen, H.~Bossi, A.~Badea, A.~Baty, C.~McGinn,
  Y.-J. Lee, and Y.~Chen, {\it {Analysis note: measurement of thrust and track
  energy-energy correlator in e+e- collisions at 91.2 GeV with DELPHI open
  data}},  \href{http://arxiv.org/abs/2510.18762}{{\tt arXiv:2510.18762}}.

\bibitem{Dixon:2019uzg}
L.~J. Dixon, I.~Moult, and H.~X. Zhu, {\it {Collinear limit of the
  energy-energy correlator}},  {\em Phys. Rev. D} {\bf 100} (2019), no.~1
  014009, [\href{http://arxiv.org/abs/1905.01310}{{\tt arXiv:1905.01310}}].

\bibitem{Korchemsky:2019nzm}
G.~P. Korchemsky, {\it {Energy correlations in the end-point region}},  {\em
  JHEP} {\bf 01} (2020) 008, [\href{http://arxiv.org/abs/1905.01444}{{\tt
  arXiv:1905.01444}}].

\bibitem{Yan:2022cye}
K.~Yan and X.~Zhang, {\it {Three-Point Energy Correlator in N=4 Supersymmetric
  Yang-Mills Theory}},  {\em Phys. Rev. Lett.} {\bf 129} (2022), no.~2 021602,
  [\href{http://arxiv.org/abs/2203.04349}{{\tt arXiv:2203.04349}}].

\bibitem{Chen:2019bpb}
H.~Chen, M.-X. Luo, I.~Moult, T.-Z. Yang, X.~Zhang, and H.~X. Zhu, {\it {Three
  point energy correlators in the collinear limit: symmetries, dualities and
  analytic results}},  {\em JHEP} {\bf 08} (2020) 028,
  [\href{http://arxiv.org/abs/1912.11050}{{\tt arXiv:1912.11050}}].

\bibitem{Yang:2022tgm}
T.-Z. Yang and X.~Zhang, {\it {Analytic Computation of three-point energy
  correlator in QCD}},  {\em JHEP} {\bf 09} (2022) 006,
  [\href{http://arxiv.org/abs/2208.01051}{{\tt arXiv:2208.01051}}].

\bibitem{Yang:2024gcn}
T.-Z. Yang and X.~Zhang, {\it {Three-point energy correlators in hadronic Higgs
  boson decays}},  {\em Phys. Rev. D} {\bf 109} (2024), no.~11 114036,
  [\href{http://arxiv.org/abs/2402.05174}{{\tt arXiv:2402.05174}}].

\bibitem{Chicherin:2024ifn}
D.~Chicherin, I.~Moult, E.~Sokatchev, K.~Yan, and Y.~Zhu, {\it {Collinear limit
  of the four-point energy correlator in N=4 supersymmetric Yang-Mills
  theory}},  {\em Phys. Rev. D} {\bf 110} (2024), no.~9 L091901,
  [\href{http://arxiv.org/abs/2401.06463}{{\tt arXiv:2401.06463}}].

\bibitem{He:2024hbb}
S.~He, X.~Jiang, Q.~Yang, and Y.-Q. Zhang, {\it {From squared amplitudes to
  energy correlators}},  \href{http://arxiv.org/abs/2408.04222}{{\tt
  arXiv:2408.04222}}.

\bibitem{Chen:2020vvp}
H.~Chen, I.~Moult, X.~Zhang, and H.~X. Zhu, {\it {Rethinking jets with energy
  correlators: Tracks, resummation, and analytic continuation}},  {\em Phys.
  Rev. D} {\bf 102} (2020), no.~5 054012,
  [\href{http://arxiv.org/abs/2004.11381}{{\tt arXiv:2004.11381}}].

\bibitem{Komiske:2022enw}
P.~T. Komiske, I.~Moult, J.~Thaler, and H.~X. Zhu, {\it {Analyzing N-Point
  Energy Correlators inside Jets with CMS Open Data}},  {\em Phys. Rev. Lett.}
  {\bf 130} (2023), no.~5 051901, [\href{http://arxiv.org/abs/2201.07800}{{\tt
  arXiv:2201.07800}}].

\bibitem{Lee:2025okn}
K.~Lee and I.~W. Stewart, {\it {Dihadron Fragmentation and the Confinement
  Transition in Energy Correlators}},  {\em Phys. Rev. Lett.} {\bf 136} (2026),
  no.~8 081902, [\href{http://arxiv.org/abs/2507.11495}{{\tt
  arXiv:2507.11495}}].

\bibitem{Chang:2025kgq}
C.-H. Chang, H.~Chen, X.~Liu, D.~Simmons-Duffin, F.~Yuan, and H.~X. Zhu, {\it
  {Quantum Scaling in Energy Correlators beyond the Confinement Transition}},
  {\em Phys. Rev. Lett.} {\bf 136} (2026), no.~8 081903,
  [\href{http://arxiv.org/abs/2507.15923}{{\tt arXiv:2507.15923}}].

\bibitem{Kang:2025zto}
Z.-B. Kang, A.~Metz, D.~Pitonyak, and C.~Zhang, {\it {Dihadron Fragmentation
  Framework for Near-Side Energy-Energy Correlators}},  {\em Phys. Rev. Lett.}
  {\bf 136} (2026), no.~8 081905, [\href{http://arxiv.org/abs/2507.17444}{{\tt
  arXiv:2507.17444}}].

\bibitem{Lee:2022uwt}
K.~Lee, B.~Me{\c{c}}aj, and I.~Moult, {\it {Conformal collider physics meets
  LHC data}},  {\em Phys. Rev. D} {\bf 111} (2025), no.~1 L011502,
  [\href{http://arxiv.org/abs/2205.03414}{{\tt arXiv:2205.03414}}].

\bibitem{Chen:2023zlx}
W.~Chen, J.~Gao, Y.~Li, Z.~Xu, X.~Zhang, and H.~X. Zhu, {\it {NNLL resummation
  for projected three-point energy correlator}},  {\em JHEP} {\bf 05} (2024)
  043, [\href{http://arxiv.org/abs/2307.07510}{{\tt arXiv:2307.07510}}].

\bibitem{FlavourLatticeAveragingGroupFLAG:2024oxs}
{\bf Flavour Lattice Averaging Group (FLAG)} Collaboration, Y.~Aoki et~al.,
  {\it {FLAG review 2024}},  {\em Phys. Rev. D} {\bf 113} (2026), no.~1 014508,
  [\href{http://arxiv.org/abs/2411.04268}{{\tt arXiv:2411.04268}}].

\bibitem{Abbate:2010xh}
R.~Abbate, M.~Fickinger, A.~H. Hoang, V.~Mateu, and I.~W. Stewart, {\it {Thrust
  at $N^{3}LL$ with Power Corrections and a Precision Global Fit for
  $\alpha_{s}(mZ)$}},  {\em Phys. Rev. D} {\bf 83} (2011) 074021,
  [\href{http://arxiv.org/abs/1006.3080}{{\tt arXiv:1006.3080}}].

\bibitem{Becher:2008cf}
T.~Becher and M.~D. Schwartz, {\it {A precise determination of $\alpha_s$ from
  LEP thrust data using effective field theory}},  {\em JHEP} {\bf 07} (2008)
  034, [\href{http://arxiv.org/abs/0803.0342}{{\tt arXiv:0803.0342}}].

\bibitem{Hoang:2015hka}
A.~H. Hoang, D.~W. Kolodrubetz, V.~Mateu, and I.~W. Stewart, {\it {Precise
  determination of $\alpha_s$ from the $C$-parameter distribution}},  {\em
  Phys. Rev. D} {\bf 91} (2015), no.~9 094018,
  [\href{http://arxiv.org/abs/1501.04111}{{\tt arXiv:1501.04111}}].

\bibitem{Hoang:2014wka}
A.~H. Hoang, D.~W. Kolodrubetz, V.~Mateu, and I.~W. Stewart, {\it
  {$C$-parameter distribution at N$^3$LL' including power corrections}},  {\em
  Phys. Rev. D} {\bf 91} (2015), no.~9 094017,
  [\href{http://arxiv.org/abs/1411.6633}{{\tt arXiv:1411.6633}}].

\bibitem{Benitez:2024nav}
M.~A. Benitez, A.~H. Hoang, V.~Mateu, I.~W. Stewart, and G.~Vita, {\it {On
  determining {\ensuremath{\alpha}}$_{s}$(m$_{Z}$) from dijets in
  e$^{+}$e$^{-}$ thrust}},  {\em JHEP} {\bf 07} (2025) 249,
  [\href{http://arxiv.org/abs/2412.15164}{{\tt arXiv:2412.15164}}].

\bibitem{Benitez:2025vsp}
M.~A. Benitez, A.~Bhattacharya, A.~H. Hoang, V.~Mateu, M.~D. Schwartz, I.~W.
  Stewart, and X.~Zhang, {\it {A Precise Determination of $\alpha_s$ from the
  Heavy Jet Mass Distribution}},  \href{http://arxiv.org/abs/2502.12253}{{\tt
  arXiv:2502.12253}}.

\bibitem{Huston:2023ofk}
J.~Huston, K.~Rabbertz, and G.~Zanderighi, {\it {Quantum Chromodynamics}},
  \href{http://arxiv.org/abs/2312.14015}{{\tt arXiv:2312.14015}}.

\bibitem{dEnterria:2022hzv}
D.~d'Enterria et~al., {\it {The strong coupling constant: state of the art and
  the decade ahead}},  {\em J. Phys. G} {\bf 51} (2024), no.~9 090501,
  [\href{http://arxiv.org/abs/2203.08271}{{\tt arXiv:2203.08271}}].

\bibitem{CMS:2024mlf}
{\bf CMS} Collaboration, A.~Hayrapetyan et~al., {\it {Measurement of Energy
  Correlators inside Jets and Determination of the Strong Coupling
  {\ensuremath{\alpha}}S(mZ)}},  {\em Phys. Rev. Lett.} {\bf 133} (2024), no.~7
  071903, [\href{http://arxiv.org/abs/2402.13864}{{\tt arXiv:2402.13864}}].

\bibitem{FCC:2018evy}
{\bf FCC} Collaboration, A.~Abada et~al., {\it {FCC-ee: The Lepton Collider}:
  {Future Circular Collider Conceptual Design Report Volume 2}},  {\em Eur.
  Phys. J. ST} {\bf 228} (2019), no.~2 261--623.

\bibitem{CEPCStudyGroup:2018ghi}
{\bf CEPC Study Group} Collaboration, M.~Dong et~al., {\it {CEPC Conceptual
  Design Report: Volume 2 - Physics \& Detector}},
  \href{http://arxiv.org/abs/1811.10545}{{\tt arXiv:1811.10545}}.

\bibitem{HaoTalkSCET}
H.~Chen, P.~F. Monni, Z.~Xu, and H.~X. Zhu, {\it Perturbative evolution of
  hadronization effects in energy correlators},  {\em talk by Hao Chen at SCET
  2024} (2024).

\bibitem{Lee:2024esz}
K.~Lee, A.~Pathak, I.~W. Stewart, and Z.~Sun, {\it {Nonperturbative Effects in
  Energy Correlators: From Characterizing Confinement Transition to Improving
  {\ensuremath{\alpha}}s Extraction}},  {\em Phys. Rev. Lett.} {\bf 133}
  (2024), no.~23 231902, [\href{http://arxiv.org/abs/2405.19396}{{\tt
  arXiv:2405.19396}}].

\bibitem{Chen:2024nyc}
H.~Chen, P.~F. Monni, Z.~Xu, and H.~X. Zhu, {\it {Scaling Violation in Power
  Corrections to Energy Correlators from the Light-Ray Operator Product
  Expansion}},  {\em Phys. Rev. Lett.} {\bf 133} (2024), no.~23 231901,
  [\href{http://arxiv.org/abs/2406.06668}{{\tt arXiv:2406.06668}}].

\bibitem{Lee:2006fn}
C.~Lee and G.~F. Sterman, {\it {Universality of nonperturbative effects in
  event shapes}},  {\em eConf} {\bf C0601121} (2006) A001,
  [\href{http://arxiv.org/abs/hep-ph/0603066}{{\tt hep-ph/0603066}}].

\bibitem{Budhraja:2024tev}
A.~Budhraja, H.~Chen, and W.~J. Waalewijn, {\it {{\ensuremath{\nu}}-point
  energy correletors with FastEEC: Small-x physics from LHC jets}},  {\em Phys.
  Lett. B} {\bf 861} (2025) 139239,
  [\href{http://arxiv.org/abs/2409.12235}{{\tt arXiv:2409.12235}}].

\bibitem{Somogyi:2006da}
G.~Somogyi, Z.~Trocsanyi, and V.~Del~Duca, {\it {A Subtraction scheme for
  computing QCD jet cross sections at NNLO: Regularization of doubly-real
  emissions}},  {\em JHEP} {\bf 01} (2007) 070,
  [\href{http://arxiv.org/abs/hep-ph/0609042}{{\tt hep-ph/0609042}}].

\bibitem{Somogyi:2006db}
G.~Somogyi and Z.~Trocsanyi, {\it {A Subtraction scheme for computing QCD jet
  cross sections at NNLO: Regularization of real-virtual emission}},  {\em
  JHEP} {\bf 01} (2007) 052, [\href{http://arxiv.org/abs/hep-ph/0609043}{{\tt
  hep-ph/0609043}}].

\bibitem{Aglietti:2008fe}
U.~Aglietti, V.~Del~Duca, C.~Duhr, G.~Somogyi, and Z.~Trocsanyi, {\it {Analytic
  integration of real-virtual counterterms in NNLO jet cross sections. I.}},
  {\em JHEP} {\bf 09} (2008) 107, [\href{http://arxiv.org/abs/0807.0514}{{\tt
  arXiv:0807.0514}}].

\bibitem{Catani:1996jh}
S.~Catani and M.~H. Seymour, {\it {The Dipole formalism for the calculation of
  QCD jet cross-sections at next-to-leading order}},  {\em Phys. Lett. B} {\bf
  378} (1996) 287--301, [\href{http://arxiv.org/abs/hep-ph/9602277}{{\tt
  hep-ph/9602277}}].

\bibitem{Catani:1996vz}
S.~Catani and M.~H. Seymour, {\it {A General algorithm for calculating jet
  cross-sections in NLO QCD}},  {\em Nucl. Phys. B} {\bf 485} (1997) 291--419,
  [\href{http://arxiv.org/abs/hep-ph/9605323}{{\tt hep-ph/9605323}}]. [Erratum:
  Nucl.Phys.B 510, 503--504 (1998)].

\bibitem{Gehrmann-DeRidder:2014hxk}
A.~Gehrmann-De~Ridder, T.~Gehrmann, E.~W.~N. Glover, and G.~Heinrich, {\it
  {EERAD3: Event shapes and jet rates in electron-positron annihilation at
  order $\alpha_s^3$}},  {\em Comput. Phys. Commun.} {\bf 185} (2014) 3331,
  [\href{http://arxiv.org/abs/1402.4140}{{\tt arXiv:1402.4140}}].

\bibitem{Aveleira:2025svg}
B.~C. Aveleira, A.~Gehrmann-De~Ridder, T.~Gehrmann, N.~Glover, G.~Heinrich, and
  C.~T. Preuss, {\it {EERAD3 version 2: QCD corrections in hadronic
  colour-singlet decays}},  {\em SciPost Phys. Codeb.} {\bf 59} (2025) 1,
  [\href{http://arxiv.org/abs/2503.20610}{{\tt arXiv:2503.20610}}].

\bibitem{Rijken:1996vr}
P.~J. Rijken and W.~L. van Neerven, {\it {O (alpha-s**2) contributions to the
  longitudinal fragmentation function in e+ e- annihilation}},  {\em Phys.
  Lett. B} {\bf 386} (1996) 422--428,
  [\href{http://arxiv.org/abs/hep-ph/9604436}{{\tt hep-ph/9604436}}].

\bibitem{Rijken:1996ns}
P.~J. Rijken and W.~L. van Neerven, {\it {Higher order QCD corrections to the
  transverse and longitudinal fragmentation functions in electron - positron
  annihilation}},  {\em Nucl. Phys. B} {\bf 487} (1997) 233--282,
  [\href{http://arxiv.org/abs/hep-ph/9609377}{{\tt hep-ph/9609377}}].

\bibitem{Rijken:1996npa}
P.~J. Rijken and W.~L. van Neerven, {\it {O (alpha-s**2) contributions to the
  asymmetric fragmentation function in e+ e- annihilation}},  {\em Phys. Lett.
  B} {\bf 392} (1997) 207--215,
  [\href{http://arxiv.org/abs/hep-ph/9609379}{{\tt hep-ph/9609379}}].

\bibitem{Mitov:2006wy}
A.~Mitov and S.-O. Moch, {\it {QCD Corrections to Semi-Inclusive Hadron
  Production in Electron-Positron Annihilation at Two Loops}},  {\em Nucl.
  Phys. B} {\bf 751} (2006) 18--52,
  [\href{http://arxiv.org/abs/hep-ph/0604160}{{\tt hep-ph/0604160}}].

\bibitem{Mitov:2006ic}
A.~Mitov, S.~Moch, and A.~Vogt, {\it {Next-to-Next-to-Leading Order Evolution
  of Non-Singlet Fragmentation Functions}},  {\em Phys. Lett. B} {\bf 638}
  (2006) 61--67, [\href{http://arxiv.org/abs/hep-ph/0604053}{{\tt
  hep-ph/0604053}}].

\bibitem{Moch:2007tx}
S.~Moch and A.~Vogt, {\it {On third-order timelike splitting functions and
  top-mediated Higgs decay into hadrons}},  {\em Phys. Lett. B} {\bf 659}
  (2008) 290--296, [\href{http://arxiv.org/abs/0709.3899}{{\tt
  arXiv:0709.3899}}].

\bibitem{Almasy:2011eq}
A.~A. Almasy, S.~Moch, and A.~Vogt, {\it {On the Next-to-Next-to-Leading Order
  Evolution of Flavour-Singlet Fragmentation Functions}},  {\em Nucl. Phys. B}
  {\bf 854} (2012) 133--152, [\href{http://arxiv.org/abs/1107.2263}{{\tt
  arXiv:1107.2263}}].

\bibitem{Chen:2020uvt}
H.~Chen, T.-Z. Yang, H.~X. Zhu, and Y.~J. Zhu, {\it {Analytic Continuation and
  Reciprocity Relation for Collinear Splitting in QCD}},  {\em Chin. Phys. C}
  {\bf 45} (2021), no.~4 043101, [\href{http://arxiv.org/abs/2006.10534}{{\tt
  arXiv:2006.10534}}].

\bibitem{Gehrmann:2023cqm}
T.~Gehrmann, A.~von Manteuffel, V.~Sotnikov, and T.-Z. Yang, {\it {Complete $
  {N}_f^2 $ contributions to four-loop pure-singlet splitting functions}},
  {\em JHEP} {\bf 01} (2024) 029, [\href{http://arxiv.org/abs/2308.07958}{{\tt
  arXiv:2308.07958}}].

\bibitem{Falcioni:2023luc}
G.~Falcioni, F.~Herzog, S.~Moch, and A.~Vogt, {\it {Four-loop splitting
  functions in QCD {\textendash} The quark-quark case}},  {\em Phys. Lett. B}
  {\bf 842} (2023) 137944, [\href{http://arxiv.org/abs/2302.07593}{{\tt
  arXiv:2302.07593}}].

\bibitem{Falcioni:2023vqq}
G.~Falcioni, F.~Herzog, S.~Moch, and A.~Vogt, {\it {Four-loop splitting
  functions in QCD {\textendash} The gluon-to-quark case}},  {\em Phys. Lett.
  B} {\bf 846} (2023) 138215, [\href{http://arxiv.org/abs/2307.04158}{{\tt
  arXiv:2307.04158}}].

\bibitem{Falcioni:2024xyt}
G.~Falcioni, F.~Herzog, S.~Moch, A.~Pelloni, and A.~Vogt, {\it {Four-loop
  splitting functions in QCD {\textendash} The quark-to-gluon case}},  {\em
  Phys. Lett. B} {\bf 856} (2024) 138906,
  [\href{http://arxiv.org/abs/2404.09701}{{\tt arXiv:2404.09701}}].

\bibitem{Falcioni:2024qpd}
G.~Falcioni, F.~Herzog, S.~Moch, A.~Pelloni, and A.~Vogt, {\it {Four-loop
  splitting functions in QCD {\textendash} the gluon-gluon case
  {\textendash}}},  {\em Phys. Lett. B} {\bf 860} (2025) 139194,
  [\href{http://arxiv.org/abs/2410.08089}{{\tt arXiv:2410.08089}}].

\bibitem{Falcioni:2025hfz}
G.~Falcioni, F.~Herzog, S.~Moch, A.~Pelloni, and A.~Vogt, {\it {Additional
  results on the four-loop flavour-singlet splitting functions in QCD}},  {\em
  Phys. Lett. B} {\bf 875} (2026) 140278,
  [\href{http://arxiv.org/abs/2512.10783}{{\tt arXiv:2512.10783}}].

\bibitem{NOGUEIRA1993279}
P.~Nogueira, {\it Automatic feynman graph generation},  {\em Journal of
  Computational Physics} {\bf 105} (1993), no.~2 279--289.

\bibitem{Vermaseren:2000nd}
J.~A.~M. Vermaseren, {\it {New features of FORM}},
  \href{http://arxiv.org/abs/math-ph/0010025}{{\tt math-ph/0010025}}.

\bibitem{Kuipers:2012rf}
J.~Kuipers, T.~Ueda, J.~A.~M. Vermaseren, and J.~Vollinga, {\it {FORM version
  4.0}},  {\em Comput. Phys. Commun.} {\bf 184} (2013) 1453--1467,
  [\href{http://arxiv.org/abs/1203.6543}{{\tt arXiv:1203.6543}}].

\bibitem{Ruijl:2017dtg}
B.~Ruijl, T.~Ueda, and J.~Vermaseren, {\it {FORM version 4.2}},
  \href{http://arxiv.org/abs/1707.06453}{{\tt arXiv:1707.06453}}.

\bibitem{vanRitbergen:1998pn}
T.~van Ritbergen, A.~N. Schellekens, and J.~A.~M. Vermaseren, {\it {Group
  theory factors for Feynman diagrams}},  {\em Int. J. Mod. Phys. A} {\bf 14}
  (1999) 41--96, [\href{http://arxiv.org/abs/hep-ph/9802376}{{\tt
  hep-ph/9802376}}].

\bibitem{Mertig:1990an}
R.~Mertig, M.~Bohm, and A.~Denner, {\it {FEYN CALC: Computer algebraic
  calculation of Feynman amplitudes}},  {\em Comput. Phys. Commun.} {\bf 64}
  (1991) 345--359.

\bibitem{Shtabovenko:2016sxi}
V.~Shtabovenko, R.~Mertig, and F.~Orellana, {\it {New Developments in FeynCalc
  9.0}},  {\em Comput. Phys. Commun.} {\bf 207} (2016) 432--444,
  [\href{http://arxiv.org/abs/1601.01167}{{\tt arXiv:1601.01167}}].

\bibitem{Shtabovenko:2020gxv}
V.~Shtabovenko, R.~Mertig, and F.~Orellana, {\it {FeynCalc 9.3: New features
  and improvements}},  {\em Comput. Phys. Commun.} {\bf 256} (2020) 107478,
  [\href{http://arxiv.org/abs/2001.04407}{{\tt arXiv:2001.04407}}].

\bibitem{Shtabovenko:2023idz}
V.~Shtabovenko, R.~Mertig, and F.~Orellana, {\it {FeynCalc 10: Do multiloop
  integrals dream of computer codes?}},
  \href{http://arxiv.org/abs/2312.14089}{{\tt arXiv:2312.14089}}.

\bibitem{Lee:2012cn}
R.~N. Lee, {\it {Presenting LiteRed: a tool for the Loop InTEgrals REDuction}},
   \href{http://arxiv.org/abs/1212.2685}{{\tt arXiv:1212.2685}}.

\bibitem{Lee:2013mka}
R.~N. Lee, {\it {LiteRed 1.4: a powerful tool for reduction of multiloop
  integrals}},  {\em J. Phys. Conf. Ser.} {\bf 523} (2014) 012059,
  [\href{http://arxiv.org/abs/1310.1145}{{\tt arXiv:1310.1145}}].

\bibitem{Smirnov:2019qkx}
A.~V. Smirnov and F.~S. Chukharev, {\it {FIRE6: Feynman Integral REduction with
  modular arithmetic}},  {\em Comput. Phys. Commun.} {\bf 247} (2020) 106877,
  [\href{http://arxiv.org/abs/1901.07808}{{\tt arXiv:1901.07808}}].

\bibitem{Meyer:2017joq}
C.~Meyer, {\it {Algorithmic transformation of multi-loop master integrals to a
  canonical basis with CANONICA}},  {\em Comput. Phys. Commun.} {\bf 222}
  (2018) 295--312, [\href{http://arxiv.org/abs/1705.06252}{{\tt
  arXiv:1705.06252}}].

\bibitem{Lee:2014ioa}
R.~N. Lee, {\it {Reducing differential equations for multiloop master
  integrals}},  {\em JHEP} {\bf 04} (2015) 108,
  [\href{http://arxiv.org/abs/1411.0911}{{\tt arXiv:1411.0911}}].

\bibitem{Lee:2020zfb}
R.~N. Lee, {\it {Libra: A package for transformation of differential systems
  for multiloop integrals}},  {\em Comput. Phys. Commun.} {\bf 267} (2021)
  108058, [\href{http://arxiv.org/abs/2012.00279}{{\tt arXiv:2012.00279}}].

\bibitem{Campbell:1997hg}
J.~M. Campbell and E.~W.~N. Glover, {\it {Double unresolved approximations to
  multiparton scattering amplitudes}},  {\em Nucl. Phys. B} {\bf 527} (1998)
  264--288, [\href{http://arxiv.org/abs/hep-ph/9710255}{{\tt hep-ph/9710255}}].

\bibitem{Catani:1998nv}
S.~Catani and M.~Grazzini, {\it {Collinear factorization and splitting
  functions for next-to-next-to-leading order QCD calculations}},  {\em Phys.
  Lett. B} {\bf 446} (1999) 143--152,
  [\href{http://arxiv.org/abs/hep-ph/9810389}{{\tt hep-ph/9810389}}].

\bibitem{Ritzmann:2014mka}
M.~Ritzmann and W.~J. Waalewijn, {\it {Fragmentation in Jets at NNLO}},  {\em
  Phys. Rev. D} {\bf 90} (2014), no.~5 054029,
  [\href{http://arxiv.org/abs/1407.3272}{{\tt arXiv:1407.3272}}].

\bibitem{Gong:2025jqi}
J.~Gong, A.~Pokraka, K.~Yan, and X.~Zhang, {\it {Toward the Analytic Bootstrap
  of Energy Correlators}},  \href{http://arxiv.org/abs/2509.22782}{{\tt
  arXiv:2509.22782}}.

\bibitem{Hahn:2004fe}
T.~Hahn, {\it {CUBA: A Library for multidimensional numerical integration}},
  {\em Comput. Phys. Commun.} {\bf 168} (2005) 78--95,
  [\href{http://arxiv.org/abs/hep-ph/0404043}{{\tt hep-ph/0404043}}].

\bibitem{Schindler:2023cww}
S.~T. Schindler, I.~W. Stewart, and Z.~Sun, {\it {Renormalons in the
  energy-energy correlator}},  {\em JHEP} {\bf 10} (2023) 187,
  [\href{http://arxiv.org/abs/2305.19311}{{\tt arXiv:2305.19311}}]. [Erratum:
  JHEP 10, 175 (2024)].

\bibitem{Gao:2026xuq}
A.~Gao, K.~Lee, and X.~Zhang, {\it {Precision Jet Substructure of Boosted Boson
  Decays with Energy Correlators}},
  \href{http://arxiv.org/abs/2601.20933}{{\tt arXiv:2601.20933}}.

\bibitem{Holguin:2026vld}
J.~Holguin, I.~Moult, A.~Pathak, M.~Procura, and S.~Sule, {\it {High precision
  heavy-boson-jet substructure with energy correlators}},
  \href{http://arxiv.org/abs/2601.20923}{{\tt arXiv:2601.20923}}.

\bibitem{Bierlich:2022pfr}
C.~Bierlich et~al., {\it {A comprehensive guide to the physics and usage of
  PYTHIA 8.3}},  {\em SciPost Phys. Codeb.} {\bf 2022} (2022) 8,
  [\href{http://arxiv.org/abs/2203.11601}{{\tt arXiv:2203.11601}}].

\bibitem{Bellm:2015jjp}
J.~Bellm et~al., {\it {Herwig 7.0/Herwig++ 3.0 release note}},  {\em Eur. Phys.
  J. C} {\bf 76} (2016), no.~4 196,
  [\href{http://arxiv.org/abs/1512.01178}{{\tt arXiv:1512.01178}}].

\bibitem{Nachtmann:1973mr}
O.~Nachtmann, {\it {Positivity constraints for anomalous dimensions}},  {\em
  Nucl. Phys. B} {\bf 63} (1973) 237--247.

\bibitem{Tackmann:2024kci}
F.~J. Tackmann, {\it {Beyond scale variations: perturbative theory
  uncertainties from nuisance parameters}},  {\em JHEP} {\bf 08} (2025) 098,
  [\href{http://arxiv.org/abs/2411.18606}{{\tt arXiv:2411.18606}}].

\bibitem{ALEPH:2013dgf}
{\bf ALEPH, DELPHI, L3, OPAL, LEP Electroweak} Collaboration, S.~Schael et~al.,
  {\it {Electroweak Measurements in Electron-Positron Collisions at
  W-Boson-Pair Energies at LEP}},  {\em Phys. Rept.} {\bf 532} (2013) 119--244,
  [\href{http://arxiv.org/abs/1302.3415}{{\tt arXiv:1302.3415}}].

\bibitem{Altarelli:1996ww}
{\it {Physics at LEP2: Vol.2}},  {\em CERN Yellow Reports} (2, 1996).

\bibitem{Kang:2016mcy}
Z.-B. Kang, F.~Ringer, and I.~Vitev, {\it {The semi-inclusive jet function in
  SCET and small radius resummation for inclusive jet production}},  {\em JHEP}
  {\bf 10} (2016) 125, [\href{http://arxiv.org/abs/1606.06732}{{\tt
  arXiv:1606.06732}}].

\bibitem{Lee:2024icn}
K.~Lee, I.~Moult, and X.~Zhang, {\it {Revisiting single inclusive jet
  production: timelike factorization and reciprocity}},  {\em JHEP} {\bf 05}
  (2025) 129, [\href{http://arxiv.org/abs/2409.19045}{{\tt arXiv:2409.19045}}].

\bibitem{Generet:2025vth}
T.~Generet, K.~Lee, I.~Moult, R.~Poncelet, and X.~Zhang, {\it {Small radius
  inclusive jet production at the LHC through NNLO+NNLL}},  {\em JHEP} {\bf 08}
  (2025) 015, [\href{http://arxiv.org/abs/2503.21866}{{\tt arXiv:2503.21866}}].

\bibitem{Chang:2013rca}
H.-M. Chang, M.~Procura, J.~Thaler, and W.~J. Waalewijn, {\it {Calculating
  Track-Based Observables for the LHC}},  {\em Phys. Rev. Lett.} {\bf 111}
  (2013) 102002, [\href{http://arxiv.org/abs/1303.6637}{{\tt
  arXiv:1303.6637}}].

\bibitem{Li:2021zcf}
Y.~Li, I.~Moult, S.~S. van Velzen, W.~J. Waalewijn, and H.~X. Zhu, {\it
  {Extending Precision Perturbative QCD with Track Functions}},  {\em Phys.
  Rev. Lett.} {\bf 128} (2022), no.~18 182001,
  [\href{http://arxiv.org/abs/2108.01674}{{\tt arXiv:2108.01674}}].

\bibitem{Chen:2022muj}
H.~Chen, M.~Jaarsma, Y.~Li, I.~Moult, W.~J. Waalewijn, and H.~X. Zhu, {\it
  {Collinear parton dynamics beyond Dokshitzer-Gribov-Lipatov-Altarelli-Parisi
  framework}},  {\em Phys. Rev. D} {\bf 111} (2025), no.~7 076021,
  [\href{http://arxiv.org/abs/2210.10061}{{\tt arXiv:2210.10061}}].

\bibitem{Jaarsma:2023ell}
M.~Jaarsma, Y.~Li, I.~Moult, W.~J. Waalewijn, and H.~X. Zhu, {\it {Energy
  correlators on tracks: resummation and non-perturbative effects}},  {\em
  JHEP} {\bf 12} (2023) 087, [\href{http://arxiv.org/abs/2307.15739}{{\tt
  arXiv:2307.15739}}].

\bibitem{Lee:2023npz}
K.~Lee and I.~Moult, {\it {Energy Correlators Taking Charge}},
  \href{http://arxiv.org/abs/2308.00746}{{\tt arXiv:2308.00746}}.

\bibitem{Lee:2023tkr}
K.~Lee and I.~Moult, {\it {Joint Track Functions: Expanding the Space of
  Calculable Correlations at Colliders}},
  \href{http://arxiv.org/abs/2308.01332}{{\tt arXiv:2308.01332}}.

\end{thebibliography}\endgroup
\bibliographystyle{JHEP}
\end{document}